\documentstyle[aps,epsfig,amstex,amssymb,psfrag]{revtex}
\newcommand{\mel}[3]{\langle #1 | #2 | #3 \rangle } 
\newcommand{\ket}[1]{| #1 \rangle } 
\newcommand{\bra}[1]{\langle #1 | }

\newcommand{\bfrho}{\mbox{\boldmath $\rho$}}
\newcommand{\tr}{\mathop{\rm tr}\nolimits}

\begin{document}
\title{Strong coupling theory for tunneling and vibrational 
relaxation in driven bistable systems}
\author{M. Thorwart$^{1,2}$, M. Grifoni$^{2}$,  
       and 
       P. H\"anggi$^1$}
\address{
              $^1$Institut f\"ur Physik,
              Universit\"at  Augsburg, Universit\"atsstr.\ 1, 86135 Augsburg,
              Germany\\
 $^2$Delft University of Technology, Lorentzweg 1, 2628 CJ Delft, 
 The Netherlands\\   
 }
\date{Date: \today}
\maketitle
\begin{abstract}
A study of 
the dynamics of a tunneling particle in a driven bistable 
potential which is moderately-to-strongly coupled to a bath is presented. 
Upon restricting the system dynamics to the Hilbert 
space spanned by the $M\/$ lowest energy eigenstates of the bare static potential, 
a set of coupled non-Markovian master equations for the 
diagonal elements of the reduced density  
matrix, within the {\em discrete variale representation\/}, is derived. 
The resulting dynamics is in good agreement with predictions of 
{\em ab-initio\/} real-time path integral simulations. Numerous 
results, analytical as  well as numerical, for the {\em quantum 
relaxation rate\/} and for the {\em asymptotic populations\/} are presented. 
Our method is particularly convenient to investigate the case of shallow,  
time-dependent 
potential barriers and moderate-to-strong damping, where both a semi-classical 
and a Redfield-type approach are inappropriate.
\\[10mm]
\end{abstract}
%
 
%
\tableofcontents

\newpage
\section{Introduction}
The sensitivity 
of tunneling to the influence of the environment has been in the focus 
of intense research over the last 
 years \cite{Leggett8187,HanggiRMP90,Weiss99,VCHBuch,Grifoni98}. 
 A 
popular model for the investigation of 
tunneling processes is  a double-well potential with an energy 
barrier that separates two energetically degenerate minima. In an idealized system, 
the barrier can be coherently traversed by a quantum mechanical particle 
({\em coherent tunneling\/}).   
A real physical system, however, experiences the influence of the surrounding 
``outer world''. This coupling disturbs the coherent tunneling process and 
it constitutes the origin of decoherence and dissipation in the quantum system.  
To model the dissipative influence, 
the environment is commonly described as an 
ensemble of harmonic oscillators  ({\em heat bath, reservoir\/}) 
being at thermal equilibrium at temperature $T$. 
A bilinear coupling between the quantum system and the bath mimics 
phenomenologically the interaction of the system with the ``rest of the world''.   
By this method, the quantum mechanical 
analogue of the generalized Langevin equation can be derived. 

The spectrum of the uncoupled symmetric bistable potential 
consists of a ladder of doublets being pairs of energetically nearly degenerate 
energy eigenstates. The degeneracy is lifted by the tunneling splittings within the 
single doublets. The doublets themselves are separated by large interdoublet 
energy gaps which are of the 
order of the related characteristic system frequency scale, the latter are 
 generally   
orders of magnitude larger than the tunneling splittings. 

By now, two different situations have been in the center of detailed  
investigations on the dissipative tunneling dynamics in a bistable 
potential: (i) 
On the one hand, one considers the regime of low 
temperatures, i.e., $k_{\rm B} T$ is of the order of the 
energy splitting of the lowest tunneling doublet. 
A common approach to simplify the {\em spatially 
continuous\/} dynamics consists then in restricting the problem to 
 the two lowest energy eigenstates, being the solely  
%
%
%
significantly thermally 
populated 
in this deep quantum regime. 
Coupling the 
two-level system to a bosonic bath of harmonic oscillators 
leads to the prominent {\em spin-boson problem\/} 
\cite{Leggett8187,Weiss99,Grifoni98}. 
(ii) 
On the other hand, the starting point is classical rate theory. 
{\em Semi-classical\/}  tunneling 
corrections to the relaxation rate are calculated by use of various 
 instanton techniques 
\cite{HanggiRMP90}. This formalism is applicable when the quantized energy levels 
lie very dense below the barrier, i.e., in cases when the energy 
barrier is large 
compared to the characteristic level splitting of the quantum system.
Moreover, a {\em local equilibrium} is 
required, restricting this approach only to  
time-{\em in\/}dependent systems. By complex-time 
path integral techniques, the free energy is calculated in a semiclassical 
steepest-descent method. This leads to the dissipative bounce solution 
which  in turn determines the semiclassical decay rate.

Modern experimental developments have paved the way to study the influence of 
time-dependent external driving forces like a laser beam or an rf-field.  
Such time-dependent driving fields have most interesting implications 
for quantum systems like, for instance, the effect of {\em coherent destruction 
of tunneling\/} 
\cite{Grossmann91,Oelschlaegel93},
 the effect of quantum stochastic resonance 
\cite{Loefstedt94,Makarov95,Grifoni96a,Thorwart97a,Pareek97,Thorwart98a,HanggiSRRMP,Goychuk99,Wellens00}, or the occurrence of quantum steps 
in hysteresis loops \cite{Thorwart98a,Thorwart98b},  
   to name but a few (for recent reviews, see 
\cite{VCHBuch,Grifoni98,HanggiSRRMP,ChemPhysDriv97}). 
Such driving fields may also be used to control and reduce decoherence 
in open quantum systems \cite{Viola99,Thorwart00}. 

The present work deals with tunneling processes in a 
time-dependent bistable potential 
in a temperature regime 
where the two-level approximation ({\em spin-boson regime\/}) is invalid.  
Likewise, the (possibly) strong time-dependent external fields 
prevents the use of {\em semiclassical\/} methods. 
%
Our analysis therefore bridges those two well established limiting 
regimes in quantum rate theory.

With this objective in mind, we release 
the restriction of the bistable potential to its two 
lowest energy eigenstates and extend the model to include 
more energy eigenstates which are populated at higher temperatures. 
This implies an interesting consequence: Since the energy splittings of 
the higher doublets are larger, tunneling becomes more favorable via the 
higher doublets. However, for the temperature being too large, tunneling 
is again hampered due to the decoherent influence of the environment. 
This interplay among tunneling, vibrational relaxation 
(i.e., transitions {\em between\/} the doublets) and thermal effects 
leads to a rich and complex dynamics. 

The specific problem we tackle is the following: 
Let us consider a quantum particle which is initially localized 
in one of the two wells of a double-well potential.  
What is then the {\em rate\/} with which 
the probability of finding the particle in this well 
 decays 
 in presence of an Ohmic-like environment? In addition, what are the asymptotic 
 well populations? 
An additional manipulation of the potential barrier, 
i.e., a static bias or a time-dependent harmonic driving 
may be applied. In this work, we provide an analytic method to solve this 
non-trivial problem - also in presence of a time-dependent driving field - 
in a very general manner. We  restrict  ourselves neither 
to a large semi-classical potential barrier, nor to a weak system-bath 
interaction  nor to  weak driving fields.  
Our analysis is based  on the real-time path integral 
technique which uses the Feynman-Vernon formulation as a starting basis. 
 By treating the bath induced correlations between quantum paths  
 within a generalized  non-interacting cluster  
approximation, a generalized master equation for the 
diagonal elements of the reduced density matrix is derived. It turns out that 
the approximation is  appropriate in the regime 
of moderate temperature and/or moderate system-bath coupling. 
  A further 
simplification of the integro-differential equation leads to a 
Markovian approximated master equation whose rate coefficients are 
obtained in the form of closed analytical expressions. 
By comparing the results of the full generalized master equation 
versus the Markovian approximated master equation and versus the numerical 
quasiadiabatic propagator path integral algorithm \cite{QUAPI}, 
we conclude that the analytical approximation permits 
correct predictions for the decay process out of the 
initially populated potential well. 
The rate governing the long-time 
dynamics of the decay is obtained as the {\em smallest eigenvalue\/} 
of the  
matrix of the (time averaged) rate coefficients. 
The dependence of this quantum relaxation rate and of the asymptotic 
population of the metastable well 
on the various physical parameters is investigated in detail.
 
We stress that the developed method is {\em not\/} restricted 
to this specific problem but can be applied to many different other 
physical situations where a potential with a discrete energy 
spectrum can be assumed. 
A short summary of this present work has been published in Ref.\ 
\cite{Thorwart00a}. 

Before we proceed, we motivate that the stated problem is not of formal 
academic nature but, in contrast, has several applications to real 
physical systems. 
For that purpose, we have collected numerous experimental works 
in the following subsection \ref{subsec.exp}. In the subsequent 
subsection \ref{subsec.theo} 
we briefly review the few existing theoretical works and discuss 
some of their 
shortcomings and inconsistencies which we attempt to 
 overcome by our techniques. 
The rest of the paper is organized as follows: In Section \ref{sec.model}   
our specific model 
is introduced. The succeeding Section \ref{sec.realtimepathint} is devoted 
to the derivation of the dissipative real-time path integral 
which is cast in the 
discrete variable representation (DVR), i.e., the eigenbasis of that 
system operator 
which couples to the reservoir. The example of the 
double-doublet system illustrates this transformation in Section 
\ref{subsec.dds}. In Section  \ref{sec.gnica} we introduce an approximation  
to the so far exact real-time path integral expressions. This 
approximative treatment of the bath induced path correlations allows 
for the derivation of a generalized master equation (GME). This 
is shown in Section  \ref{sec.gme}, where also the lowest order 
expressions for the 
integral kernels of the GME are given. In Section \ref{sec.rate} 
 we extract the leading rate for the decay out of one of the 
 two potential wells. This is possible if one applies an additional 
 Markovian approximation to the GME. A detailed study of the dependence 
 of the quantum relaxation rate on the various model parameter is 
 put forward in Section \ref{sec.results}. Moreover, an investigation of the 
 asymptotic well population is presented. 
%
%
 Finally, our conclusions together with an outlook 
 are presented in Section  \ref{sec.conclusio}. 
\subsection{Experiments}
\label{subsec.exp}
Several experiments where dissipative multilevel systems are involved 
have been performed in many different physical systems. We report on 
four timely examples to motivate the importance and the need for a 
consistent and general theory for the above stated problem. 

The first set of experiments deals with quantum tunneling of magnetization 
in nanomagnets \cite{MacroTunnel95}.  
A macroscopic sample of molecular magnets consists of a large number 
(typically $10^4 -10^{11}$) of 
chemically identical magnetic clusters of the same magnetic size. 
They are regularly arranged on a crystal lattice. 
The single molecules have usually a large spin quantum number, typically 
$S \simeq 10$. Experiments 
(see below) indicate a strong uniaxial magnetocrystalline anisotropy.  
It favors a doubly-degenerate spin alignment along the c-axis of the 
crystal, $m_S=\pm S$, and generates  an energy 
barrier  
for the reversal of magnetization. This implies two-fold 
degenerate excited states corresponding to the spin-projections 
$m_s=\pm (S-1), \pm (S-2), ..., 0$ in a double-well potential 
\cite{vanHemmen86}. 
At sufficiently low temperatures, the spins can tunnel through the 
anisotropy barrier. 
Two such materials are currently studied in detail: The first is 
referred 
to as $\mathrm{Mn_{12}}$-acetate. It possesses a tunneling barrier of 
 $\Delta U / k_{\rm B} \simeq 62 \mathrm{K}$ 
($k_{\rm B}$ denotes the Boltzmann constant). 
Resonant tunneling 
of magnetization reveals itself as quantum steps in hysteresis loops 
which go along with maxima in the relaxation rate for specific 
values of an external magnetic field 
\cite{Mn12}. 
The second candidate is 
known as 
$\mathrm{Fe_8}$ and has the advantage that the anisotropy 
barrier is approximately three times smaller than in  $\mathrm{Mn_{12}}$ 
($\Delta U / k_{\rm B} \simeq 22 \mathrm{K}$). This property 
enhances the 
observed effect by several orders of magnitude as compared to the case 
with $\mathrm{Mn_{12}}$. 
For the $\mathrm{Fe_8}$ samples several experiments on quantum tunneling 
of magnetization have been reported as well 
\cite{Fe8,Wernsdorfer00}. Especially interesting for us are the measurements 
by Wernsdorfer {\em et al.\/} \cite{Wernsdorfer00};  
those are performed  at non-adiabatic driving fields and at 
temperatures where many doublets contribute to the 
dynamics.  

%
%
A second class of experiments addresses tunneling of the magnetic flux 
in superconducting quantum interference devices (SQUIDs) 
\cite{Devoret88,Han91,Rouse95,Han96,Silvestrini97,Han00,Friedmann00,Mooij99,vanderWal00}. 
The dynamics of the total flux threaded through the 
SQUID (or the phase difference across a current 
biased Josephson junction) obeys a collective motion of a 
macroscopic number 
of quasiparticles. The classical equation of motion for the flux 
dynamics maps to that of a particle moving dissipatively 
in a (symmetric) double-well potential. Its lowest 
left (right) well corresponds 
to one of the two fluxoid states 0 (1) of the SQUID. For sufficiently low 
temperature, the transition between these states occurs via tunneling 
through the potential barrier.  
Measurements of the relaxation of a fluxoid state initially prepared in an 
rf-SQUID have addressed two different physical situations: 
The results in Ref.\ \cite{Han91} have been 
interpreted as {\em incoherent tunneling\/}  
 in a  {\em macroscopic two-state system\/} 
 and those in Ref.\   \cite{Rouse95} have been explained 
 as {\em resonant tunneling\/} between two quasi-degenerate localized states 
 in different fluxoid wells. 
The rate of tunneling out of the metastable well {\em vs.\/} the applied 
external flux exhibits a series of local maxima. These occur 
at those values of the external flux where the adiabatic energy levels 
of the biased SQUID potential form avoided level crossings.  
By applying a resonant time-dependent external rf-field, 
Han {\em et al.\/} \cite{Han96} created a population inversion 
between the two 
adjacent fluxoid wells. Furthermore, Silvestrini {\em et al.\/} 
\cite{Silvestrini97} 
reported the observation of energy level quantization in underdamped 
Josephson junctions {\em above\/} the crossover temperature which separates 
 the classical from the quantum regime. 
 Han {\em et al.\/} \cite{Han00} recently presented evidence for transitions between the 
 fluxoid wells due to cascaded, two-photon processes. 
In the latest work of this group, 
Friedman {\em et al.\/} \cite{Friedmann00} report on the 
realization of a quantum superposition of macroscopic states in an rf-SQUID. 
Similar observations were recently made by the group of Mooij 
\cite{Mooij99,vanderWal00} 
where symmetric and antisymmetric quantum superpositions 
of macroscopic states of a dc-SQUID have been created.  

Another set of experiments concerns the tunneling dynamics 
of substitutional 
 defects in solids \cite{DefectExp}. For instance, in a crystalline environment 
tunneling arises from defect ions which do not fit properly in the 
sites offered by the host lattice. The symmetry of the host crystal determines 
a complicated potential energy landscape with several degenerate 
minima for the defect ion. Golding {\em et al.\/} \cite{Golding92} 
studied the relaxation rates of 
individual microscopic defects in a mesoscopic disordered Bi-metal. 
Since the sample dimensions were comparable to the phase-breaking 
length for quantum transport \cite{VCHBuch} at low temperatures, the sample's 
conductance was highly sensitive to the positions of the scattering 
centers. Their observations were found to be consistent with 
predictions of the dissipative two-level system 
\cite{Leggett8187,HanggiRMP90,Weiss99} at  
low temperatures. However, measurements at higher temperatures 
\cite{Chun93} have indicated the failure of the two-level theory 
\cite{Cukier95}. Furthermore, the study of thermally assisted 
tunneling of atomic hydrogen and deuterium in boron-doped 
crystalline silicon reveals \cite{Noya97} 
that the relaxation rate calculated 
by a path integral centroid formalism differs from experimental 
 measurements by two orders of magnitude. Finally, Enss and Hunklinger 
\cite{Enss97} have pointed out several discrepancies between 
predictions of a semiclassical tunneling model, where the two wells 
are approximated by harmonic oscillators ({\em harmonic-well 
approximation\/}, see also Appendix \ref{app.harmwell}), 
and experimental measurements at low temperatures. 
They proposed an improved approach by taking into account elastic interactions  
among the tunneling systems to resolve these discrepancies. 

The last class of experiments concerns systems in  
chemical physics with the goal of controlling of chemical reactions 
\cite{ChemExp,Doslic98,Naundorf99}. 
The hydrogen pair transfer in the hydrogen-bonded cyclic 
dimers of numerous 
carboxylic acids is used as a prototype system to study the 
relation between quantum tunneling and chemical kinetics. The measurements 
show that the free hydrogen-bonded dimers possess two energetically 
degenerate equilibrium configurations. They correspond to the two 
minima of a double-well potential. Both quantum tunneling and vibrational 
excitation are important for the transfer of the hydrogen pair. This 
has been studied ex\-pe\-ri\-mentally in detail in Refs.\ 
\cite{ChemExp}. 
A specific control scheme (``Hydrogen-Subway'') has been proposed 
\cite{Doslic98,Naundorf99} to 
steer intramolecular hydrogen transfer reactions in malonaldehyde 
by ultrashort laser pulses. The conventionally proposed method for the 
transfer consists in applying a laser pulse that lifts an 
initially localized wavepacket in the reactant region {\em over\/} 
the barrier thus allowing propagation towards the final 
product configuration. 
%
The new approach in Ref.\  \cite{Naundorf99} 
is to drive the wave packet not over 
but {\em through\/} the barrier. 
This is achieved by exciting higher lying doublets where 
tunneling occurs on a much shorter time-scale than in the lower doublets. 
The advantage of this new proposal is that it 
 requires laser intensities which are considerably smaller 
than those used in the conventional approach.

\subsection{Prior theoretical approaches}
\label{subsec.theo}
Previous theoretical works dealing with dissipative spatially continuous 
quantum systems, being driven or undriven, naturally fall into  two classes: 
Approaches that are more of a numerical or analytical flavor, respectively. 

In Ref.\ \cite{Oelschlaegel93}, the harmonically driven double-well potential 
has been  
investigated numerically in presence of dissipation. For that 
purpose, a master equation for the reduced density matrix has been derived 
 on the basis of the standard Born-Markov assumption 
\cite{LouisellHaake73}. Subsequently, an analytical Floquet approach is used  
to derive  the master equation. In doing so, 
an improved master equation has been obtained 
in Ref.\  \cite{Kohler97}. Here, the Floquet 
theory is applied on the level of the Schr\"odinger equation and the 
Born-Markov approximation is made for the quasienergy spectrum. In both cases, 
the system-bath coupling is treated perturbatively. This restricts the 
method to the {\em weak-coupling regime}.  
The same regime of a weak system-bath coupling 
was treated by Naundorf {\em et al.\/}  \cite{Naundorf99}. 
 Also, standard 
Redfield (i.e., weak-coupling) techniques have been applied to derive a 
master equation. 
The specific shape of a laser pulse is determined in order 
to control hydrogen tunneling in a dissipative environment \cite{Naundorf99}.  
In the strong coupling regime, 
the harmonically driven double-well potential has been 
studied in the context of quantum hysteresis and quantum 
stochastic resonance \cite{Thorwart97a}. 
In this work, the system has been iterated numerically using the 
tensor multiplication scheme within the 
quasiadiabatic propagator path integral technique developed by Makri and 
Makarov \cite{QUAPI}. 

%
More analytical oriented 
works in the context of dissipative multilevel bistable systems 
have been performed by several groups 
\cite{SilbeyGroup,MorilloGroup,DekkerGroup,Larkin86,Larkin88,Ovchi94,Silvestrini90,Silvestrini96a,Silvestrini96b}. 
%
%


The starting point in Refs.\  \cite{SilbeyGroup,MorilloGroup} is a multilevel 
system with interdoublet transition terms (vibrational relaxation) which are 
{\em not\/} strictly derived from a {\em continuous\/} double-well potential;  
these are constructed phenomenologically. 
This leads to the assumption that the vibrational 
coupling occurs only between vibrational states located inside the same well.

The group of Silbey 
\cite{SilbeyGroup} considered a {\em static\/} multilevel system.  
Additionally, only tunneling states differing by one quantum of 
vibrational excitation are assumed to be connected. 
Finally, it is assumed that the vibrational 
coupling within each well is the same for both wells. This {\em a priori} 
excludes the case with a static asymmetry of the potential. 

The group of Morillo and Cukier \cite{MorilloGroup} 
started out from a similar Hamiltonian like in \cite{SilbeyGroup}. 
They restricted the model further and included only the {\em two doublets\/} 
with the lowest energy, i.e., the so-called {\em double-doublet 
system\/}. The authors for the first time included 
 a time-dependent driving which couples to a phenomenologically constructed 
 dipole operator of the multilevel system.  The system-bath 
interaction is treated perturbatively within a generalized Redfield approach. 

In a sequence of articles \cite{DekkerGroup}, 
Dekker analyzed the real-time dynamics 
of a quantum particle in the dissipative static 
double-well potential {\em ab initio\/} 
by means of a multisite spin-hopping model. He derives the reduced quantum 
Liouville equation for the particle, thereby not restricting 
the dynamics to the lowest doublet only. 
The interdoublet vibrational dynamics is approximated by 
coarse-graining the density matrix elements on a time scale of many vibrational 
periods. It is further assumed that the localized states 
in the wells are approximated by the eigenfunctions of a harmonic oscillator 
({\em harmonic-well approximation\/}). This latter assumption 
can  be justified as long as the barrier height is large compared to the 
interdoublet energy gap. In this parameter regime, however, the application 
of the standard semiclassical rate theory \cite{HanggiRMP90} 
is appropriate, and even 
simpler to apply. In the deep quantum 
regime with low to intermediate barrier heights, this assumption  
increasingly becomes invalid and 
leads to considerable deviations of the approximated wave functions 
from the exact ones (cf.\ also Appendix \ref{app.harmwell}). 
Also, the eigenenergies of the harmonic potential are 
considerably different from the exact ones for a shallow energy barrier. 

%
A related  problem has  been investigated in a series of theoretical 
works by Ovchinnikov and co-workers 
\cite{Larkin86,Larkin88,Ovchi94,Silvestrini90,Silvestrini96a,Silvestrini96b} 
by applying {\em semiclassical\/} techniques.  
In Ref.\ \cite{Larkin86} Larkin and Ovchinnikov developed 
a method to calculate the decay rate of metastable voltage states of 
Josephson junctions. They constructed a kinetic equation for the 
probabilities of population of many energy levels. The transition 
probabilities are determined for a cubic potential in  
 semiclassical approximation for  {\em weak\/} system-bath coupling. 
This procedure assumes a decay into the continuum via quantum tunneling 
or thermal hopping. However, within confining potentials such as a 
double-well this assumption may be not justified. 
The effect of time-dependent driving is included 
within an approximation. The low temperature 
case where tunneling prevails is considered in Ref.\ \cite{Larkin88} for 
vortices moving in a washboard potential being weakly coupled 
to the environment. Also quasiclassical conditions have been assumed. 
The problem of divergent expressions for the decay rate at avoided 
level crossings is cured in Ref.\ \cite{Ovchi94} where a {\em two-level 
approximation\/} at the avoided level crossings is invoked. The 
authors treat the problem within the harmonic well 
(i.e., quasiclassical) approximation 
for a constant spectral density of the bath modes, and for  a 
weak system-bath coupling. The semiclassical expressions of Ref.\ 
\cite{Larkin86} are applied to Josephson junctions (i) in 
Ref.\  \cite{Silvestrini90} to calculate numerically the decay rate of the 
zero-voltage state for non-stationary conditions, and (ii) in Ref.\ 
\cite{Silvestrini96a} to study the influence of temperature for 
resonant macroscopic quantum tunneling. Finally, the theory 
is adapted to SQUIDs in Ref.\ \cite{Silvestrini96b} to explain the 
experimental findings of Ref.\ \cite{Rouse95}. However, the 
theoretical results follow qualitatively those obtained from 
the standard WKB-approximation. 
The calculated decay rate differs from 
the experimental results by two-to-four orders of magnitude for small 
static potential 
asymmetries, i.e., with  still large barriers, where the semiclassical treatment 
should yield rather good agreement. In contrast,  for large 
bias asymmetries, one of the two barrier heights becomes rather small so that 
 the semiclassical 
approximation is expected to yield worse results. The 
agreement with the experimental data turned out to be of the same 
order of magnitude. This inconsistency may be mainly traced back to the 
fact that the semiclassical treatment is not appropriate for a 
system in the deep quantum regime when only two to six 
levels lie below the energy barrier. 

In summary, {\em no\/} analytic treatment exists in the prior 
literature where tunneling and 
vibrational relaxation is investigated consistently 
in the  regime where a finite number of discrete 
 energy eigenstates rules the dissipative 
dynamics. This is so even for the situation that no time-dependent 
driving acts upon the system.  While 
standard Redfield theory for a weak system-bath coupling is used 
frequently, the theory for the strong coupling regime for 
static as well as for driven multilevel systems 
is still in its infancy. The main objective of this work is 
to fill this gap in deriving analytical schemes that cover the physics in this 
prominent regime of a moderate-to-strong system-bath coupling. 
%

%
\section{The driven dissipative bistable system}
\label{sec.model}
We consider a quantum particle with mass ${\cal M}$, 
position operator ${\bf q}$ and 
momentum operator ${\bf p}$  moving in a  
one-dimensional double-well  
potential $V_0({\bf q})$ which may include a static asymmetry. 
The potential experiences  a time-dependent 
 external force, $s \sin (\Omega t)$,  with 
field strength $s$ and frequency $\Omega$. It is  described by 
the Hamiltonian
\begin{equation}
 {\bf H}_{\rm S}(t) = {\bf H}_{0}- {\bf q} \,  s \sin \Omega t = 
\frac{{\bf p}^2}{2 \cal M} + V_0({\bf q}) 
- {\bf q} \, s \sin \Omega t\, , 
 \label{system}
\end{equation}
with 
\begin{equation}
V_0({\bf q})={{\cal M}^{2}\omega_{0}^{4} \over 64\Delta U}{\bf q}^{4} - {{\cal M}
 \omega_{0}^{2} \over 4}{\bf q}^{2} -{\bf q} \,  \varepsilon \label{staticpotential}
\end{equation}
being the asymmetric double-well potential. The quantity 
$\varepsilon$ denotes the static bias force. In absence of the 
asymmetry ($\varepsilon=0$), $\Delta U$ denotes the barrier height,   
and $\omega_0$ is the angular frequency  of classical oscillations around the 
well minima. 

The energy spectrum of ${\bf H}_{0}$ follows from the time-independent 
Schr\"odinger equation with a static double-well potential $V_0({\bf q})$, 
i.e., 
${\bf H}_{0} \ket{n} = {\cal E}_n \ket{n}, n=1,2, \dots$. 

In absence of a static bias ($\varepsilon=0$) and 
for energies well below the barrier, the spectrum   
 consists of a ladder of pairs of energy 
eigenstates ({\em doublets}). The energy gaps within each doublet 
generally are several orders of magnitude smaller than the inter-doublet 
energy gaps and are responsible for the {\em tunneling dynamics\/} 
 between the two wells. 
The large energy gaps are of the order of the harmonic oscillator 
energy gap $\hbar \omega_0$ associated with each well. For energies above the 
barrier, the energy gaps are also of the order of $\hbar \omega_0$. 
Transitions between those largely separated energy 
eigenstates are termed {\em vibrational relaxation\/}. 
In presence of a static tilt ($\varepsilon \ne 0$), no general statement 
can be made. Spectra with typical avoided level crossings can occur as well 
as such with almost equally separated energy levels, cf.\  
Fig.\ \ref{fig.rateasy1} a.). 

Following the  common approach 
\cite{Leggett8187,HanggiRMP90,Weiss99,VCHBuch,Vernon63} to model the influence 
of the environment by an ensemble of harmonic oscillators, the bath 
Hamiltonian ${\bf H}_{\rm B}$ (including the interaction with the 
system) is given by 
\begin{equation}
{\bf H}_{\rm B}=
\sum_{j=1}^{\cal N} \frac{1}{2}\Big 
[\frac{{\bf p}_j^2}{m_j}+m_j\omega_j^2
\Big ({\bf x}_j-\frac{c_j}{m_j\omega_j^2} {\bf q} \Big ) ^2 \Big ]\, . 
 \label{hamilton}
\end{equation}
The whole system is thus described by the Hamiltonian 
${\bf H}(t)={\bf H}_{\rm S}(t)+{\bf H}_{\rm B}$. In the case of a 
thermal equilibrium bath,  its influence on the 
system is fully characterized by the spectral density 
\begin{equation}
J(\omega) = 
\frac{\pi}{2}\sum_{j=1}^{\cal N} \frac{c_j^2}{m_j \omega_j} \delta(\omega - \omega_j).
\end{equation}
With the number ${\cal N}$ of harmonic oscillators approaching infinity, we 
arrive at a continuous spectral density. Throughout this work, we choose 
 an  Ohmic spectral density with an exponential 
 cut-off, i.e.,  
\begin{equation}
J(\omega)=  \eta \, \omega \exp (-\omega/\omega_c). \label{specdens}
\end{equation}
Here, $\eta = {\cal M} \gamma$, with $\gamma$ 
being the strength of the coupling to the heat bath. Moreover,  
 $\omega_c \gg (\omega_0,\Omega,\gamma)$ denotes a cut-off  frequency being 
 the largest frequency in the model. 

We choose a factorizing initial condition of Feynman-Vernon form  
\cite{Vernon63}. This means that at 
time $t=t_0$, the full density operator ${\bf W}(t_0)$ is given as 
a product of the initially prepared  
system density operator ${\bfrho}_{\rm S}(t_0)$
and the canonical bath density operator
at temperature $T=1/k_{\rm B} \beta$, i.e.,
\begin{equation}
{\bf W}(t_0) = {\bfrho}_{\rm S}(t_0)\, \, Z_{\rm B}^{-1}\, 
\exp (-\beta {\bf H}_{\rm B}^0)\ ,
\label{ic}
\end{equation}
where $Z_{\rm B} = {\rm tr} \, \exp (-\beta {\bf H}_{\rm B}^0)$ and  
%
${\bf H}_{\rm B}^0 =
\sum_{j=1}^{\cal N} \frac{1}{2}
\Big [\frac{{\bf p}_j^2}{m_j}+m_j\omega_j^2\, {\bf x}_j ^2 \Big ]$. 

In order to describe the dynamics of the system of interest 
we focus on the time evolution of the reduced 
density matrix. In position representation it reads  
\begin{eqnarray} 
\rho (q_f,q_f';t)&=&\tr_{\rm res} 
\langle q_f \Pi_j x_j | {\bf U}(t,t_{0})\, {\bf W}(t_{0})\, {\bf U}^{-1} 
(t,t_{0}) | q_f' \Pi_j x_j' \rangle \ , \nonumber \\ 
{\bf U}(t,t_{0})&=&{\cal T}\exp \left \{ -i/\hbar \int_{t_{0}}^{t} {\bf H}(t') 
 dt' \right \} \  . 
\label{3.1new} 
\end{eqnarray} 
Here, ${\cal T}$ denotes the time ordering operator, ${\bf W}(t_{0})$ is the 
full density operator at the initial time $t_{0}$ and $\tr_{\rm res}$ 
indicates the 
partial trace over the harmonic bath oscillators $x_j$. 

\section{The reduced density matrix in the discrete variable representation DVR}
\label{sec.realtimepathint}

\subsection{The Feynman-Vernon influence functional}

Due to our 
assumption of a factorizing initial condition in Eq.\  (\ref{ic}), 
 the partial trace over the bath can be performed 
and the reduced density operator be recast according to 
Feynman and Vernon \cite{Vernon63} as 
\begin{equation} 
\rho (q_f,q_f^\prime ,t) = \int dq_0 \int dq_0^\prime \ G(q_f,q_f^\prime,t 
;q_0,q_0^\prime,t_{0}) \rho_{\rm S} (q_0, q_0^\prime, t_{0}) \ ,
\label{2.1} 
\end{equation} 
with the propagator $G$ given by 
\begin{equation} 
G(q_f,q_f^\prime, t ;q_0,q_0^\prime,t_{0}) = 
\int_{q(t_0)=q_{0}}^{q(t)=q_f} {\cal D}q
\int_{q'(t_0)=q_{0}'}^{q'(t)=q_{\rm f}'} {\cal D}q'
{\cal A}[q] {\cal A}^*[q^\prime]
{\cal F}_{\rm FV}[q,q^\prime ] \, . 
\label{2.2} 
\end{equation} 
Here, ${\cal A}[q]= \exp \left \{ i  S_{\rm S}[q] / \hbar \right \}$ 
denotes the bare system amplitude,  with 
$S_{\rm S}[q]$ being the classical action functional of the 
system variable $q$ along a path $q(t)$. 
${\cal F}_{\rm FV} [q,q^\prime ]=\exp (-\phi_{\rm FV}[q,q^\prime ]/\hbar)$ 
denotes 
the Feynman-Vernon influence functional. 
%
%
For our purpose, 
it is convenient to write the influence phase $\phi_{\rm FV}[q,q^\prime ]$ 
in terms of relative coordinates 
$\xi(t')=q(t')-q'(t')$ and center of mass coordinates 
$\chi(t')=q(t')+q'(t')$, respectively; it reads 
\begin{eqnarray}
\phi_{\rm FV}[\chi,\xi] & = &  \int_{t_0}^t dt' \int_{t_0}^{t'} dt'' 
\{ \dot{\xi}(t') S(t'-t'')\dot{\xi}(t'') + 
i \dot{\xi}(t') R(t'-t'')\dot{\chi}(t'') \} \nonumber \\
&& \mbox{} + \xi(t) \int_{t_0}^t dt' \{ \dot{\xi}(t') S(t-t') + 
i \dot{\chi}(t') R(t-t') \} \nonumber \\
&& \mbox{}+ \xi(t_0) \{ \xi(t) S(t-t_0) - \int_{t_0}^t dt' 
\dot{\xi}(t') S(t'-t_0) \} \nonumber \\ 
&&\mbox{}+ i \chi(t_0) \{ \xi(t) R(t-t_0) - \int_{t_0}^t dt' 
\dot{\xi}(t') R(t'-t_0) \} \, .
\label{infl}
\end{eqnarray}
Herein, $S(t)$ and $R(t)$ denote the real and imaginary part, respectively, 
 of the 
 bath correlation function $Q(t)$, i.e., \cite{Weiss99} 
\begin{equation}
Q(t) = S(t) + i R(t) = \frac{1}{\pi} \int_0^{\infty} 
d\omega \frac{J(\omega)}{\omega^2} 
\left\{ \coth \frac{\hbar \omega \beta}{2} \left(1- \cos \omega t\right) + 
i \sin \omega t\right\} \, .
\label{twintcorr}
\end{equation}
We evaluate in the following the reduced density matrix explicitly. 
It turns out that this is conveniently performed in the {\em discrete} 
eigenbasis of the position operator ${\bf q}$. 
This representation is the so-termed {\em discrete variable 
representation (DVR)\/} 
\cite{Harris65}. 
The reason for this basis transformation is that only then can  the 
influence phase, Eq.\  (\ref{infl}),  be evaluated at the 
eigenvalues $q_{\mu}$ of ${\bf q}$. This is shown in the subsequent section. 

%
%
\subsection{Real-time paths in the DVR basis} 
\label{subsec.dvr}
The time-independent double-well potential $V_0({\bf q})$, Eq.\  
(\ref{staticpotential}),   
possesses a discrete energy spectrum. 
The interesting  temperature regime for us is that in which only 
a finite and small number of energy eigenstates is  
thermally  significantly populated. 
A quantum mechanical description would not be necessary if 
the temperature is very large compared to the 
natural energy scale of the system.   
We assume furthermore that the time-dependent driving does not
 excite  arbitrary high lying energy eigenstates of the static problem. 
Then, it is appropriate to consider only the  $M$-dimensional
Hilbert space spanned by the $M$ lowest lying energy eigenstates 
of the static potential. 
 The problem of a spatially continuous double-well potential is 
then reduced to a problem of a finite dimensional $M$-level system ($M$LS). 
The case of $M=2$ (with $\varepsilon$ and $s$ being sufficiently small) 
is the well-known (driven) spin-boson problem 
\cite{Leggett8187,Weiss99,Grifoni98},  while $M=4$ constitutes, for instance,  
the double-doublet system \cite{MorilloGroup}. 
This reduction has been shown to be sensible for the case 
of the parametrically driven dissipative quantum harmonic oscillator 
\cite{ThorwartPRE00}. There, the spatially continuous potential is 
appropriately described  
by a discrete $M$-level system with $M=3$ to $M=6$.

Next we perform a basis transformation to the 
so-called {\em discrete variable representation} (DVR) \cite{Harris65}. 
The new basis is chosen as 
 the eigenbasis of that operator which couples the bare system to 
 the harmonic bath. In our case this is   
 the position operator ${\bf q}$. We define the DVR basis 
 $\{ \ket{q_{\mu}} \}$ according to 
\begin{equation}
\mel{q_{\mu}}{{\bf q}}{q_{\nu}} 
= q_{\mu} \delta_{\mu \nu}\, , \mbox{\hspace{2ex}} \mu, \nu = 1, \dots , M\, . 
\label{dvrtrafo}
\end{equation}
This basis follows from the energy eigenbasis $\{ \ket{m} \}$ 
by inserting the identity ${\mathbb I}=\sum_{m=1}^M 
\ket{m} \bra{m}$ yielding 
$\ket{q_{\mu}} = \sum_{m=1}^M \langle m | q_{\mu} \rangle \ket{m}$.
  This step allows to transform 
the description of the dynamics as transitions between energy 
eigenstates  to a hopping among the $M$ discrete 
position eigenvalues $q_{\mu}$ of the spatial grid. 
While for the static symmetric case, $\varepsilon=0$, the position 
eigenvalues $q_{\mu}$ are located 
symmetrically on the $q$-axis with respect to $q=0$, this is no 
longer the case  
in presence of a static bias $\varepsilon \ne 0$. 
%
%
%

To describe the dynamics in the DVR basis, we define a 
quantum mechanical path $q(t')$ along which the system evolves in time. 
It starts out at time $t'=t_0$ in the state 
$q(t'=t_0) = q_{\mu_0}$ and evolves via $\tilde{N}$ jumps 
between the $M$ discrete states into 
the final state $q(t'=t_{\tilde{N}}) = q_{\mu_{\tilde{N}}}$. 
The full time interval is split  
into $\tilde{N}$ short time intervals such that the jumps 
happen at  times $t'=t_j$. 
The intermediate states are labeled by $q_{\mu_j}$, where 
$\mu_j=1, \dots, M$ is the  quantum state index,  
and $j=1, \dots ,\tilde{N}-1$ denotes the time index.
The full path is assumed to be 
a sequence of constant paths segments according to
\begin{equation}
q(t')  =  - q_{\mu_0} \Theta(t'-t_1) + 
\sum_{j=1}^{\tilde{N}-1} q_{\mu_j} [\Theta(t'-t_j)-\Theta(t'-t_{j+1})] 
+ q_{\mu_{\tilde{N}}} \Theta(t'-t_{\tilde{N}}) \, , 
\end{equation}
%
where $\Theta(t)$ is the Heaviside function. 
Thus, upon switching to the center-of-mass and relative 
coordinates $\chi(t')=q(t')+q'(t')$  and $\xi(t')=q(t')-q'(t')$, respectively,  
(cf.\  Eq.\  (\ref{xi}) and  Eq.\  (\ref{relativepaths}) below),   
the double path integral 
over the $M$-state paths $q(t')$ and $q'(t')$ in Eq.\  (\ref{2.2})
is visualized  as  an integral 
over a {\em single\/} path that jumps between the $M^2$ states of the 
reduced density matrix in the $(q,q')$-plane. The total number $N$ of jumps 
is given by the sum of the number of jumps for the paths $q$ and $q'$, 
i.e., $N=\tilde{N}+\tilde{N}'$. 

\begin{figure}[t]
\begin{center}
\epsfig{figure=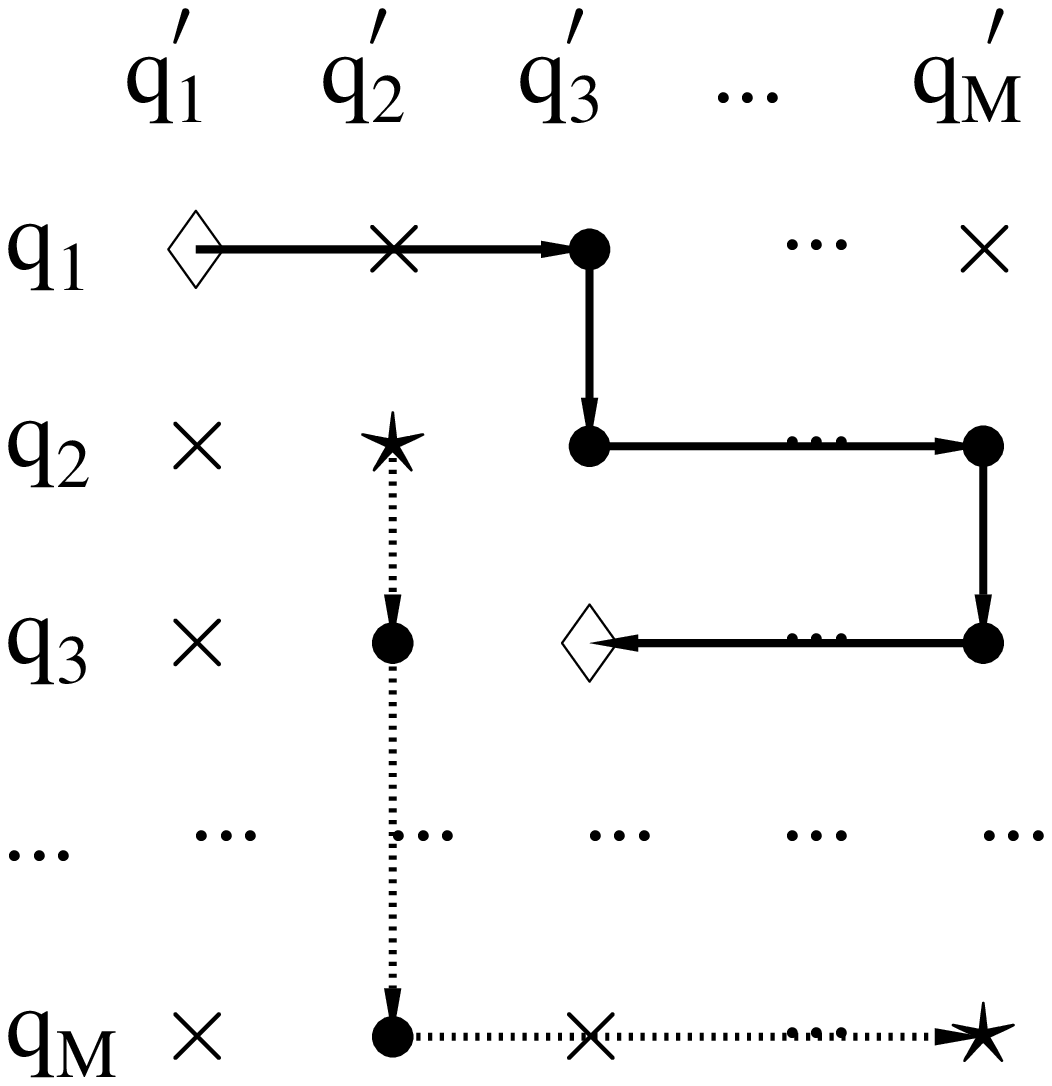,width=60mm,height=60mm,angle=0}
\caption{The $M^2$ states of the reduced density matrix 
of an $M$-level system. Shown are two examples of paths that travel 
between two diagonal states of the density matrix (see text). 
One path (solid line) connects the diagonal states $\Diamond$ 
and the other (dashed line) travels between the diagonal states 
$\star$. The intermittently visited off-diagonl states are marked 
by $\bullet$.   
 \label{fig.doublepath}}
\end{center}
\end{figure}

Fig.\  \ref{fig.doublepath} illustrates this idea for a general $M$-state 
system described by its $M\times M$ density matrix. 
Two paths are depicted: one 
(full line) starts in the diagonal 
state $(q_1,q'_1=q_1)$  and jumps in $\tilde{N}'=3$ 
horizontal jumps and in $\tilde{N}=2$ vertical 
jumps to the final diagonal state $(q_3,q'_3=q_3)$.  
It visits four intermediate off-diagonal states (filled circles). 
The second path (dashed line) starts 
in the diagonal state $(q_2,q'_2=q_2)$ and travels via two 
intermediate states 
to the final diagonal state $(q_M,q'_M=q_M)$. 

The paths in the relative and center of mass coordinates read
\begin{eqnarray}
\xi(t') & = & q(t') - q'(t') \nonumber \\
& = & - \xi_{\mu_0 \nu_0} \Theta(t'-t_1) + 
\sum_{j=1}^{N-1} \xi_{\mu_j \nu_j} 
[\Theta(t'-t_j)-\Theta(t'-t_{j+1})] \nonumber \\
& & \mbox{} + \xi_{\mu_{{N}} \nu_{{N}}} \Theta(t'-t_N) \, , \label{xi} 
\end{eqnarray}
and
\begin{eqnarray}
\chi(t') & = & q(t') + q'(t') \nonumber \\
& = & - \chi_{\mu_0 \nu_0} \Theta(t'-t_1) + 
\sum_{j=1}^{N-1} \chi_{\mu_j \nu_j} [\Theta(t'-t_j)-\Theta(t'-t_{j+1})] \nonumber \\
& & \mbox{} + \chi_{\mu_{{N}} \nu_{{N}}} \Theta(t'-t_N) \, . 
\label{relativepaths}
\end{eqnarray}
Herein, the path weights are given as
\begin{equation}
\xi_{\mu_j \nu_j} \equiv q_{\mu_j} - q'_{\nu_j}
\label{twoindexweightxi}
\end{equation}
and
\begin{equation}
\chi_{\mu_j \nu_j} \equiv q_{\mu_j} + q'_{\nu_j} \, .
\label{twoindexweightchi}
\end{equation}
In this discrete notation, the index $\mu$ refers to the path $q$ and the 
index $\nu$ to the primed path $q'$. 
The time intervals in which the system is in a diagonal state 
of the reduced density matrix are called 
{\em sojourns\/}. They are characterized by $\xi(t')=0$ and $\chi(t') \ne 0$. 
The time spans in which the system is in an off-diagonal state 
are called {\em clusters\/}. The clusters  are characterized by $\xi(t')\ne 0$ 
and $\chi(t') \ne 0$. This is different from the 
spin-boson problem \cite{Leggett8187,Weiss99,Grifoni98} 
where the off-diagonal states ({\em blips\/}) 
 are characterized by  $\xi(t')\ne 0$ and $\chi(t') = 0$. 
Upon determining the derivatives of the paths with respect 
to the time variable $t'$, 
we find 
\begin{equation}
\dot{\xi}(t') = \sum_{j=1}^{N} \xi_{j} 
\delta (t'-t_j) 
\label{pathsderivativexi}
\end{equation}
and
\begin{equation}
\dot{\chi}(t') = \sum_{j=1}^{N} \chi_{j} 
\delta (t'-t_j) \, .
\label{pathsderivativechi}
\end{equation}
Thereby, we have introduced new paths weights according to
\begin{equation}
\xi_{j} \equiv \xi_{\mu_j \nu_j} 
- \xi_{\mu_{j-1} \nu_{j-1}}
\label{fourindexweightxi}
\end{equation}
and
\begin{equation}
\chi_{j} \equiv \chi_{\mu_j \nu_j} 
- \chi_{\mu_{j-1} \nu_{j-1}} \, .
\label{fourindexweightchi}
\end{equation}
%
%
with $j=1,...,N$. For $j=0$, we define 
$\xi_0 \equiv \xi_{\mu_0 \nu_0}$ and $\chi_{0} \equiv \chi_{\mu_0 \nu_0}$. 
Hence, a path with $N$ transitions at times $t_1, t_2, \dots, t_N$ can be 
parametrized by two sets of path weights 
$\{ \chi_0, \chi_1, \chi_2, \dots, \chi_N\}$  
and $\{ \xi_0, \xi_1, \xi_2, \dots, \xi_N\}$. In the influence functional the paths 
are coupled. The situation mimics the case of interacting electrical charges. 
Thus,  the paths weights in Eqs.\  (\ref{fourindexweightxi}), 
(\ref{fourindexweightchi}) are termed {\em charges\/}. 
In the discrete notation, the real-time path integral expression (\ref{2.1}) 
assumes the form 
\begin{eqnarray}\label{rhodvr}
\rho_{\mu_{N} \nu_{N}} (t) & = & 
\mel{q_{\mu_N}}{\bfrho (t)}{q_{\nu_N}} \nonumber \\
& = &  \sum_{\mu_0 \nu_0} \int_{\xi(t_0)=\xi_0}^{\xi(t)=\xi_{N}} 
{\cal D}\xi  
\int_{\chi(t_0)=\chi_{0}}^{\chi(t)=\chi_{N}} {\cal D}\chi  \, 
{\cal B}[\chi,\xi] \, {\cal F}_{\rm FV} [\chi,\xi] \, \rho_{\mu_0 \nu_0} \, .
\end{eqnarray}
Here, 
${\cal B}[\chi,\xi] = {\cal A}[q] {\cal A}^*[q']$, and 
the influence phase takes on the form
\begin{eqnarray} \label{discrfeynman}
\phi_{\rm FV}[\chi,\xi] & = & 
- \sum_{l=1}^{N} \sum_{j=0}^{l-1} \xi_{l} 
S(t_l-t_j) \xi_{j}  
- i \sum_{l=1}^{N} \sum_{j=0}^{l-1} \xi_{l} 
R(t_l-t_j) \chi_{j} \, .
\end{eqnarray}
\subsection{The population of the left well} 
\label{subsec.boundcond}
Since we are interested in the decay of the population of one (metastable) 
well of the bistable potential, say the left well, 
we define the quantity of interest to be the sum of the populations 
of those $L$ DVR-states $\ket{q_{\mu}}, \mu=1, ..., L$, 
which belong to the {\em negative 
position eigenvalues\/} $q_{\mu}$, i.e., those which are located to the 
 left from the 
zero. This yields 
\begin{equation}
P_{\rm left} (t) = \sum_{\mu=1}^{L} \rho_{\mu \mu}(t) \, .
\label{observable}
\end{equation}

In absence of a static bias, i.e., $\varepsilon=0$ in Eq.\  
(\ref{system}), 
the energy eigenfunctions occur in pairs of symmetric and 
antisymmetric wave functions.  
This implies a choice for an even number $M$ of states. 
Then,  half of the position 
eigenvalues is on the left side and the other half is on the 
right side of the position point of reflection symmetry, being at 
$q=0$ for $V_0({\bf q})$ in Eq.\  (\ref{staticpotential}). The consequence is  
that for the population $P_{\rm left} (t)$ of the left well, 
usually $L=M/2$ DVR-states are relevant. 
However, in the case of a finite 
static asymmetry, no such general statement can be made. 

To determine $P_{\rm left} (t)$ in Eq.\  
(\ref{observable}) we focus on the case 
that the final state $(\mu_{N}, \nu_{N})$ 
of the system will be a diagonal state, i.e., 
\begin{equation}
\nu_{N} =  \mu_{N} \, .
\label{diagfinalstate}
\end{equation}
Since then $q(t)=q'(t)$,  
it follows that $\xi(t) \equiv 0$ in Eq.\  (\ref{xi}). 

The initially localized 
wave packet is assumed to be {\em a superposition of energy eigenstates\/}. 
The transformation to the DVR-basis generates an initial system  
density matrix $\rho_{\mu_0 \nu_0}$ which generally is non-diagonal, i.e., 
\begin{equation}
\nu_0 \ne  \mu_0 \, .
\label{nondiaginistate}
\end{equation}
Accordingly, we keep the general initial conditions 
$\rho_{\mu_0 \nu_0} \ne 0$ in Eq.\  (\ref{rhodvr}).

We proceed to the explicit evaluation of the path integral 
in Eq.\  (\ref{rhodvr}) 
with the boundary conditions given in Eqs.\  
(\ref{diagfinalstate}) and (\ref{nondiaginistate}).

%
To determine the transition amplitudes of the bare system 
we consider a discrete path starting in a general initial state that ends 
 in a diagonal state. 
It is described by a sequence of pairs of state labels 
\begin{equation}
(\mu_0, \nu_0) \rightarrow (\mu_1,\nu_1) \rightarrow (\mu_2,\nu_2) \rightarrow \dots 
\rightarrow (\mu_{{N}}, \nu_{{N}})=(\mu_{{N}}, \mu_{{N}}) \, .
\label{onepath}
\end{equation}
The first symbol of each pair belongs to the horizontal direction and labels the rows 
of the reduced density matrix. The second symbol corresponds to 
the vertical direction and labels the columns.  
This implies that for a horizontal jump the first index remains constant, 
i.e., $(\mu_{j},\nu_{j}) \rightarrow (\mu_{j+1},\nu_{j+1})=(\mu_{j},\nu_{j+1})$, 
while for a vertical jump the second index is unchanged meaning 
  $(\mu_{j},\nu_{j}) \rightarrow (\mu_{j+1},\nu_{j+1})=(\mu_{j+1},\nu_{j})$. 

We are interested in the probability amplitude of finding the system in state 
$(\mu_{j+1},\nu_{j+1})$ after a time $\Delta t = t_{j+1} - t_{j}$ having 
started from $(\mu_{j},\nu_{j})$. This quantity is given by the time evolution 
operator of the bare system. We find for a vertical jump, i.e., 
 $\nu_{j+1}= \nu_{j}$,  
the amplitude 
$\mel{q_{\mu_{j+1}}}{\exp\{-i {\bf H}_0 \Delta t / \hbar\}}{q_{\mu_{j}}}$ 
and for 
a horizontal jump, i.e., $\mu_{j+1}= \mu_{j}$, 
$\mel{q_{\nu_{j+1}}}{\exp\{+i {\bf H}_0 \Delta t / \hbar\}}{q_{\nu_{j}}}$, 
respectively. 
The relevant part of the system Hamiltonian ${\bf H}_{\rm S} (t)$ 
in Eq.\  (\ref{system}) 
is the time-independent part ${\bf H}_{0}$  since we 
are interested in the cases $q_{\mu_{j+1}} \ne q_{\mu_{j}}$ and 
$q_{\nu_{j+1}} \ne q_{\nu_{j}}$. Taking into account 
 the exponential operator up to linear order in the argument, i.e., 
 $\exp\{\pm i {\bf H}_0 \Delta t / \hbar\} \approx {\mathbb I} 
 \pm i {\bf H}_0 \Delta t / \hbar$, and using the orthogonality relation 
 $\langle q_l | q_m \rangle = \delta_{lm}$, the result for the 
 transition probability amplitude per unit time $\Delta t$ is obtained 
 as $\pm i \Delta_{j}/2$. Here, the factor 
of $1/2$ is extracted to have the same convention as in the 
spin-Boson-problem. The factors $\Delta_{j}$ for a horizontal jump 
are defined according to 
\begin{equation}
\Delta_{j} 
= \Delta_{\nu_{j+1}\nu_{j}} \equiv \frac{2}{\hbar} \mel{q_{\nu_{j+1}}}{
{\bf H}_{0}}{q_{\nu_{j}}} \, ,
\label{horjump}
\end{equation}   
and for a vertical jump 
\begin{equation}
\Delta_{j} 
= \Delta_{\mu_{j+1}\mu_{j}} \equiv \frac{2}{\hbar} \mel{q_{\mu_{j+1}}}{
{\bf H}_{0}}{q_{\mu_{j}}} \, , 
\label{verjump}
\end{equation}   
respectively.  The $+$ ($-$) sign belongs to a horizontal (vertical) 
transition in the reduced density matrix. The different signs for 
horizontal and vertical direction reflect the fact that 
the bare transition amplitude ${\cal A}[q]$ belongs to 
the vertical direction, while the complex conjugate transition amplitude 
${\cal A}^*[q']$ belongs to the horizontal direction of the reduced 
density matrix. 

The amplitude to stay in the $j$-th off-diagonal state 
lasting from $t_{j}$ to 
$t_{j+1}$ depends on the time-dependent diagonal elements of the 
bare system Hamiltonian in the DVR-basis. It is given by  
the so-called {\em bias factor\/} 
$\exp\left(i \int_{t_{j}}^{t_{j+1}} dt' [E_{\mu_{j}}(t') - E_{\nu_{j}}(t')]
\right)$, 
where 
\begin{equation}
E_{\mu_{j}}(t') = \frac{1}{\hbar} 
 \mel{q_{\mu_{j}}}{{\bf H}_{\rm S}(t')}{q_{\mu_{j}}} 
= \frac{1}{\hbar} (F_{\mu_j} - q_{\mu_j} s \sin \Omega t')  
 \label{diagele}
\end{equation} 
with $F_{\mu_j} \equiv \mel{q_{\mu_{j}}}{{\bf H}_{0}}{q_{\mu_{j}}}$. 
 For the entire 
 evolution from $t_0$ to $t_N$, $N$ of these factors are multiplied, yielding 
 the overall contribution
$\exp\{i \, \sum_{j=0}^{N-1} 
\int_{t_{j}}^{t_{j+1}} dt'  [ E_{\mu_{j}}(t') - E_{\nu_{j}}(t')]\}$. 
This defines the transition probability amplitudes of the bare system 
in a unique way. 
%
%
%
%

The functional integration over all {\em continuous} paths 
in Eq.\  (\ref{rhodvr}) turns into a {\em discrete} sum 
over all possible path configurations 
$\{ \mu_j \nu_j\}$  
in the DVR basis and  an integration 
over all intermediate times $\{t_j\}$. In formal terms this implies 
\begin{equation}
\int {\cal D}\xi  \int {\cal D}\chi \curvearrowright \int {\cal D} \{t_j\} 
\sum_{\{\mu_j \nu_j \}} \, , 
\end{equation}
where we have introduced a compact notation according to 
\begin{equation}
\int_{t_0}^t {\cal D} \{t_j\} \equiv \int_{t_0}^t dt_{N}
\, \int_{t_0}^{t_{N}} dt_{N-1} \, \dots \int_{t_0}^{t_3} dt_{2}  \, 
\int_{t_0}^{t_2} dt_{1}   
\end{equation}
for the time ordered integration over the $N$ transition times $t_j$ in Eq.\ 
(\ref{rhodvr}).

Collecting all parts 
we obtain the dissipative real-time path integral 
for the diagonal elements of the 
reduced density matrix of an $M$-level system in the DVR-basis, i.e.,  
\begin{eqnarray}
\rho_{\mu_N \mu_N} (t) & = & \mel{q_{\mu_N}}{\bfrho (t)}{q_{\mu_N}} \nonumber \\
& = & \sum_{\mu_0, \nu_0 = 1}^M \rho_{\mu_0 \nu_0} 
\sum_{N=1}^{\infty} \int_{t_0}^t {\cal D} \{t_j\} 
\sum_{\{\mu_j \nu_j \}} 
\exp \left\{i \, \sum_{j=0}^{N-1} 
\int_{t_{j}}^{t_{j+1}} dt'  [ E_{\mu_{j}}(t') - E_{\nu_{j}}(t')]\right\} 
\nonumber \\
& & \mbox{} \times \prod_{j=0}^{N-1} (-1)^{\delta_j} \left(\frac{i}{2}\right)^N 
\Delta_{j} \nonumber \\
& & \mbox{} \times  \exp \left\{
 \sum_{l=1}^{N} \sum_{j=0}^{l-1} \xi_{l} 
S(t_l-t_j) \xi_{j} + 
 i \sum_{l=1}^{N} \sum_{j=0}^{l-1} \xi_{l} 
R(t_l-t_j) \chi_{j}\right\} \, .
\label{fullpathint}
\end{eqnarray}
In this expression, the sum over all possible  path configurations 
$\{\mu_j \nu_j \}$ in the spirit of Eq.\  (\ref{onepath}) 
has to be performed with $\delta_j=0 (1)$ for a horizontal (vertical) jump.  

Several comments on this quite comprehensive 
path integral expression are apposite: 
First, 
the path integral in Eq.\  (\ref{fullpathint}) 
  is given in its most general form and is formally {\em exact\/} because no 
approximations, neither on the form of the system Hamiltonian nor on the 
type of the 
system-bath interaction, are made. This method  
could be applied to {\em any} problem where a potential with 
a discrete spectrum is given, and where the coupling to the heat bath 
is mediated via the position operator. The main ingredients are the 
matrix elements of the system Hamiltonian, being represented in 
the DVR-basis, and the position eigenvalues via the paths weights. 
No specific requirements on the shape of the external 
driving have been made; even a stochastic driving force (such as 
multiplicative noise) can be included. 

In the case of only two levels, i.e., 
$M=2$, Eq.\  (\ref{fullpathint}) reduces to the 
well-known expression for the (driven) spin-boson problem 
\cite{Leggett8187,Weiss99,Grifoni98}.  
There, the problem simplifies due to the fact that the path 
weights during the time evolution take on only  two values, 
corresponding to the two states localized in the left and in the right 
well of the potential. This means 
that the path flips between a sojourn and a 
blip at each jump. This implies that the spin-boson path integral 
assumes the form of 
 a power series in  the tunneling splitting 
 $\Delta^{\cal E}_1 \equiv {\cal E}_2 - {\cal E}_1$ of the two 
lowest levels. This is 
not necessarily the case for a general $M$-level system 
 where a path can travel around, 
visiting many off-diagonal states, before ending in a diagonal state. 
Certainly, such a 
path becomes less likely the longer it remains off-diagonal. This 
is due to damping. 
%

The path integral  is not tractable in its most general form 
 without assuming further  approximations. Such 
 an approximation is developed in the following Section \ref{sec.gnica}. 
 However, to gain insight into the physics behind the formal 
 expression (\ref{fullpathint}), we introduce in Section \ref{subsec.dds} 
 the example of the so-termed {\em double-doublet system\/} and discuss 
 the transformation to the DVR-basis. It refers to  the case where 
 two  doublets in a symmetric double-well potential, Eq.\ 
  (\ref{staticpotential}),  are localized below 
 the barrier, i.e., the case $M=4$.  
\subsection{An example: The symmetric double-doublet system} 
\label{subsec.dds}
We illustrate the general method  with  
the example of two doublets below the barrier 
in the double-well potential, Eq.\  
(\ref{staticpotential}). Choosing $M=4$ generates 
the first non-trivial extension to the familiar spin-boson problem. 

\begin{figure}[t]
\centerline{\hbox{
\psfrag{E}{\small ${\cal E}$}
\epsfig{figure=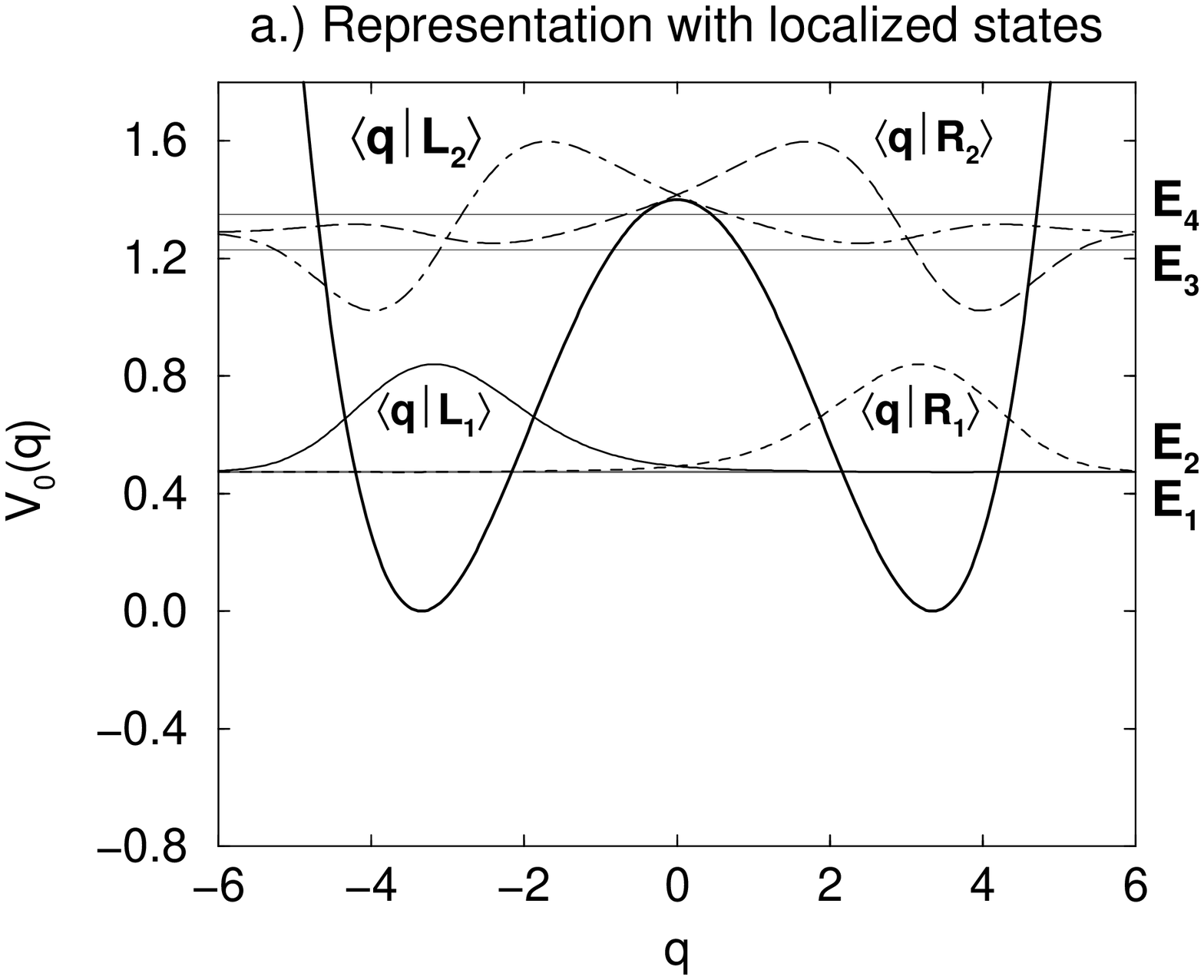,width=70mm,height=70mm,angle=0} 
\hfill \hspace{2ex}
\epsfig{figure=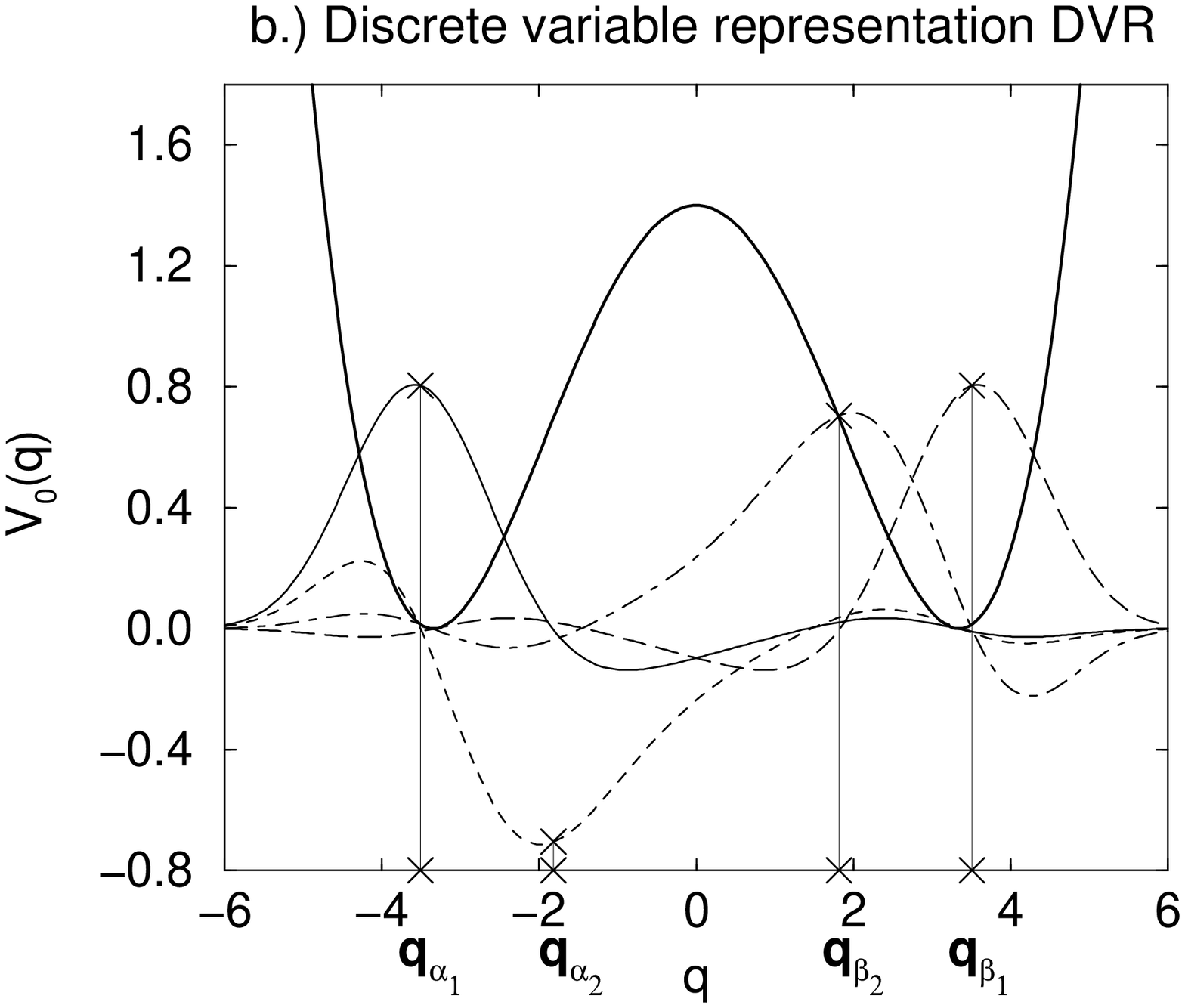,width=70mm,height=73.5mm,angle=0} 
}
}
\caption{a.) The first four localized states $\langle q|L_1\rangle, ..., 
\langle q|R_2\rangle$
of the static symmetric 
double-well potential, Eq.\ (\ref{staticpotential}) with $\varepsilon=s=0$, 
are shown in position space. They are defined 
according to Eq.\  (\ref{locbasis})
The barrier height is chosen to be $E_{\rm B} = \Delta U / \hbar 
\omega_0 = 1.4 $ (we use here and in the following figure captions 
dimensionless quantities according 
to the standard scaling defined in Eq.\  (\ref{dwscale}) ).
 The energy 
eigenvalues ${\cal E}_1, ..., {\cal E}_4$ are marked by thin solid horizontal lines. 
 b.) The corresponding four DVR-states 
are shown, i.e., 
$\langle q|\alpha_1\rangle$ (solid line),  
$\langle q|\alpha_2\rangle$ (dashed line),
$\langle q|\beta_2\rangle$ (dashed-dotted line), and   
$\langle q|\beta_1\rangle$ (long-dashed line). 
 On the $q$-axis, the exact eigenvalues $q_{\mu}$ are  marked by crosses. 
 \label{fig.dvrstates}}
\end{figure}

For the sake of simplicity, but without loss 
of generality, we consider the symmetric potential, i.e., we set 
$\varepsilon=0$ in Eq.\  (\ref{system}).  
For the isolated  system the energy spectrum follows from the 
 time-independent Schr\"odinger equation  as ${\bf H}_0|n\rangle ={\cal E}_n|n\rangle$,
  $n=1,2,...$. 
The two lowest 
  doublets  $\hbar\Delta^{\cal E}_1={\cal E}_2 -{\cal E}_1$, 
  and $\hbar\Delta^{\cal E}_2={\cal E}_4-{\cal E}_3$ 
  are separated by the energy 
 gap $\hbar \overline{\omega}_0 = \frac{1}{2} ({\cal E}_4+{\cal E}_3) - 
 \frac{1}{2} 
({\cal E}_2+{\cal E}_1) \gg \hbar \Delta^{\cal E}_i$. The interdoublet frequency 
$\overline{\omega}_0$ is of the 
order of the classical oscillation frequency ${\omega}_0$ and becomes 
equal to it in the limit of high barriers when the two intrawell oscillators 
approach harmonic oscillator potentials.  
 With the objective of the decay of a localized state in mind, 
 we start from 
 the so-called {\em localized basis\/}. 
 It is this basis which is favorably used to describe the tunneling dynamics. 
 It follows from the energy eigenbasis by a unitary transformation 
 according to 
\begin{eqnarray} 
\begin{array}[t]{cclcccl}
|L_1\rangle & = &\frac{1}{\sqrt 2}\left(|1\rangle - |2\rangle\right)\, , & &
|R_1\rangle & = &\frac{1}{\sqrt 2}\left(|1\rangle + |2\rangle\right)\, ,  \\ 
|L_2\rangle & = &\frac{1}{\sqrt 2}\left(|3\rangle - |4\rangle\right)\, , & & 
|R_2\rangle & = &\frac{1}{\sqrt 2}\left(|3\rangle + |4\rangle\right)\, . 
\end{array} \label{locbasis}
\end{eqnarray}
These states are localized in the left ($|L_j\rangle$) and 
in the right ($|R_j\rangle$) well  with lower ($j=1$) and higher 
($j=2$) energy, respectively.  The localized states are depicted in Fig.\ 
\ref{fig.dvrstates} a.) in position space. Shown is the double-well 
potential (thick solid line) for a barrier height of $E_{\rm B}
=\Delta U / \hbar 
\omega_0 = 1.4 $ (we use in the figures dimensionless quantities according 
to the standard scaling defined in the Appendix \ref{app.scaling}, 
Eq.\  (\ref{dwscale}) ). The energy 
eigenvalues ${\cal E}_1, ..., {\cal E}_4$ are marked by thin solid horizontal lines. 
The wave functions $\langle q|L_1\rangle$ (solid line) and 
$\langle q|L_2\rangle$ (dashed-dotted line) are localized in the 
left well,  and the wave functions $\langle q|R_1\rangle$ (dashed line) and 
$\langle q|R_2\rangle$ (long dashed line) are localized in the 
right well. 
In the literature \cite{DekkerGroup}, these localized states in Eq.\  
(\ref{locbasis})  are sometimes 
approximated by the eigenstates of  harmonic potentials shifted to the 
position of the well minima, cf.\  Appendix \ref{app.harmwell}. 
This approximation is justified for large 
barrier heights where, however, semiclassical techniques \cite{HanggiRMP90} to 
determine the quantum relaxation rate are already applicable. 
By use of basic algebra, the matrix for the 
bare system Hamiltonian of the double-doublet 
system in the localized basis is calculated to be
\begin{eqnarray}
  {\bf H}_{\rm DDS}^{\rm loc} & = & \sum_{i=1,2}-\frac{\hbar \Delta^{\cal E}_i}{2}
  (|R_i\rangle\langle L_i| +
 |L_i\rangle\langle R_i|) \nonumber \\
 & & 
+ \mbox{} \hbar \overline{\omega}_0 ( |R_2\rangle\langle R_2|+|L_2\rangle
\langle L_2|) \, ,
 \label{hamddsloc}
\end{eqnarray}
with frequencies $\Delta^{\cal E}_i$ and $\overline{\omega}_0$ defined above. 
The position operator in this 
localized representation then reads    
\begin{eqnarray}
  {\bf q}^{\rm loc} & = & \sum_{i,j=1,2}a_{ij}(|R_i\rangle\langle R_j| -
 |L_i\rangle\langle L_j|) + b(|L_1\rangle\langle R_2| \nonumber \\
 & & 
+ \mbox{} |R_2\rangle\langle L_1|-|R_1\rangle
\langle L_2|-|L_2\rangle\langle R_1|) \, , 
 \label{posoploc}
\end{eqnarray}
where  $a_{11}=\langle 1| {\bf q}| 2 \rangle$,  
 $a_{22}=\langle 3|{\bf q}| 4 \rangle$, $a_{12}=
 a_{21}=(\langle 1| {\bf q}| 4 \rangle 
 + \langle 2| {\bf q}| 3 \rangle)/2$ and 
 $b=(\langle 1| {\bf q}| 4 \rangle - \langle 2| {\bf q}| 3 \rangle)/2 
\ll a_{ij}$. 
 Note that, in clear contrast to the 
 spin-boson case $M=2$, the position operator in the 
 localized basis is {\em nondiagonal\/}. 
Since the energies  in the Hamiltonian are of different orders of 
magnitude, i.e., $\hbar \Delta^{\cal E}_1 \ll \hbar \Delta^{\cal E}_2 \ll 
\hbar \overline{\omega}_0$, the general 
time evolution of an initial state proceeds on 
different time scales.
The coherent dynamics exhibits transitions between the wells due to {\em 
tunneling\/}. 
It occurs in the lower doublet on a time scale $(\Delta^{\cal E}_1)^{-1}$ 
and in the upper doublet on a much shorter time scale 
$(\Delta^{\cal E}_2)^{-1}$, being 
 still long compared to the time scale $\overline{\omega}_0^{-1}$ 
 of the interdoublet dynamics.  
The coupling to the heat bath  is mediated by the position operator while   
the interdoublet transitions  are responsible for vibrational relaxation.

For the following analytical treatment, we simplify the approach by 
setting  $b=0$ in Eq.\  
(\ref{posoploc}). 
 This is  for the sake of an illustrative purpose only and has 
 no impact on the path integral formalism introduced above. 
 For specific results, the diagonalization of the position operator 
 is performed numerically on the computer with $b \ne 0$.   
By means of ordinary diagonalization performed for 
the matrix in Eq.\ (\ref{posoploc}) the DVR-states 
read  
\begin{eqnarray}  
\begin{array}[t]{cclcccl}
|\alpha_1\rangle & = &v(|L_1\rangle - u | L_2 \rangle)\, , & &
|\beta_1\rangle & = &v(|R_1\rangle - u | R_2 \rangle)\, ,  \\ 
|\alpha_2\rangle & = &v(u|L_1\rangle +| L_2 \rangle)\, , & & 
|\beta_2\rangle & = &v(u|R_1\rangle +| R_2 \rangle)\, , 
\end{array} \label{dvrbasis}
\end{eqnarray}
 with $| \alpha_j \rangle$  ($| \beta_j \rangle$)  being
  localized in the
 left (right) well, respectively. 
 Here, $v=1/\sqrt{1+u^2}$ and $u=(a_{11}+q_{\alpha_1})/a_{12}=-(a_{22}+
 q_{\alpha_2})/a_{12}$, 
 and $q_{\alpha_i}=-q_{\beta_i}$ denote the position eigenvalues:
\begin{equation} 
q_{\alpha_{1,2}}=\left[-(a_{11}+a_{22})\mp \sqrt{
 (a_{11}-a_{22})^2+4a_{12}^2}\right]/2 \, . \label{poseigvaldds} 
\end{equation}
The four DVR-states are depicted in Fig.\  \ref{fig.dvrstates} b.) for a  
barrier height of $E_{\rm B} = \Delta U / \hbar \omega_0 = 1.4$, i.e., 
$\langle q|\alpha_1\rangle$ (solid line),  
$\langle q|\alpha_2\rangle$ (dashed line),
$\langle q|\beta_2\rangle$ (dashed-dotted line), and   
$\langle q|\beta_1\rangle$ (long-dashed line). 
 On the $q$-axis, the exact eigenvalues $q_{\mu}$ are  marked by crosses 
 (the eigenvalues are obtained by  numerical diagonalization 
 of the position operator in Eq.\  (\ref{posoploc})). 
 As expected, the DVR-states 
are localized around their corresponding position eigenvalue $q_{\mu}$.
 
 It is suggestive to call transitions between the left and right well, i.e., 
 between 
 $| \alpha_i \rangle$ 
 and $| \beta_j \rangle$ as 
 {\em DVR-tunneling\/}. These are characterized by the 
 effective tunneling matrix elements  
\begin{eqnarray}
\Delta_{\alpha_1\beta_1} &\equiv &v^2(\Delta^{\cal E}_1+u^2\Delta^{\cal E}_2)\;,
\; \Delta_{\alpha_2\beta_2}\equiv v^2(u^2\Delta^{\cal E}_1+\Delta^{\cal E}_2)\;,
\; \nonumber\\
\Delta_{\alpha_1\beta_2}& =& \Delta_{\alpha_2\beta_1} 
\equiv v^2u(\Delta^{\cal E}_1-\Delta^{\cal E}_2)\, , 
\label{dvrtunnel}
\end{eqnarray}
 which constitute   
 a {\em linear combination\/} of the bare 
 tunneling splittings $\Delta^{\cal E}_1$ and 
$\Delta^{\cal E}_2$. 
On the other hand, transitions within one well, i.e., 
between $| \alpha_i \rangle$ and $| \alpha_j \rangle$ and between 
$| \beta_i \rangle$ and $| \beta_j \rangle$, may be 
termed {\em DVR-vibrational relaxation\/}.  Those can be characterized 
by the transition matrix elements 
\begin{equation}
 \Delta_{\alpha_1\alpha_2}=\Delta_{\alpha_2\alpha_1}=
 \Delta_{\beta_1\beta_2}=\Delta_{\beta_2\beta_1} 
 \equiv \Delta_{\rm R} = 2v^2u\overline{\omega}_0\, .
 \label{dvrvibra}
\end{equation}
Due to parity symmetry, they assume equal values. 
The Hamiltonian of the double-doublet system in the DVR-basis 
can thus be written as
\begin{eqnarray}
{\bf H}_{\rm DDS}^{\rm DVR} & = & -\sum_{i,j=1,2}
\frac{1}{2}\hbar\Delta_{\alpha_i\beta_j}
(|\alpha_i\rangle\langle \beta_j|+|\beta_j\rangle\langle \alpha_i|)
  \mbox{} - \frac{1}{2}\hbar \Delta_{\rm R} {\bf R} \nonumber \\
  & & \mbox{} + \sum_{i=1,2}\hbar\, (F_{\alpha_i}|\alpha_i\rangle\langle \alpha_i|+
F_{\beta_i}|\beta_i\rangle\langle \beta_i|) \, ,
 \label{hamddsdvr}
 \end{eqnarray} 
 with $F_{\alpha_1}=F_{\beta_1}=u^2v^2\overline{\omega}_0$,  
 $F_{\alpha_2}=F_{\beta_2}=v^2\overline{\omega}_0$. The operator ${\bf R}$
 accounts for DVR-vibrational relaxation, i.e., 
 ${\bf R} =|\alpha_1\rangle\langle \alpha_2|+|\alpha_2\rangle\langle 
 \alpha_1|+|\beta_1\rangle\langle \beta_2|+|\beta_2\rangle\langle \beta_1|$. 
Thereby, the time-independent problem is fully characterized. 
The time-dependent driving $s(t)=s \, \sin (\Omega t)$ couples 
to the position operator ${\bf q}$. The total system Hamiltonian 
in the DVR-basis thus reads 
\begin{equation}
{\bf H}_{\rm S}^{\rm DVR} (t) = {\bf H}_{\rm DDS}^{\rm DVR} - 
s \, \sin (\Omega t) \sum_{i=1,2} q_{\alpha_i} 
(\ket{\alpha_i} \bra{\alpha_i} - \ket{\beta_i} \bra{\beta_i}) \, .
\label{tdhamddsdvr}
\end{equation}
Note that in the DVR-basis the time-dependence enters only in the 
diagonal elements of the Hamiltonian.
So far, we have provided all required parameters for the 
path integral formalism developed in the previous Section  
\ref{subsec.boundcond}, namely the 
matrix elements in Eq.\  (\ref{hamddsdvr}) of the Hamiltonian in the 
DVR-basis, and the position operator eigenvalues in Eq.\  (\ref{poseigvaldds}). 
In the following Section, we return to the dissipative real-time 
path integral formalism and 
develop a suitable approximation scheme to the exact expression in 
Eq.\  (\ref{fullpathint}). 
%
\section{The generalized non-interacting cluster approximation}
\label{sec.gnica}
In the context of dissipative real-time path integrals, a common 
strategy of approximative treatment is as follows: It concerns 
the treatment of 
the interactions between different paths which are induced by the 
coupling to the heat bath and  which are described by the 
influence functional. The working idea behind the strategy 
\cite{Leggett8187,Weiss99} 
is to neglect some of those 
correlations in order to get  tractable expressions. 
One possible approximation within the spin-boson problem ($M=2$) is the 
so-termed {\em non-interacting blip approximation (NIBA)\/} \cite{Leggett8187}. 
There,    
the interactions between off-diagonal states (blips) are neglected. 
In the NIBA the sojourn-blip interactions are disregarded except 
neighboring ones, and even those are  treated approximately.  Within the NIBA, 
the influence phase simplifies drastically and the path integral series in Eq.\  
(\ref{fullpathint}) reduces to terms which are of lowest order in the level 
splitting $\Delta^{\cal E}_1$. 
The NIBA can be justified in the case of Ohmic damping 
for high enough temperatures and/or large dissipation strength. 
In this regime, the average blip length is small compared to the 
average sojourn length, and the blip-blip and the blip-sojourn interactions 
can consequently be neglected. 
In fact, for finite temperatures and Ohmic damping, 
long blips are exponentially suppressed by the intrablip interactions. 
The NIBA fails for systems with asymmetry when the friction becomes 
weak; however, it becomes a systematic weak-damping approximation down to zero 
temperature for the case of a symmetric spin-boson system 
\cite{Leggett8187,HanggiRMP90,Weiss99,VCHBuch,Grifoni98}. 

Improved approximations take into account some of the correlations between 
the blips. 
One such step of  an improved approximation has been denoted as the 
{\em interacting blip chain approximation (IBCA)\/} \cite{Winterstetter97}. 
There,  the interactions 
of all nearest-neighbor blip pairs and the full interactions of the 
nearest-neighbor sojourn-blip pairs are taken into account in addition. 
This improved approximation confirms the validity of the NIBA in 
the stated parameter regime. It 
is valid also in an extended parameter regime where 
the NIBA already breaks down.  

The NIBA being applicable for a spin-boson system, i.e., a system 
with two tight binding sites 
has been generalized to the case with 
arbitrary many tight binding 
sites 
by Egger, Mak and Weiss \cite{Egger94}. In  their work 
only tunneling transitions between {\em nearest-neighbor sites\/} are 
considered. The multisite paths along the discrete states of the 
reduced density matrix result in a sequence of {\em sojourns} (time intervals 
with the system being in a diagonal state) and {\em clusters} (time intervals 
with the system being in an off-diagonal state). 
It turns out that the corresponding path weights 
of the clusters sum up to zero. 
Consequently, the clusters can be considered as 
neutral objects. This suggests to neglect all interactions between the 
clusters  
yielding the {\em non-interacting cluster approximation (NICA)\/}. 
In the time-{\em in\/}dependent problem considered in Ref.\  \cite{Egger94}, the 
time-integrations over the sojourn times in the path integral appear as 
convolutions. This feature makes the expression solvable  
by means of Laplace transforms. 

Motivated by the NICA, we here generalize it to the case with many levels 
by observing that the multilevel problem can be mapped onto the multisite one 
 when also 
tunneling between non-nearest neighbor sites are considered. 
Moreover, we generalize the approach in Ref.\  \cite{Egger94} by taking 
into account a time-dependent system Hamiltonian, together with a 
general initial reduced density matrix. 

The key argument in Ref.\ \cite{Egger94}  refers to the overall neutrality 
 of a cluster because the cumulative charge is zero. 
 This is also the case for 
a general multi-level path integral in Eq.\  (\ref{fullpathint}). 
To show this, we 
consider a general cluster at a diagonal state 
$(\mu_k, \nu_k=\mu_k)$ at time $t_k$. It subsequently  travels 
around among arbitrary many 
off-diagonal states and re-enters at time $t_l$ a diagonal state 
$(\mu_l, \nu_l=\mu_l)$. The cumulative charge $W_{\rm cl}$ 
of this cluster is the sum over all individual path weights defined in Eq.\ 
 (\ref{fourindexweightxi}), i.e., 
\begin{eqnarray}
W_{\rm cl}& = & \sum_{j=k+1}^l \xi_{j}\nonumber \\
& = & (q_{\mu_l} - q_{\nu_l} - q_{\mu_{l-1}} + q_{\nu_{l-1}}) + 
(q_{\mu_{l-1}} - q_{\nu_{l-1}} - q_{\mu_{l-2}} + q_{\nu_{l-2}}) \nonumber \\
& & \mbox{} + \dots +  (q_{\mu_{k+2}} - q_{\nu_{k+2}} 
- q_{\mu_{k+1}} + q_{\nu_{k+1}})+  (q_{\mu_{k+1}} - q_{\nu_{k+1}} 
- q_{\mu_{k}} + q_{\nu_{k}}) \nonumber \\
& = & q_{\mu_l} - q_{\nu_l} - q_{\mu_{k}} + q_{\nu_{k}} \nonumber \\[4mm]
& = & 0\, ,
\label{clusterneutrality}
\end{eqnarray}
because $\nu_k=\mu_k$ and $\nu_l=\mu_l$. In this language, any path 
is just a sequence of sojourns and clusters, where the $\xi$-charges 
within each cluster sum up to zero. 

In general, the influence functional in Eq.\  (\ref{discrfeynman}) couples the 
$\chi$-charges {\em inside\/} each cluster with the $\xi$-charges {\em inside\/}  
all other clusters. Similarly, all $\xi$-charges are coupled 
to each other. These interactions consequently 
render the path summation intractable. 

Since the entire cluster can 
be seen as a neutral object which is only weakly interacting with 
all other clusters which are themselves neutral, 
it is suggestive to {\em neglect all  the intercluster interactions\/} in the 
influence phase in Eq.\  (\ref{fullpathint}). However, all,  the 
{\em intra\/}cluster interactions, as well as all interactions of a cluster 
with the preceding sojourn are fully taken into account (see below). 
For a path starting in an off-diagonal state, we call  
 that part of a path which precedes the first sojourn a 
 {\em semi-cluster\/}. Within our approximative 
 description, we neglect the interaction of this semi-cluster with 
all the later clusters but take into account the correlation 
of the first sojourn with the preceding semi-cluster.  
This ``coarse-graining'' is 
performed for general transitions. 
%
We call this the {\em generalized non-interacting cluster 
approximation (gNICA)\/}. 

Before we exploit the consequences of this approximation, 
we discuss its regime of validity. 
The gNICA is justified 
when the average cluster length is small compared to the average sojourn 
length. This is fulfilled for high temperatures and/or strong damping.  
Far excursions from the diagonal state are damped exponentially, see  
Eq.\  (\ref{infl}) for the 
influence phase. 
As such, the gNICA becomes {\em exact\/} in the limit of high temperatures. 
%
%

For the case of the spin-boson-system  
 at low temperature $T$, small friction $\gamma$ and 
no bias ($\varepsilon=0$), the 
interblip correlations are only of second order 
in the coupling strength $\gamma$ while the {\em intra\/}blip 
ones are of linear order in $\gamma$. Hence, the gNICA is a good 
approximation down to zero temperature \cite{Weiss99}. 
However, with $M>2$, lowest-order 
contributions to interblip correlations arise due to the 
non-zero diagonal elements $F_{\nu}$ in Eq.\  (\ref{hamddsdvr}). 
This yields a rough condition of validity for the 
gNICA; it reads: $\hbar \Delta^{\cal E}_{\rm max} < k_{\rm B} T$ 
or $\Delta^{\cal E}_{\rm max} 
 \lesssim  \gamma$, where 
 $\Delta^{\cal E}_{\rm max} = {\rm max} \{\Delta^{\cal E}_1, \Delta^{\cal E}_2, 
 \dots \}$. On the other hand,  
an upper limit for the allowed values of the damping constant can 
be extracted by the following argumentation: The damping 
leads to a level broadening of the unperturbed eigenenergies. 
This is seen best  in the  form in Eq.\  (\ref{infl}) for the 
influence phase. The damping could be viewed as an 
additional contribution to the bare system propagator. The  
contribution is of stochastic nature and implies the level broadening. 
In order that a tunneling description makes sense, this 
frictional level broadening should not exceed the 
bare interdoublet level spacing, i.e., $\gamma \ll \omega_0$. 
%
  This condition is not really restrictive because 
 friction strengths of the order of the oscillator frequency, 
 i.e., $\gamma\approx \omega_0$,  would indeed 
  strongly suppress  quantum effects.

%
\section{The generalized master equation in the 
discrete variable representation}
\label{sec.gme}
%
\subsection{General derivation} 
\label{subsec.generalgme}
%
 
First, we address the non-driven case. 
We start by observing that every path 
which begins and ends in a diagonal state 
can be seen as a sequence of $p$ clusters punctuated by 
sojourns. 
For paths starting out at an off-diagonal and ending in a diagonal state,  
also the initial semi-cluster appears. Within the gNICA prescription, 
it is now straightforward to see that 
the integrations over the sojourn times in 
Eq.\  (\ref{fullpathint}) {\em appear as convolutions\/}! To use this property 
effectively, 
we switch to the Laplace transform $\rho_{\mu_N \mu_N}(\lambda)$. 
It then follows that the integration over each sojourn 
contributes a factor  $\lambda^{-1}$, 
while each cluster yields a factor which depends on the number of 
charges and on their configuration inside that particular 
cluster according to Eq.\  (\ref{fullpathint}). This very  
point is elucidated with an example in Appendix \ref{app.examplepath};  
there, we present 
%
%
%
%
%
in detail the contribution of 
one specific path to the full path sum. In a second step,  
we generalize this idea. 

We consider transitions from the initial 
state $(\mu_0,\nu_0)$ at time $t_0$ to 
the final state $(\mu_N,\mu_N)$ at time $t_N$. In doing so, we must  
distinguish between two cases: (i) the initial state is a diagonal state, i.e., 
$\mu_0 = \nu_0$ and (ii) the initial state is an off-diagonal state, i.e., 
$\mu_0 \ne \nu_0$. We separate the contributions to the path 
sum and obtain then for the Laplace transform 
$\rho_{\mu_N \mu_N} (\lambda) = \int_0^{\infty} dt \, e^{-\lambda t} 
\rho_{\mu_N \mu_N} (t)$ the expression 
\begin{equation}
\rho_{\mu_N \mu_N} (\lambda) = \sum_{\mu_0=1}^M \rho_{\mu_0 \mu_0} 
\rho_{\mu_N \mu_N, {\rm D}} (\lambda) + 
\sum_{\stackrel{\mu_0, \nu_0 =1}{\mu_0 \ne \nu_0 }}^M \rho_{\mu_0 \nu_0} 
\rho_{\mu_N \mu_N, {\rm O}} (\lambda)\, , 
\label{rhosep}
\end{equation}
where $\rho_{\mu_N \mu_N, {\rm D}}$ ($\rho_{\mu_N \mu_N, {\rm O}}$) 
denotes the contribution of the diagonal (off-diagonal) initial part, 
respectively. 

To proceed, we need to consider an arbitrary cluster which 
%
begins   
in the diagonal state $(\mu_i,\nu_i=\mu_i)$   
at time $t_i$ 
and ends in the 
diagonal state $(\mu_j,\nu_j=\mu_j)$ at time $t_j$. 
We sum over {\em all\/} the path configurations 
 and denote this collected contribution 
the {\em cluster function\/} $h_{\mu_j \mu_i}(\lambda)$.
Proceeding as in 
Appendix \ref{app.examplepath} yields for 
the cluster function
\begin{eqnarray}
h_{\mu_j \mu_i}(\lambda) & = &  
\sum_{m=|i-j|}^{\infty} \int_{0}^{\infty} d\tau_1 \dots  
\int_{0}^{\infty} d\tau_{m-1}   
\exp\{ - \lambda (\tau_1 + \dots + \tau_{m-1})\} \nonumber \\
&& \times \sum_{\stackrel{ \{ \mu_k \nu_k \} }{\mu_k \ne \nu_k}} 
\exp \left\{i \, \sum_{k=1}^{m-1} 
\int_{\sum_{l=1}^{k} \tau_l}^{\sum_{l=1}^{k+1} \tau_l} 
dt'  [ E_{\mu_{k+i}}(t') - E_{\nu_{k+i}}(t')]\right\} 
\nonumber \\ 
& & \times 
\prod_{k=0}^{m} (-1)^{\delta_k} \left(\frac{i}{2}\right)^m 
\Delta_{k+i} \nonumber \\
& & \mbox{} \times  \exp \left\{
 \sum_{l=2}^{m} \sum_{k=1}^{l-1} \xi_{l+i} 
S\left(\sum_{n=k}^{l-1} \tau_n\right) \xi_{k+i} + 
 i \sum_{l=2}^{m} \sum_{k=1}^{l-1} \xi_{l+i} 
R\left(\sum_{n=k}^{l-1} \tau_n\right) \chi_{k+i}\right\} \nonumber \, . \\
\label{fullpsum}
\end{eqnarray}
with the difference times $\tau_k = t_k - t_{k-1}$ and 
with the conventions and notations taken from Eq.\  (\ref{fullpathint}). 

Each contribution to 
$\rho_{\mu_N \mu_N, {\rm D}} (\lambda)$ can be viewed as a sequence of 
sojourns punctuated by clusters. Thus, in the first case (i), we sum up the 
contributions of {\em all\/} paths which 
start in $(\mu_0, \nu_0=\mu_0)$ and end in $(\mu_N, \nu_N=\mu_N)$ 
and which contain $p$ clusters starting in some intermediate diagonal states
$(\sigma_k,\sigma_k)$ and ending in $(\sigma_{k+1},\sigma_{k+1})$, i.e., 
\begin{equation}
\rho_{\mu_N \mu_N, {\rm D}}^{(p)} 
(\lambda) = \sum_{\sigma_1, \sigma_2, \dots, \sigma_p} 
\frac{1}{\lambda} h_{\sigma_1,\mu_0}(\lambda)
\frac{1}{\lambda} h_{\sigma_2,\sigma_1} (\lambda)
\dots 
 h_{\mu_N,\sigma_p} (\lambda) \frac{1}{\lambda} \, ,
 \label{pclustersum}
\end{equation}
where the sum runs over all possible intermediate diagonal states 
$\sigma = 1, \dots, M$. The factors $1/\lambda$ are the results of the 
integration over the sojourns, see Appendix \ref{app.examplepath}. 

In the second case (ii), where the initial state is an off-diagonal state, 
we assume that the path travels among off-diagonal ones and hits after $d$ 
transitions  {\em for the 
first time\/} a diagonal state $(\kappa_d, \kappa_d)$ at time $t_d$. This part 
of the path is termed {\em semi-cluster\/} and the interaction 
with all the other clusters is neglected according to the gNICA.
The sum of all such semi-clusters that start in $(\mu_0, \nu_0)$ 
and end in $(\kappa_d, \kappa_d)$ results in a semi-cluster function 
$f_{\kappa_d \kappa_d, \mu_0\nu_0} (\lambda)$. From time 
$t_d$ on, the formalism from (i) is applied. Summing over 
all possible diagonal states $\kappa_d = 1, \dots, M$  yields the 
contribution to the path sum
\begin{equation}
\rho_{\mu_N \mu_N, {\rm O}}^{(p)} (\lambda) = \sum_{\kappa_d=1}^M 
\left\{
f_{\kappa_d \kappa_d, \mu_0\nu_0} (\lambda) 
\sum_{\sigma_1, \sigma_2, \dots, 
\sigma_p} 
\frac{1}{\lambda} h_{\sigma_1, \kappa_d}(\lambda)
\frac{1}{\lambda} h_{\sigma_2,\sigma_1}(\lambda) 
\dots 
 h_{\mu_N, \sigma_p}(\lambda) \frac{1}{\lambda} \right\} \, ,
\label{semiclustersum}
\end{equation}
where the semi-cluster function is given as  
\begin{eqnarray}
f_{\kappa_d \kappa_d, \mu_0\nu_0} (\lambda) & = &  
\sum_{m=1}^{\infty} \int_{0}^{\infty} d\tau_1 \dots  
\int_{0}^{\infty} d\tau_{m-1}   
\exp\{ - \lambda (\tau_1 + \dots + \tau_{m-1})\} \nonumber \\
&& \times \sum_{\stackrel{ \{ \mu_k \nu_k \} }{\mu_k \ne \nu_k}} 
\exp \left\{i \, \sum_{k=1}^{m-1} 
\int_{\sum_{l=1}^{k} \tau_l}^{\sum_{l=1}^{k+1} \tau_l} 
dt'  [ E_{\mu_{k}}(t') - E_{\nu_{k}}(t')]\right\} 
\nonumber \\ 
& & \times 
\prod_{k=1}^{m} (-1)^{\delta_k} \left(\frac{i}{2}\right)^m 
\Delta_{k} \nonumber \\
& & \mbox{} \times  \exp \left\{
 \sum_{l=2}^{m} \sum_{k=1}^{l-1} \xi_{l} 
S\left(\sum_{n=k}^{l-1} \tau_n\right) \xi_{k} + 
 i \sum_{l=2}^{m} \sum_{k=1}^{l} \xi_{l} 
R\left(\sum_{n=k}^{l-1} \tau_n\right) \chi_{k}\right\} \nonumber \, . \\
\end{eqnarray}
%
Defining the {\em cluster matrix\/} ${\cal H}(\lambda)$ with the matrix 
elements $h_{\mu_j \mu_i}(\lambda)$, we can rewrite Eq.\  (\ref{pclustersum})
and the inner sum in Eq.\ (\ref{semiclustersum}) 
as a matrix product, i.e., 
\begin{eqnarray}
\rho_{\mu_N \mu_N, {\rm D}}^{(p)} (\lambda) & = & 
\frac{1}{\lambda} \left\{ \left[ \frac{{\cal H}(\lambda)}{\lambda}  \right]^p
\right\}_{\mu_N \mu_0} \, , \label{clusterd}\\
 \rho_{\mu_N \mu_N, {\rm O}}^{(p)} (\lambda) & = &
  \sum_{\kappa_d=1}^M \left\{ f_{\kappa_d \kappa_d, \mu_0\nu_0} (\lambda)  
\frac{1}{\lambda} \left\{ \left[ \frac{{\cal H}(\lambda)}{\lambda}  \right]^p
\right\}_{\mu_N \kappa_d}  \right\} \, .
 \label{clustero}
\end{eqnarray}
In a last step,  
the summation over all possible numbers $p$ of clusters within a path 
has to be performed. This last sum can be formally recast, 
yielding 
\begin{eqnarray}
\rho_{\mu_N \mu_N, {\rm D}} (\lambda) & = & 
\left\{ \frac{1}{\lambda -{\cal H}(\lambda)}  \right\}_{\mu_N \mu_0}
 \, , \label{calH1}\\
 \rho_{\mu_N \mu_N, {\rm O}} (\lambda) & = &
  \sum_{\kappa_d=1}^M \left\{ f_{\kappa_d \kappa_d,\mu_0\nu_0} (\lambda)  
\left\{ \frac{1}{\lambda -{\cal H}(\lambda)}  \right\}_{\mu_N \kappa_d} \right\}\, .
 \label{calH2}
\end{eqnarray}
We insert this result into Eq.\  (\ref{rhosep}), exchange the order 
of summation in the second term of the r.h.s.\ and  end up with
\begin{equation}
\rho_{\mu_N \mu_N} (\lambda) = \sum_{\mu_0=1}^M \rho_{\mu_0 \mu_0} 
\left\{ \frac{1}{\lambda -{\cal H}(\lambda)}  \right\}_{\mu_N \mu_0} 
+
\sum_{\kappa_d=1}^M 
\left\{ \frac{1}{\lambda -{\cal H}(\lambda)}  \right\}_{\mu_N \kappa_d} 
i_{\kappa_d \kappa_d} (\lambda) \,  
\label{rhosep2}
\end{equation}
with
\begin{equation}
i_{\kappa_d \kappa_d} (\lambda) = 
\sum_{\stackrel{\mu_0, \nu_0 =1}{\mu_0 \ne \nu_0 }}^M \rho_{\mu_0 \nu_0} 
f_{\kappa_d \kappa_d, \mu_0\nu_0} (\lambda) \, .
\label{rhosep3}
\end{equation}
Eq.\  (\ref{rhosep2}) can be viewed as a vector equation with two 
vector-matrix products on the r.h.s..  For convenience we introduce 
a vector-matrix notation. $\rho_{\mu_N \mu_N} (\lambda)$ 
then appear as elements of a vector $\vec{\rho}(\lambda)$. The 
initial populations $\rho_{\mu_0 \mu_0}$ are arranged in the vector 
$\vec{\rho}_0$ and the initial off-diagonal elements are contained 
in the vector $\vec{I}(\lambda)$ with the elements 
$i_{\kappa_d \kappa_d} (\lambda)$. 
In this notation, Eq.\  (\ref{rhosep2}) reads
\begin{equation}
\vec{\rho}(\lambda) = \frac{1}{\lambda -{\cal H}(\lambda)} \vec{\rho}_0 + 
\frac{1}{\lambda -{\cal H}(\lambda)} \vec{I}(\lambda) \, .
\label{vector1}
\end{equation}
Multiplying  Eq.\  (\ref{vector1}) with the inverse 
of the matrix $\frac{1}{\lambda -{\cal H}(\lambda)}$ and rearranging 
the equation, we find
\begin{equation}
\lambda \vec{\rho}(\lambda) -\vec{\rho}_0 = {\cal H}(\lambda) \vec{\rho}(\lambda) 
+ \vec{I}(\lambda) \, .
\label{vector2}
\end{equation}
Finally, we perform the inverse Laplace transform and end up with the 
equation
\begin{equation}
\dot{\vec{\rho}} (t) = \int_{t_0}^t dt' \, {\cal H} (t - t') \vec{\rho}(t') 
+ \vec{I}(t-t_0) \, 
\label{vecgme}
\end{equation}
or in the original notation for the single components
\begin{equation}
\dot{\rho}_{\mu\mu} (t) = \sum_{\nu=1}^{M} \, \int_{t_0}^t dt' \, 
{\cal H}_{\mu \nu} (t- t') \rho_{\nu \nu} (t') + I_{\mu} (t-t_0) \, ,  
\mbox{\hspace{3ex}} \mu=1, ..., M \,  .
\label{thegme} 
\end{equation}
The overdot denotes the derivative with respect to time $t$. 
The initial conditions for Eq.\  (\ref{thegme}) are $\rho_{\mu\nu} (t_0) = 
\rho_{\mu_0\nu_0}$. 
Eqs.\  (\ref{vecgme}) and (\ref{thegme}) are of convolutive form 
since a time-{\em in\/}dependent Hamiltonian was assumed. 
A similar, although technically more involved line of reasoning 
must be used for the {\em driven\/} case. We find equations 
similar to  Eqs.\  (\ref{vecgme}) and (\ref{thegme}) with 
${\cal H}_{\mu \nu} (t- t') \curvearrowright 
{\cal H}_{\mu \nu} (t, t')$ and 
$I_{\mu} (t-t_0) \curvearrowright I_{\mu} (t,t_0)$. To be explicit, 
the elements of the rate matrix ${\cal H}_{\mu \nu} (t, t')$ are  
in the general time-dependent case given as 
\begin{eqnarray}
{\cal H}_{\mu \nu} (t, t') & = & 
\sum_{N=2}^{\infty} \int_{t'}^t {\cal D} \{t_j\} 
\sum_{\stackrel{ \{ \mu_j \nu_j \} }{\mu_j \ne \nu_j}} 
\exp \left\{i \, \sum_{j=0}^{N-1} 
\int_{t_{j}}^{t_{j+1}} dt''  [ E_{\mu_{j}}(t'') - E_{\nu_{j}}(t'')]\right\} 
\nonumber \\
& & \mbox{} \times 
\prod_{j=0}^{N-1} (-1)^{\delta_j} \left(\frac{i}{2}\right)^N 
\Delta_{j} \nonumber \\
& & \mbox{} \times  \exp \left\{
 \sum_{l=1}^{N} \sum_{j=0}^{l-1} \xi_{l} 
S(t_l-t_j) \xi_{j} + 
 i \sum_{l=1}^{N} \sum_{j=0}^{l-1} \xi_{l} 
R(t_l-t_j) \chi_{j} \right\} \, .
\label{kernels}
\end{eqnarray}
The inhomogeneity $I_{\mu} (t,t_0)$ arises because of the contributions of the 
non-diagonal initial states; its explicit form reads 
\begin{eqnarray}
I_{\mu} (t,t_0) &=&  
\sum_{\stackrel{\mu_0, \nu_0 =1}{\mu_0 \ne \nu_0 }}^M \rho_{\mu_0 \nu_0} 
f_{\mu_0\nu_0, 
\mu \mu } (t,t_0) \nonumber \\
& = & 
\sum_{\stackrel{\mu_0, \nu_0 =1}{\mu_0 \ne \nu_0 }}^M \rho_{\mu_0 \nu_0} 
\sum_{m=1}^{\infty} \int_{t_0}^t {\cal D} \{t_j\} 
\sum_{\stackrel{ \{ \mu_j \nu_j \} }{\mu_j \ne \nu_j}} 
\exp \left\{i \, \sum_{j=0}^{m-1} 
\int_{t_{j}}^{t_{j+1}} dt''  [ E_{\mu_{j}}(t'') - E_{\nu_{j}}(t'')]\right\} 
\nonumber \\
& & \mbox{} \times 
\prod_{j=0}^{m-1} (-1)^{\delta_j} \left(\frac{i}{2}\right)^m 
\Delta_{j} \nonumber \\
& & \mbox{} \times  \exp \left\{
 \sum_{l=1}^{m} \sum_{j=0}^{l-1} \xi_{l} 
S(t_l-t_j) \xi_{j} + 
 i \sum_{l=1}^{m} \sum_{j=0}^{l-1} \xi_{l} 
R(t_l-t_j) \chi_{j} \right\} \, .
\label{inhomogenities}
\end{eqnarray}
The integro-differential equation (\ref{thegme}) is called the 
{\em generalized master equation (GME)\/}; it constitutes 
one {\em central result\/} of this work. 

In the following, we will see that the inhomogeneity in 
Eq.\  (\ref{inhomogenities}) 
plays an important role at short times. However, it will become
 exponentially suppressed   at long 
times reflecting the fact that the asymptotic state is independent 
of the initial preparation.

We note that this integro-differential equation 
(\ref{thegme}) is represented in the DVR-basis for the 
diagonal elements of the reduced density matrix $\bfrho (t)$. 
For all practical 
calculations, the kernels in Eq.\  (\ref{kernels}) and the inhomogeneities 
in Eq.\  (\ref{inhomogenities}) have to be determined up to 
a certain order ${\cal O}(\Delta^N)$. In practice,  
this means that $N=\infty$ as the upper limit of the 
summations has to be replaced by a finite value. 

Some comments to elucidate the physical content of the GME (\ref{thegme}) 
are in order: 
 The transformation of the problem 
from the localized basis to the DVR-basis maps the dynamics of 
the particle in the spatially continuous potential  onto 
a hopping process of the particle on a spatially discrete grid. 
The grid points are 
the discrete positions characterized by the 
eigenvalues $q_{\kappa}, \kappa=1,..., M$, of the position operator 
${\bf q}$, according to Eq.\  (\ref{dvrtrafo}). 

Next, we consider the example of the double-doublet system which has already 
been introduced in Section \ref{subsec.dds}. We give the explicit 
expressions for the kernels in Eq.\  (\ref{kernels}) in the 
GME up to  second order in $\Delta_j$ 
and illustrate the damping mechanism further.
%
\subsection{The leading order approximation} 
\label{subsec.lowestorder}
In this Section we investigate the GME, Eq.\  (\ref{thegme}), with the 
kernels in Eq.\  (\ref{kernels}) and the inhomogeneities in Eq.\  
(\ref{inhomogenities})   
derived to lowest order in the Hamiltonian matrix elements $\Delta_{j}$. 
To illustrate the general scheme, we describe the method for the case of the 
double-doublet system with $M=4$ levels. 

As we want to evaluate the 
relaxation rate of an initially localized wave packet in one of the wells, say,  
the left well, we 
prepare the system in an equally weighted superposition of symmetric 
and antisymmetric wave function belonging to the {\em lowest\/} doublet, i.e., 
\begin{equation}
\bfrho (t_0) = \ket{L_1} \bra{L_1} \, , 
\label{initialstate}
\end{equation} 
where $|L_1\rangle  = \frac{1}{\sqrt 2}(|1\rangle - |2\rangle)$ and 
$\ket{n}, n=1,2$ are the nearly degenerate energy eigenstates of the 
static, symmetric Hamiltonian ${\bf H}_0$ (cf.\ Eq.\ (\ref{staticpotential}) 
 and Eqs.\  (\ref{locbasis})). Transforming this initial state 
to the DVR-basis via Eq.\  (\ref{dvrbasis}), we find 
\begin{equation}
\bfrho (t_0)=v^2(|\alpha_1\rangle\langle \alpha_1|
 +u^2|\alpha_2\rangle\langle \alpha_2| +u
 |\alpha_1\rangle\langle \alpha_2|+u |\alpha_2\rangle\langle \alpha_1|) 
\label{initialstatedvr}
\end{equation} 
with the parameters $u$ and $v$ defined below Eq.\  (\ref{dvrbasis}). 
We note that the initially prepared localized state is characterized 
by a {\em nondiagonal\/} density matrix in the DVR basis. 
The diagonal elements in Eq.\  
(\ref{initialstatedvr}) enter as initial conditions for the first 
part of the r.h.s.\ of Eq.\  (\ref{thegme}), while the 
off-diagonal elements determine the inhomogeneity. 
 
To first order, i.e., with one jump, no transition from an initial diagonal 
state to a final diagonal state is possible. To achieve this, at least two jumps 
are necessary. However, transitions starting in an off-diagonal 
state and ending in a diagonal state are possible within one jump. 
This means that 
a first order contribution appears only in the inhomogeneity of Eq.\  
(\ref{thegme}). The relevant transitions are the jumps ending in the 
diagonal state $(\alpha_1,\alpha_1)$, i.e., 
$(\alpha_1,\alpha_2) \rightarrow (\alpha_1,\alpha_1)$ and 
$(\alpha_2,\alpha_1) \rightarrow (\alpha_1,\alpha_1)$, 
and the jumps ending in $(\alpha_2,\alpha_2)$, i.e., 
$(\alpha_1,\alpha_2) \rightarrow (\alpha_2,\alpha_2)$ and 
$(\alpha_2,\alpha_1) \rightarrow (\alpha_2,\alpha_2)$, respectively. 
From  Eq.\ (\ref{fullpathint}),  
it follows that each path traveling ``above'' the diagonal has a 
corresponding mirror path traveling ``below'' the diagonal. The mirror 
path yields a contribution to the path sum being the complex conjugate 
of the upper path. Using this feature and the fact that 
$\rho_{\alpha_1 \alpha_2} (t_0)= \rho_{\alpha_2 \alpha_1}^*(t_0) = u v^2$, we 
obtain for the inhomogeneity the following first-order expression
 \begin{eqnarray} 
 I_{\mu}^{(1)}(t,t_0)& =& (\delta_{\mu\alpha_1}-\delta_{\mu\alpha_2}) \;  
 u v^2 \; 
 \Delta_{\alpha_1\alpha_2} 
 \exp \left\{- (q_{\alpha_1}-q_{\alpha_2})^2 \; S(t-t_0)\right\} \nonumber \\
 &  & \mbox{} \times 
 \sin\left\{\int_{t_0}^t dt'\, [E_{\alpha_2}(t')-E_{\alpha_1}(t')]
-(q_{\alpha_1}-q_{\alpha_2})^2 \; R(t-t_0)\right\} \; .
 \label{ddsinhomogenity} 
\end{eqnarray}
Here, $\Delta_{\mu\nu} = 
\mel{\mu}{{\bf H}_{\rm DDS}^{\rm DVR}}{\nu}, \mu\ne\nu$,  are   
the  off-diagonal matrix elements of 
the system Hamiltonian in Eq.\  (\ref{hamddsdvr}), 
see also Eq.\  (\ref{dvrvibra}). Note that   
{\em only\/} vibrational transitions {\em within\/} the initially 
populated well contribute in Eq.\  (\ref{ddsinhomogenity}).   
Moreover, the $q_{\kappa}$ $(\kappa=\alpha_1,\alpha_2)$ denote the position 
eigenvalues, see Eq.\ 
 (\ref{poseigvaldds}), and $E_{\kappa}(t')=F_{\kappa} - q_{\kappa} s\, \sin 
(\Omega t')$ are the time-dependent  diagonal elements of the system 
Hamiltonian, see in above  Eqs.\  (\ref{tdhamddsdvr}), (\ref{hamddsdvr}) 
and (\ref{diagele}). The influence of the bath 
enters via the real and imaginary part of the twice integrated 
 bath correlation function, i.e., 
$S(t)$ and $R(t)$, respectively, see Eq.\  (\ref{twintcorr}) and Appendix 
\ref{app.corr}. The conservation of probability is reflected 
with the opposite signs of the Kronecker symbols $\delta_{\nu\kappa}$.  
From Eq.\  (\ref{ddsinhomogenity}) it clearly follows that 
the contribution of the initial off-diagonal states are damped 
exponentially  on a time scale determined by the 
damping constant $\gamma$ and the temperature $T$. 
We recall that the lowest order 
of the contribution of the integral part of the GME, Eq.\  (\ref{thegme}),  
is of second order. This implies that  the 
contribution of second order to the inhomogeneity 
should also be taken into account for a consistent treatment. 
However, we refrain from writing 
down the complicated second order term which would yield only minor 
physical insight. It can be neglected anyhow when investigating 
the long-time dynamics in the following Sections. 

The lowest order for the kernels  in the integral 
part of the GME (\ref{thegme}) is the second order, because  
at least two jumps are required 
starting in a diagonal state to end again in a diagonal state. 
We use once more the feature that each path traveling above the diagonal has 
a mirror path traveling below the diagonal, yielding the complex 
conjugate of the upper path contribution. We then obtain for the 
GME kernels the leading order results 
\begin{eqnarray} 
{\cal H}_{\mu\nu}^{(2)}(t,t')&=&\frac{\Delta^2_{\mu\nu}}{2}
 \exp \left\{ -(q_{\mu}-q_{\nu})^2 \, S(t-t')\right\} \nonumber \\ 
&& \mbox{} \times \cos \left\{ \int_{t'}^t dt'' \, [ E_{\nu}(t'')-E_{\mu}(t'') ] 
 -(q_{\mu}-q_{\nu})^2 R(t-t') \right\}\, , \mbox{\hspace{3ex}} 
 \hfill \mu \ne \nu \, .\nonumber \\  
 \label{ddskernel}
\end{eqnarray}
The conservation of probability implies for the diagonal kernels the 
condition 
\begin{equation}
{\cal H}_{\nu \nu}^{(2)} (t, t')=-\sum_{ \kappa=1 \atop \kappa\ne\nu}^M
{\cal H}_{\kappa \nu }^{(2)}
(t, t') \, . %
\label{probconserv}
\end{equation} 
%
%

We emphasize here that the lowest-order expression in Eq.\ (\ref{ddskernel}) 
is  applicable to 
a {\em general\/} number $M$ of levels. The explicit example 
of the double-doublet system with $M=4$ is used for illustrative purpose only. 

Note  that the 
${\cal H}_{\mu \nu}^{(2)}$ represents the transition probability 
for a path starting in the diagonal state $(\nu, \nu)$, then jumping to 
the off-diagonal state $(\nu, \mu)/(\mu, \nu)$,  
and finally ending in the diagonal state $(\mu, \mu)$. 
In clear contrast to Eq.\  (\ref{ddsinhomogenity}),
 now tunneling and vibrational 
relaxation both contribute in Eq.\  (\ref{ddskernel}). 

The structure of the GME with the kernels ${\cal H}_{\mu \nu }^{(2)}$ 
restricted to leading, i.e., 
second, order is similar to that one obtained for the driven spin-boson 
system within the non-interacting blip approximation (NIBA)  
\cite{Leggett8187,Weiss99,Grifoni98}, 
and to that one for the dissipative  
tight-binding model  within the non-interacting 
cluster approximation (NICA) performed to lowest order 
\cite{Egger94}.  
The main difference to these GMEs is that in our case  
the factors $(q_{\mu}-q_{\nu})^2$  
enter as {\em prefactors\/} for the damping 
constant $\gamma$ in $S(t)$ and $R(t)$, respectively.
Since they arise from {\em non-nearest neighbor hopping\/} on a 
non-equally spaced grid of DVR eigenvalues, they  are {\em not equal\/} for 
all transitions. This means that 
transitions between far away lying DVR-states are stronger damped and 
therefore less probable compared to those 
lying close to each other.   This insight is especially relevant 
for the tunneling transitions from one well to the adjacent. Then, the 
main contribution to the dynamics comes from those two DVR-states which 
lie closest to the barrier within each well. 

One remark on the notation should be made: In the following, we use the 
superscript in the kernels ${\cal H}_{\mu\nu}^{(2)}$ when they are 
utilized  in 
second order. Whenever this superscript is omitted, the respective formula is 
valid to any order of ${\cal H}_{\mu\nu}$.

The generalized master equation (\ref{thegme}) is clearly not 
solvable analytically in closed form, 
not even with the kernels and the inhomogeneities 
approximated to lowest order. Thus, in Appendix \ref{app.numgme}, we provide  
a numerical iteration algorithm to obtain a numerical solution. 
\subsection{Comparison with numerical {\em ab-initio\/} path integral simulations} 
\label{subsec.quapicomp}
In this subsection we compare the results for $P_{\rm left} (t)$ 
obtained from the numerical solution of the GME (\ref{thegme}) with 
those of the numerical iterative algorithm using the method of 
the quasiadiabatic 
propagator path integral QUAPI of  Makri \cite{QUAPI}. It is known that 
the QUAPI technique yields reliable results for time-dependent 
spatially continuous confining potentials \cite{ThorwartPRE00}. Hence, we 
use it here as a reference in order to check the gNICA.

\begin{figure}[t]
\centerline{
\epsfig{figure=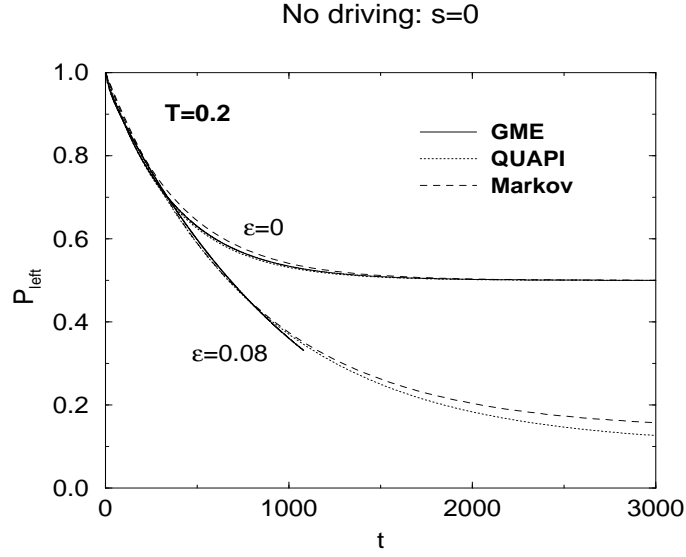,width=90mm,height=75mm,angle=0}}
\caption{ 
\label{fig.pleft1}
The probability $P_{\rm left}$ 
of finding the particle in the left well as 
a function of time for the symmetric ($\varepsilon=0$) and the 
asymmetric ($\varepsilon=0.08$) case. 
Considered is the system of two doublets, i.e.,  $M=4$. We 
 start from an initially fully localized state in the left well. 
The barrier height is set to $E_{\rm B}=1.4$. 
The temperature is $T=0.2$, the damping constant is $\gamma=0.1$ and 
the cut-off frequency is $\omega_c=10.0$. For this set of parameters, the 
dynamics is fully incoherent and the Markov approximation to the GME 
(\ref{thegme}) is rather satisfactory. 
}
\end{figure}

We present results for the double-doublet system $M=4$. 
Fig.\ \ref{fig.pleft1} depicts the outcome for $P_{\rm left} (t)$ for the 
symmetric ($\varepsilon=0$) and for the asymmetric ($\varepsilon=0.08$) system. 
Each figure contains three lines: (i) the results of the 
full generalized master equation (full line), (ii) findings of the 
QUAPI algorithm  which are used 
as a reference, and (iii) the outcome of a Markovian master equation which 
is introduced in the following Section \ref{sec.rate}. We postpone the entire 
discussion of the Markovian results  to the following Section. 
We find a very good agreement, both 
for the symmetric as well as the asymmetric system. We note that for the 
asymmetric case, the full GME is solved only up to $t=1000$ due to the 
 necessary choice of a very small $\Delta t=5 \times 10^{-3}$ 
 (for a detailed discussion see Appendix \ref{app.numgme}).  
The QUAPI results have been obtained with $K=4$ (the number of memory 
time steps, see Refs.\ \cite{QUAPI,ThorwartPRE00} for details) 
and $\Delta t=0.1$ for 
the symmetric  and $\Delta t=0.35$ for the asymmetric case. 

\begin{figure}[t]
\centerline{
\epsfig{figure=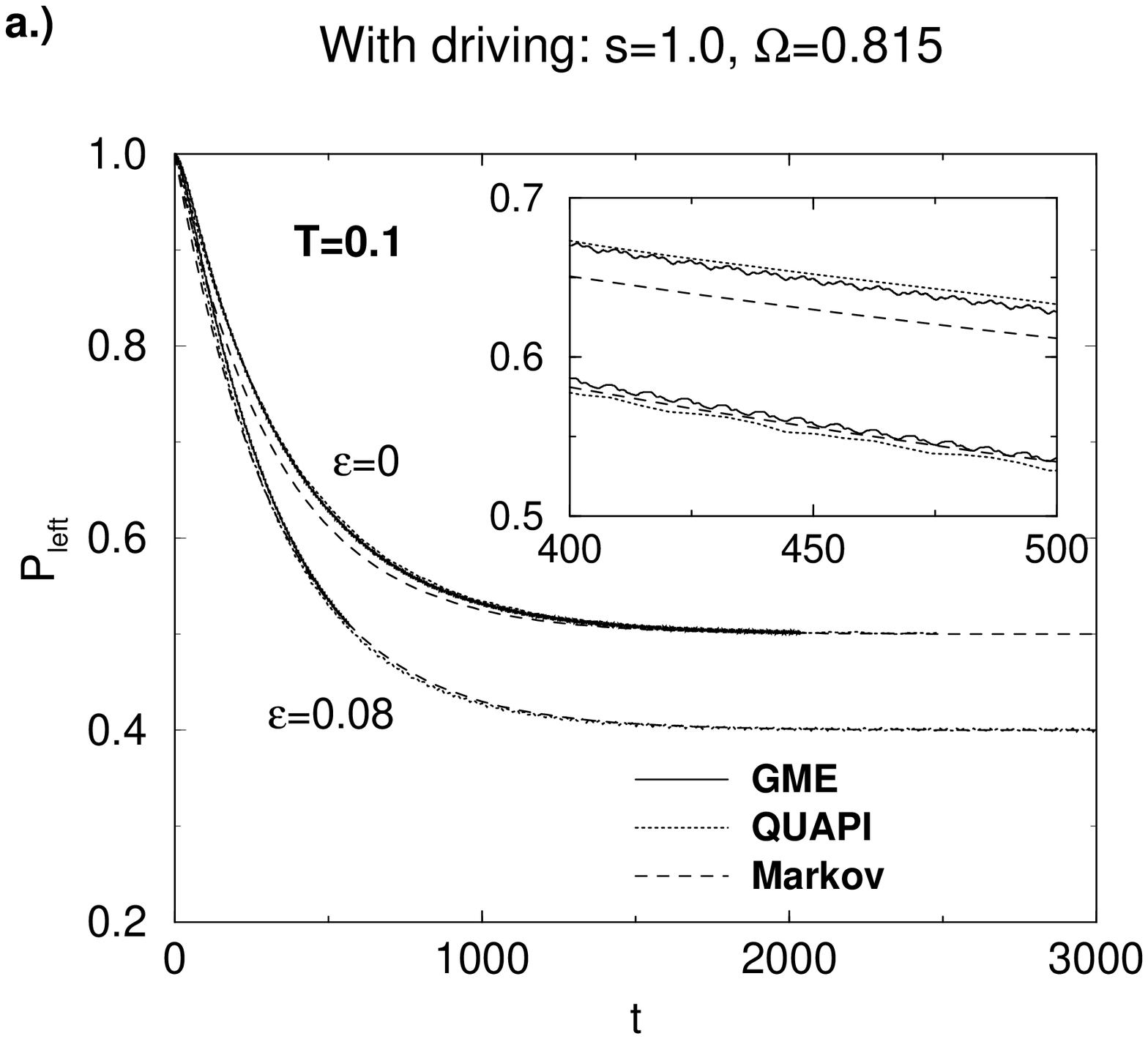,width=90mm,height=75mm,angle=0}
\epsfig{figure=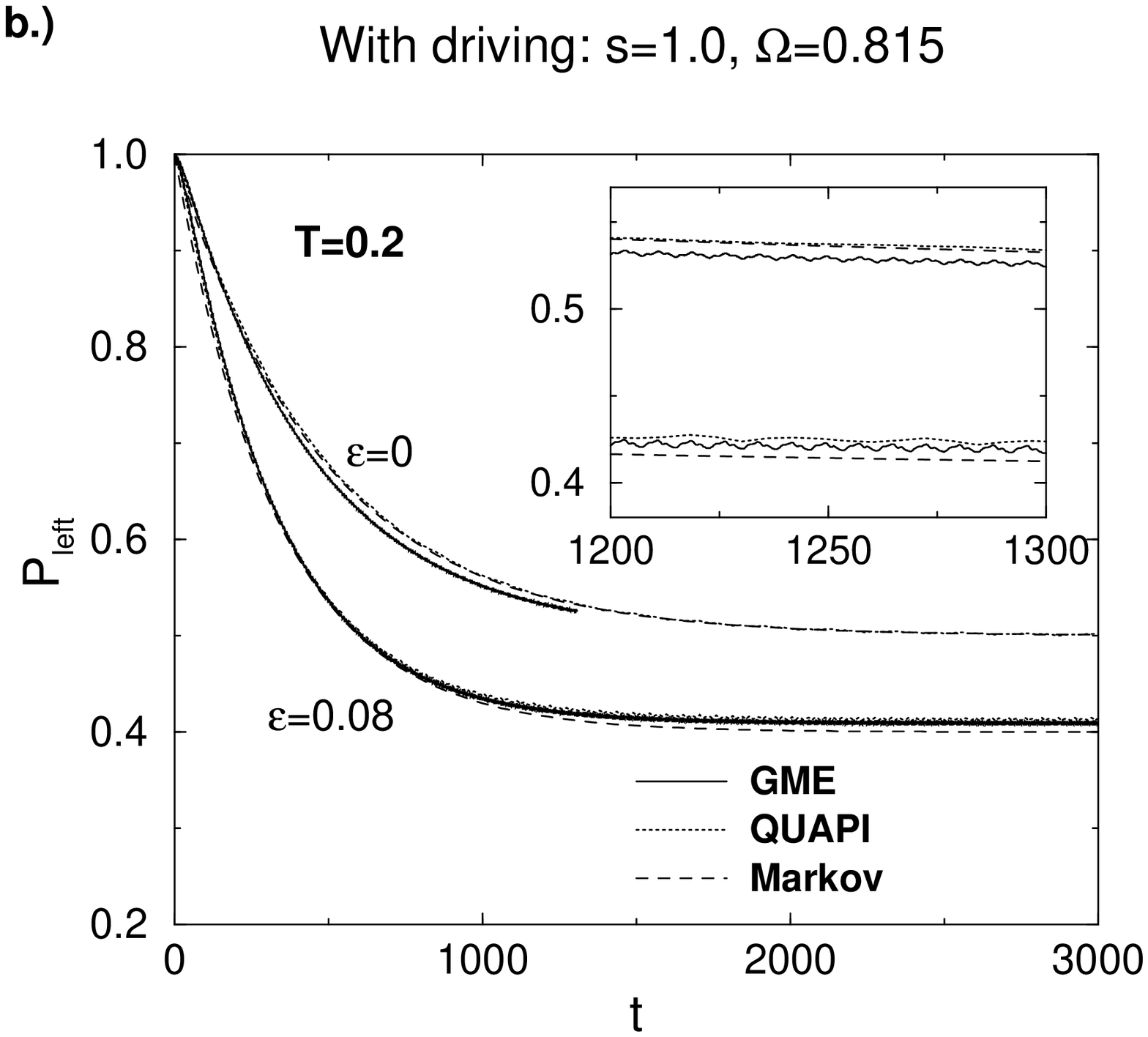,width=90mm,height=75mm,angle=0}
}
\caption{ 
\label{fig.pleft2} 
Same as Fig.\ \ref{fig.pleft1} in 
presence of resonant driving $s(t)=s \, \sin (\Omega t)$ 
($s=1.0, \Omega=\overline{\omega}_0 = 0.815$)  
for $T=0.1$ (a.) and for $T=0.2$ (b.). 
Insets: enlarged parts of the figures.}
\end{figure}

The same very good agreement is found for the case with resonant driving, 
i.e.,  
$s=1.0, \Omega=\overline{\omega}_0=0.815$ which is depicted in 
 Fig.\  \ref{fig.pleft2} a.)  
for $T=0.1$. The inset reveals that  the 
agreement is satisfactory also  on a shorter time scale. 
The QUAPI-parameters are $K=4$ and 
$\Delta t=0.75$ for the symmetric,  and $\Delta t=0.3$ for the asymmetric case, 
respectively. 
Also for a higher temperature $T=0.2$ the agreement is very good, 
see Fig.\ \ref{fig.pleft2} b.). 

All results exhibit a {\em single exponential decay\/} 
at long times. This reveals that 
the bath parameters have been chosen such that the dynamics is indeed 
{\em incoherent\/} and no quantum coherent oscillations can be observed. 
In absence of a static asymmetry $\varepsilon=0$, 
the (averaged) 
asymptotic population of the left well is clearly 0.5. This holds for the 
undriven ($s=0$) as well as for the driven case (resonant driving $s=1.0, 
\Omega=0.815$). However, in presence of a bias $\varepsilon=0.08$, 
the (averaged) asymptotic population of the left well falls below 0.5. 
 The effect of the additional time-dependent driving 
 is to increase the asymptotic population on the left, see Figs.\  
 \ref{fig.pleft1} and \ref{fig.pleft2} b.). The quantum relaxation rate 
 and the asymptotic population of the left well are studied in 
 greater detail in the subsequent Section \ref{sec.rate}.

%
%
\section{The quantum relaxation rate}
\label{sec.rate}
The generalized master equation (\ref{thegme}) is an 
integro-differential equation that governs the decay of the 
population out of one (metastable) well. However, 
{\em to extract analytically 
one single rate\/}, which rules the interesting dynamics on the 
largest time scale, 
requires further approximations. Motivated by the numerical 
fact that the decay 
of the population is observed to be exponential with one single exponent 
(see Section \ref{subsec.quapicomp}), we 
proceed  by invoking a Markovian approximation for the 
GME (\ref{thegme}). This approximation yields a set of coupled 
ordinary first-order differential equations. In absence of 
 external driving, the corresponding 
 coefficients are time-independent. For a driven 
system, they depend on the actual time variable $t$. 
When the frequency of the 
periodic external driving is of the order of the frequency 
associated with the interdoublet energy gap or larger,   
$\Omega \gtrsim \overline{\omega}_0$, the  
averaging of the dynamics over a full  
driving period is appropriate. 
After averaging, the coefficients of the set of 
coupled first-order differential equations then assume time-independent 
values. 
They form the (time-averaged) 
rate matrix. {\em The smallest real part of the 
eigenvalues of this rate matrix\/} yields the relevant rate, 
{\em the quantum relaxation rate\/},  
which  rules the dynamics on the largest time scale.  

%

It is this novel expression for the quantum relaxation rate which constitutes 
a second major result of this work. On one side, we consider shallow barriers 
$\Delta U \gtrsim \hbar \omega_0$ as well as high barriers 
$\Delta U \gg \hbar \omega_0$. In the latter case, the condition for the 
validity of a semiclassical treatment  is met. Put differently, since 
we deal in our 
approach  with discrete energy eigenvalues, 
the semiclassical limit is reached when the number 
of levels below the barrier becomes large. In this case, however, the 
{\em numerical\/} solution of the GME becomes intractable. 
On the other hand, we may consider 
 temperatures $k_{\rm B}T \approx \hbar \omega_0$,  
such that the higher 
lying energy doublets cannot be neglected, as well as lower temperatures 
$k_{\rm B}T \ll \hbar \omega_0$. In fact, our analysis contains 
the spin-boson solution, being the appropriate limit when 
$k_{\rm B}T\ll \hbar \omega_0$ and in the absence of strong resonant driving. 
Finally, we can allow for large 
driving amplitudes and interdoublet resonant driving frequencies. 
In this latter case, both the restriction to a two-level system as well as an 
equilibrium semiclassical analysis is prohibited. 

In the following Section \ref{subsec.markov}, we describe the 
Markov approximation 
 for the generalized  master equation. In Section 
 \ref{subsec.rate},  the quantum relaxation rate 
 is determined as the smallest real part of the eigenvalues of the 
 rate matrix. 
\subsection{Markovian approximation} 
\label{subsec.markov}
The starting point is the generalized master equation (\ref{thegme}). 
The inhomogeneities $I_{\mu}(t,t_0)$ on the r.h.s.\ do not contribute to the 
long-time dynamics since they decay exponentially with time on a rather 
short time scale, see Eq.\  (\ref{ddsinhomogenity}) for the 
inhomogeneity in the case of the double-doublet system determined within 
lowest order in $\Delta_j$. Hence, this term can be neglected. 

We assume furthermore that the characteristic memory time $\tau_{\rm mem}$ 
of the kernels of Eq.\ (\ref{thegme}) is the smallest time scale of the 
problem ({\em Markovian limit\/}). This means that we can substitute the 
argument of $\rho_{\nu \nu} (t')$ under the integral by the time $t$, and 
draw $\rho_{\nu \nu} (t)$ in front of the integral. Moreover, the upper 
limit $t$ of the integral can be replaced by $\infty$. We then  obtain 
the {\em Markovian approximated generalized master equation\/}
\begin{equation}
\dot{\rho}_{\mu\mu} (t) = \sum_{\nu=1}^{M} \,  
\Gamma_{\mu \nu} (t) \rho_{\nu \nu} (t) \, 
\label{themarkovgme} 
\end{equation}
with the time-dependent rate coefficients  
\begin{equation}
\Gamma_{\mu \nu} (t) = \int_{0}^{\infty} \, d\tau \, 
{\cal H}_{\mu \nu} (t, t-\tau) \, .
\label{markovgmetimedeprates} 
\end{equation}
The explicit time-dependence of the rate coefficients 
reflects the explicit time-dependent 
external forcing.  In the case without external driving, the rate 
 coefficients in Eq.\  (\ref{markovgmetimedeprates}) become time-independent. 
\subsubsection{Analytic result for the case without driving}
To obtain specific results, we investigate the lowest order for the 
kernels ${\cal H}_{\mu \nu}$. 
The time independent rate  coefficients then read, 
to lowest second order, 
\begin{eqnarray} 
\Gamma^{(2)}_{\mu\nu} 
&=& \frac{\Delta^2_{\mu\nu}}{2} \int_{0}^{\infty}  \!\! d\tau \,
 \exp \left\{ -(q_{\mu}-q_{\nu})^2 \, S(\tau)\right\}  
 \cos \big[(F_{\nu}-F_{\mu})\tau 
 -(q_{\mu}-q_{\nu})^2 R(\tau)\big] \, , \nonumber \\
 &&\mbox{\hspace{59ex}} \mu \ne \nu \, .
\label{markovgmerates} 
\end{eqnarray}
The used quantities have been introduced in 
Eqs.\  (\ref{twintcorr}), (\ref{dvrtrafo}), (\ref{horjump}), (\ref{verjump}) and 
(\ref{diagele}). 
The conservation of probability requires for the diagonal elements of the 
second order rate  coefficients that  $\Gamma^{(2)}_{\nu \nu} 
=-\sum_{ 
\kappa\ne\nu
}
\Gamma^{(2)}_{\kappa \nu}$. 
The  integral in Eq.\ (\ref{markovgmerates}) can be solved numerically 
by standard integration routines \cite{NumRec}. However, an analytical solution 
can also be derived. 
 In the  
limit $\omega_c t \rightarrow \infty$  \cite{Weiss99}, 
the correlation functions $S(t)$ and $R(t)$ assume the form in 
 Eq.\ (\ref{scalinglimits}) for the real part $S(\tau)$,  and 
in Eq.\  (\ref{largeomc}) for the imaginary part $R(\tau)$.
%
After some basic algebra, we obtain for 
the Markovian approximated rate  coefficients the expression  
\begin{eqnarray} 
\Gamma^{(2)}_{\mu\nu} 
&=&
\frac{\Delta^2_{\mu\nu}}{4 \, \omega_c} 
\exp \left\{(F_{\nu}-F_{\mu})\, \frac{\hbar\beta}{2} \right\} \nonumber \\
& & \mbox{} \times 
\left( \frac{\hbar\beta\omega_c}{2 \pi} \right)^{1-(q_{\mu}-q_{\nu})^2 
\eta/\pi} 
\frac{
\Big|
 {\Gamma} \left[ (q_{\mu}-q_{\nu})^2\eta/2\pi +i \,  
\hbar \beta (F_{\nu}-F_{\mu})/2\pi \right]\Big|^2}{
\Gamma \left[(q_{\mu}-q_{\nu})^2\eta/\pi\right]
}
\,  ,
\label{matrixel3} 
\end{eqnarray}
with $\Gamma(z)$ being the $\Gamma$--function \cite{Gradshteyn}. 
\subsubsection{High-frequency-driving}
To extract an average long-time relaxation rate in the case with driving, 
we choose 
the external driving frequency $\Omega$ to be  
of the order of the 
interdoublet level spacing $\overline{\omega}_0 \approx \omega_0$.   
This assumption is met, for instance, if one wishes to pump population 
from the lower to the upper doublet by
 an interdoublet resonant field. A time-average of 
 the time-dependent rates (Krylov-Bogoliubov-scheme)  
 in Eq.\  (\ref{markovgmetimedeprates})  over the driving period 
 is then appropriate.   
In general,  this averaging procedure is reasonable when 
the driving frequency is much larger than the time scales related 
to tunneling, i.e., when 
$\Omega \gg \Delta^{\cal E}_1, \Delta^{\cal E}_2, ...$ where the 
$\Delta^{\cal E}_i$ are the tunneling splittings of the doublets. 

We insert the explicit shape of the periodic driving 
$s(t)= s \sin (\Omega t)$ in the second order kernels in Eq.\  
(\ref{ddskernel}). 
%
The averaging with respect to the driving frequency reads
  $\langle \Gamma_{\mu\nu}(t) \rangle_{\Omega} = 
(\Omega/2\pi) \int_0^{2\pi/\Omega} dt \, \Gamma_{\mu\nu}(t)$. 
The integration over $t$ can be performed if one represents the 
sine- and cosine-function in terms of Bessel functions $J_n(x)$
\cite{Gradshteyn}. 
The only remaining non-zero 
part is the one which contains the zeroth Bessel function 
$J_0(x)$.  
The time-independent averaged Markovian rate matrix elements
to second order emerge as 
\begin{eqnarray} 
\Gamma^{{\rm av},(2)}_{\mu\nu} &\equiv& 
\langle \Gamma_{\mu\nu}^{(2)}(t)\rangle_{\Omega} \nonumber \\ 
&=& \frac{\Delta^2_{\mu\nu}}{2} \int_{0}^{\infty}  d\tau \,
 \exp \left\{ -(q_{\mu}-q_{\nu})^2 \, S(\tau)\right\}  
J_0\left(\frac{2s}{\Omega} (q_{\mu}-q_{\nu}) \sin \left(\frac{\Omega}{2}\tau
\right)\right)
\nonumber \\ 
&& \mbox{} \times 
 \cos \big[(F_{\nu}-F_{\mu})\tau 
 -(q_{\mu}-q_{\nu})^2 R(\tau)\big]  \, , \mbox{\hspace{3ex}} 
 \hfill \mu \ne \nu \, .
 \label{avkernels}
\end{eqnarray}
Like in the non-Markovian case, the conservation of probability 
implies for the diagonal matrix elements the condition 
$\Gamma^{{\rm av},(2)}_{\nu \nu} =-\sum_{ 
\kappa\ne\nu
}
\Gamma^{{\rm av},(2)}_{\kappa \nu}$.  
This expression reveals that the influence of driving is {\em different} 
 for each pair of DVR-states since the explicit distance $q_{\mu}-q_{\nu}$ 
 enters in the argument of the Bessel functions. 
The averaged rate matrix elements cannot be calculated in closed analytical form 
 as in the undriven case; 
however, they can be obtained numerically by standard  
integration routines \cite{NumRec}. 
\subsection{The quantum relaxation rate} 
\label{subsec.rate}
Since the diagonal elements $\rho_{\mu\mu}(t)$ obey Eq.\   
(\ref{themarkovgme}), the long-time dynamics in this regime is ruled by a 
{\em single exponential decay\/}. In the case without driving, the rate 
matrix $\Gamma_{\mu \nu}$ is already time-independent, and equivalently 
for the case of high-frequency driving after the averaging procedure. 
 Both cases reduce to a structure  
\begin{equation}
\dot{\rho}_{\mu\mu} (t) = \sum_{\nu=1}^{M} \,  
\Gamma_{\mu \nu}^{({\rm av})} \rho_{\nu \nu} (t)\, , 
\end{equation}
where the superscript $({\rm av})$ means that the formula holds for the 
averaged as well as for the time-independent case.  
This set of coupled ordinary first-order 
differential equations can be decoupled via a diagonalization procedure. 
If one denotes the elements of the transformation matrix by 
$S_{\mu \nu}$ and the eigenvalues of the (averaged) rate matrix 
by $\Lambda_\mu$, the diagonalized (averaged) rate matrix reads
\begin{equation}
\sum_{\kappa_1,\kappa_2=1}^M (S^{-1})_{\mu \kappa_1} 
\Gamma_{\kappa_1 \kappa_2}^{({\rm av})} S_{\kappa_2 \nu} = \Lambda_{\mu} 
\delta_{\mu \nu} \, .
\end{equation}
The general solution of the (averaged)  Markov approximated GME is 
  obtained to be
\begin{equation}
\rho_{\mu \mu}(t) = 
\sum_{\nu, \kappa=1}^M S_{\mu \nu} (S^{-1})_{\nu \kappa} 
e^{\Lambda_{\nu} (t-t_0)} \rho_{\kappa \kappa} (t_0) \, .
\label{markovpop}
\end{equation}
Since $\Gamma_{\mu \nu}^{({\rm av})}$ is a stochastic matrix, i.e., 
the diagonal elements of the (averaged) rate matrix 
are the negative sum of the 
 matrix elements of the corresponding columns, one eigenvalue 
 equals zero, i.e., $\Lambda_{1} = 0$ (conservation of probability). 
 Therefore, 
\begin{equation}
\rho_{\mu \mu}(t) = \rho_{\mu \mu}^{\infty} + 
\sum_{\nu=2}^M \sum_{\kappa=1}^M S_{\mu \nu} (S^{-1})_{\nu \kappa} 
e^{\Lambda_{\nu} (t-t_0)} \rho_{\kappa \kappa} (t_0) \, ,
\end{equation}
with $\rho_{\mu \mu}^{\infty}=\sum_{\kappa=1}^M S_{\mu, 1} (S^{-1})_{1,\kappa} 
 \rho_{\kappa \kappa} (t_0)$ being the asymptotic population of the 
 DVR-state $\ket{q_{\mu}}$. 
   
The rate which determines the dynamics on the largest time-scale 
is the smallest non-zero  absolute value of the 
real part of the eigenvalues of the 
(averaged) rate matrix, i.e., 
\begin{equation}
\Gamma^{({\rm av})} \equiv \min  \{ |\mbox{\rm Re} \Lambda_{\nu}|; \nu=2,..., M\} \, .
\label{qrate}
\end{equation}
It is called the {\em quantum relaxation rate\/}.
%
%

Likewise, the asymptotic population $P_{\rm left}^{\infty}$ of the 
left well is readily obtained from Eq.\  (\ref{markovpop}). It reads  
%
\begin{equation} 
P_{\rm left}^{\infty}
 = \sum_{\mu=1}^{L} \rho_{\mu \mu}^{\infty} \, .
 \label{pleftinf}
\end{equation} 

To compare the predictions of the Markovian approximated master equation 
(\ref{themarkovgme})  
with the results of the  generalized master equation (\ref{thegme}),  
 we recall the outcomes  
 presented in Figs.\ \ref{fig.pleft1} 
and \ref{fig.pleft2}  of the previous Section \ref{subsec.quapicomp}. 
The Markovian results are indicated by the dashed lines. 
In all investigated parameter combinations, 
the agreement between the generalized master equation, the predictions 
of the QUAPI algorithm and the Markovian master equation 
is very good apart from minor differences. The rate of the 
decay is described accurately by $\Gamma^{({\rm av})}$,   
as well as the asymptotic population 
of the left well. Also in presence of a time-dependent driving, the 
averaging yields the correct averaged dynamics. This allows 
for the conclusion that, 
in the investigated range of parameters, 
 the driven dissipative multi-level system 
is adequately described by the Markovian approximated master equation 
with the eigenvalues determined from second order gNICA. 
\section{Results: Quantum relaxation rate and asymptotic populations}
\label{sec.results}
In this Section, we present results for the quantum relaxation rate 
$\Gamma^{({\rm av})}$ and the asymptotic population $P_{\rm left}^{\infty}$ 
inside the left well for the (driven) double-well potential, Eq.\  
(\ref{staticpotential}). 
Throughout the following Sections, 
we choose a set of typical dimensionless 
parameter values. The corresponding 
dimensionful values follow from the standard scaling 
procedure described in Appendix \ref{app.scaling}.  
 The barrier height is consistently chosen to be $E_{\rm B}=1.4$. 
This implies that two doublets lie below the energy barrier and the other 
energy states lie above the barrier, 
see Fig.\  \ref{fig.dvrstates}. Moreover the lower tunneling 
splitting is $\Delta^{\cal E}_1=3.60 \times 10^{-3}$, 
the upper tunneling splitting 
is $\Delta^{\cal E}_2=0.121$ and the energy gap between the two doublets is 
$\overline{\omega}_0=0.815$. This choice is mainly motivated by 
the fact that we explicitly want to investigate the intermediate regime 
between the two-level approximation and the semiclassical regime. 
Furthermore, such a  shallow barrier height is convenient for 
numerical reasons. The splitting 
 of the lowest doublet decreases exponentially with increasing barrier 
 height.  

In the following subsections, we investigate on the one hand the 
double-doublet system, i.e., $M=4$, for the barrier height $E_{\rm B}=1.4$. 
We expect that the results are qualitatively similar for larger barrier heights 
when more than two doublets lie {\em below\/} the barrier because the 
spectrum is then similar to the double-doublet case. 
On the other hand, we study the question of {\em convergence\/}  
with increasing number 
$M$ of energy levels for the case of  $E_{\rm B}=1.4$. 
 For $M\rightarrow\infty$, the 
 multi-level systems is equivalent to the spatially continuous 
 potential. 

We consider two typical situations: The 
{\em unperturbed\/} energy spectrum in the 
symmetric case ($\varepsilon=0$) exhibits {\em avoided level crossings\/}, 
see Fig.\  
\ref{fig.rateasy1} a.) below. The second case refers to a  tilted 
potential with $\varepsilon=0.08$ where the energy levels are rather 
{\em strongly separated\/},  see Fig.\  \ref{fig.rateasy1} a.) below. 

The parameters for the time-dependent driving are typically chosen in 
such a way that the regimes of a 
weak (s=$0.1$) and a strong ($s=1.0$) driving amplitude   
are covered both. 
In the first case, the potential stays permanently bistable while 
in the second case the potential assumes intermediate monostable 
configurations. With  $E_{\rm B}=1.4$, the critical amplitude where 
bistability vanishes is $s_{\rm crit}=0.64$. The driving frequency is 
typically chosen either to be 
in resonance with the interdoublet energy gap, 
i.e., $\Omega=\overline{\omega}_0 
=0.815$, or off resonance, i.e., $\Omega=0.2$. 

The typical choice for the temperature is $T=0.1$, being a low to intermediate 
temperature. We note that the semiclassical expression for the 
cross-over temperature \cite{HanggiRMP90} yields $T_{\rm co}=0.12$,  
keeping in mind, however, that our choice of the 
 barrier height does not obey the semiclassical condition. 
 
We use an Ohmic spectral density with an exponential cut-off, see 
 Eq.\  (\ref{specdens}).
 The damping constant is  chosen to be $\gamma=0.1$; it represents 
 an intermediate damping strength. We note that the dependence of the results 
 on the damping strength and temperature is exponential.  
 The cut-off frequency is always fixed to be $\omega_c=10.0$.  

Finally, we note that in the following symbols such as 
$\bullet$ and $\Box$ are used  to label individual plots in the 
particular figures. Their number is not related to the 
number of calculated data points, the latter being much larger. 
\subsection{Absence of external driving} 
\label{subsec.nodriv}
\begin{figure}[t]
\begin{center}
\epsfig{figure=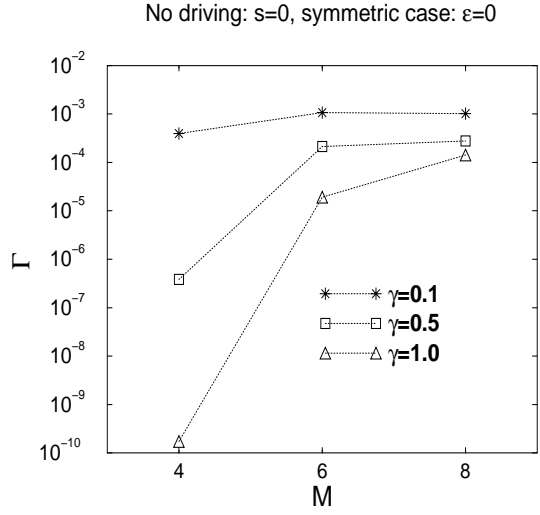,width=70mm,height=70mm,angle=0} 
\end{center}
\caption{Quantum relaxation rate $\Gamma$ for the static symmetric 
double-well potential with barrier height $E_{\rm B}=1.4$ in 
dependence of the number $M$ of energy eigenstates 
for the damping constants $\gamma=0.1 (\ast), \gamma=0.5 (\Box)$ 
and $\gamma=1.0 (\bigtriangleup)$. The temperature 
is chosen to be $T=0.1$ and the cut-off frequency is $\omega_c=10.0$. 
\label{fig.raten1}} 
\end{figure}

We start with the simplest case of the undriven symmetric 
double-well potential ($s=0, \varepsilon=0$) and consider the 
dependence of the quantum relaxation rate on the number $M$ of 
energy eigenstates. $M=4$ denotes the double-doublet system. 
 Fig.\  \ref{fig.raten1} shows the result for different 
damping constants $\gamma$. A convergence of the rate $\Gamma$ can be observed 
for an intermediate damping strength $\gamma=0.1$ ($\ast$). 
We recall that the 
lowest tunneling splitting is two orders of magnitude smaller than $\gamma$ 
and that 
the upper tunneling splitting is of the same order of magnitude as 
$\gamma$. For 
a larger damping constant $\gamma=0.5$ ($\Box$), 
convergence is also obtained. 
However, for the case of very strong damping $\gamma=1.0$ ($\bigtriangleup$), 
the result for $M=8$ does not coincide with that for $M=6$. This fact is due 
to the following feature: 
 As it can be seen from Eq.\  (\ref{ddskernel}) for the 
second order kernels of the generalized master equation, the damping 
constant $\gamma$, which enters via $S(t)$ and $R(t)$, 
is multiplied by $(q_{\mu} - q_{\nu})^2$ being the square of the 
tunneling distance between the two involved DVR-states. Upon increasing the 
number $M$ of energy levels, the DVR-eigenvalues lie more dense in 
position space. Hence, some distances become small and the multilevel system 
effectively {\em flows to weak damping\/}. The small effective damping is no 
longer sufficient to  suppress 
long intervals in the off-diagonal states. Thus, the gNICA in {\em second\/} 
 order 
is no longer applicable and  contributions of  higher orders of 
the  integral kernels in the GME have to be taken into account. 
A more detailed discussion of the effect of the flow to weak damping 
is postponed to the Appendix \ref{app.outlook}.

The question of convergence of the quantum relaxation rate 
with increasing $M$ in presence of time-dependent driving 
is investigated in the sections below. 
\subsection{The influence of external (time-dependent) driving forces}
\label{subsec.withdriv}
In this Subsection we investigate the role of  external driving. 
This external perturbation can be either 
a static potential asymmetry ({\em bias\/}) 
or a time-dependent periodic driving ({\em ac-driving\/}), or 
simultaneously both parts are present.
\subsubsection{Dependence on a static bias, no ac-driving}
\label{subsubsec.drivdc}
\begin{figure}[t]
\centerline{
\psfrag{E}{\small ${\cal E}$}
\epsfig{figure=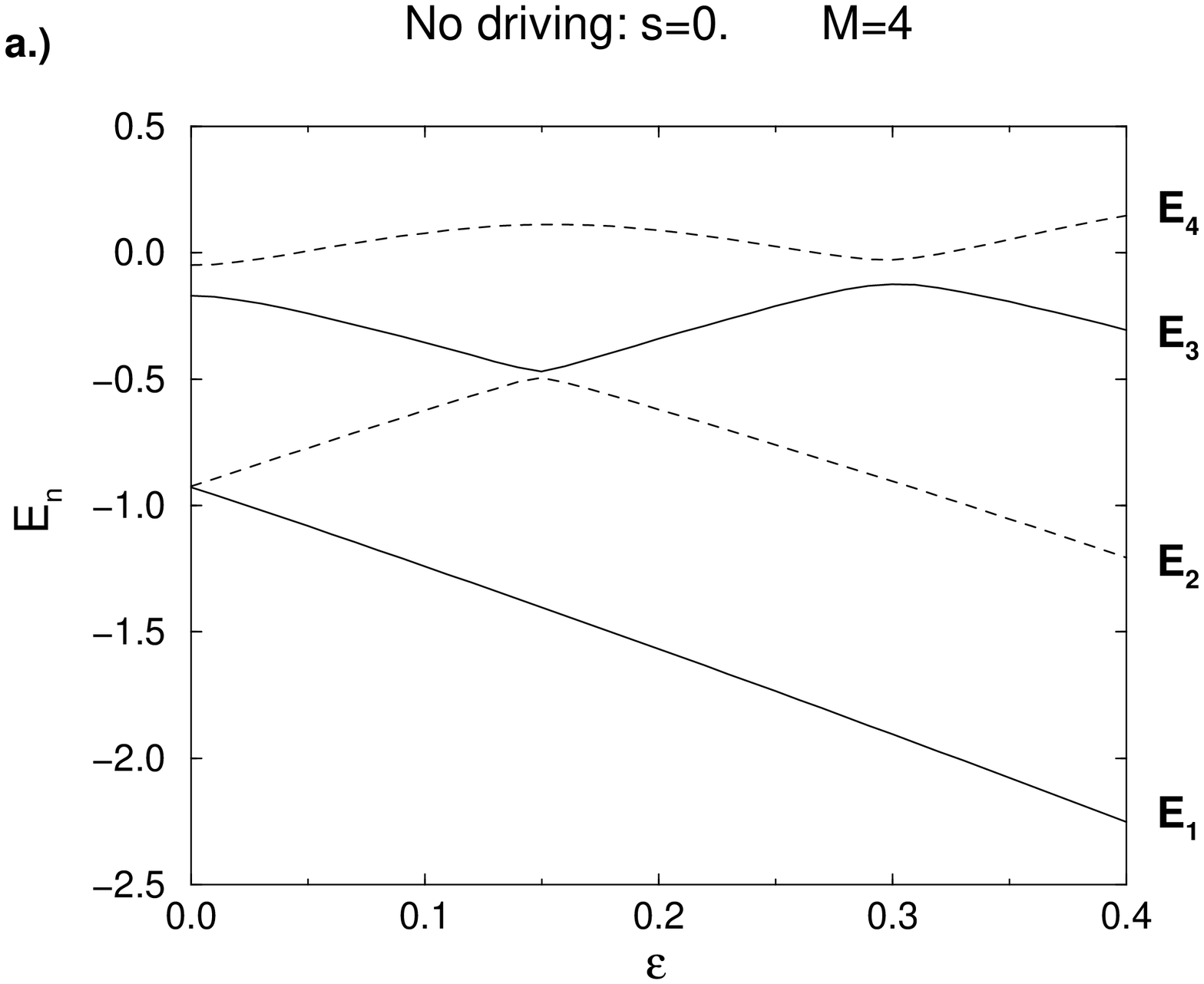,width=92.5mm,height=75mm,angle=0}
\epsfig{figure=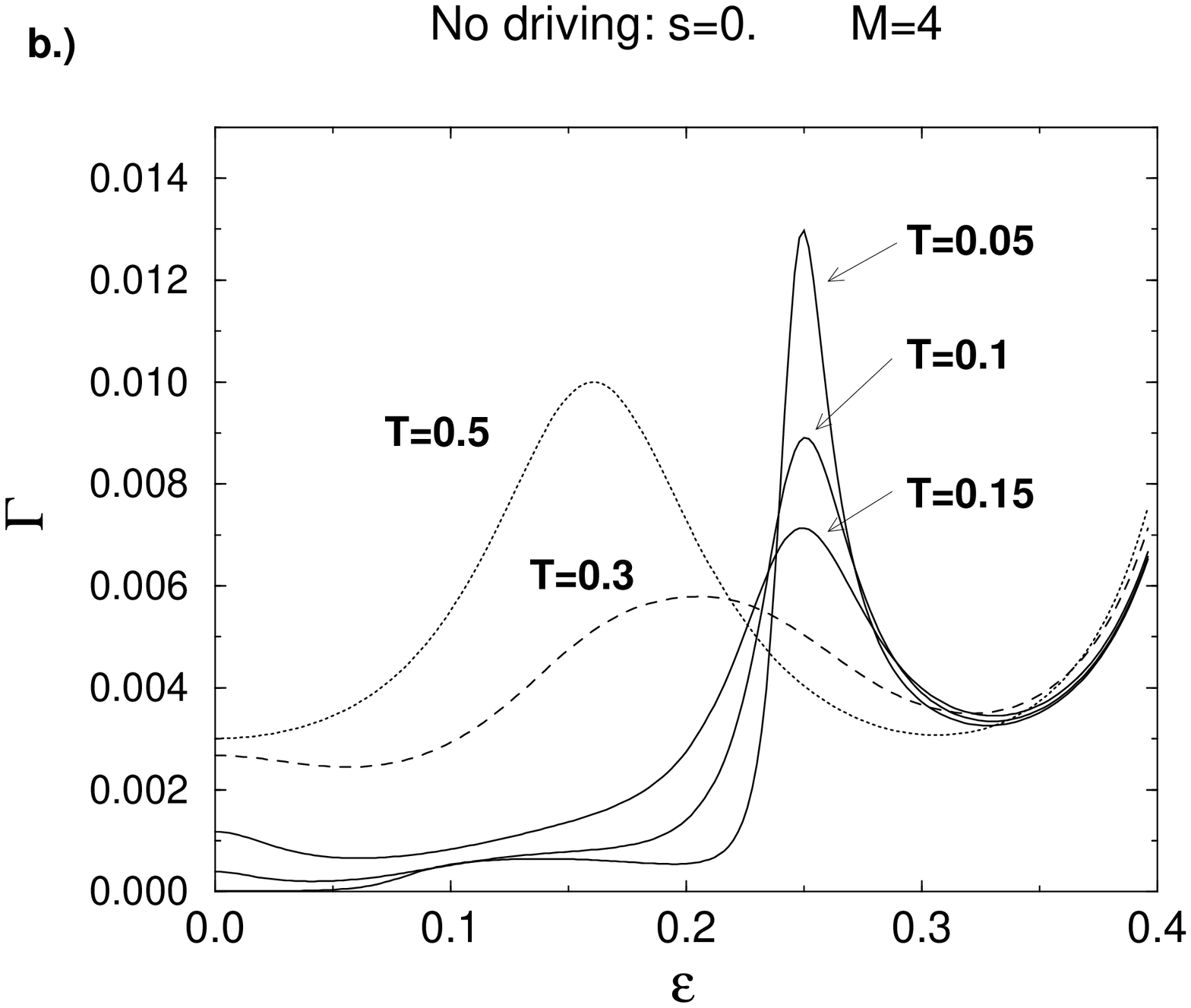,width=90mm,height=75mm,angle=0}
}
\caption{ 
\label{fig.rateasy1}
a.) Spectrum of the unperturbed static ($s=0$) 
system Hamiltonian (\ref{system}) 
for a barrier height of $E_{\rm B}=1.4$ as a function of the static 
bias $\varepsilon$. 
b.) Quantum relaxation rate $\Gamma$ according to Eq.\  (\ref{qrate}) 
as a function of the static bias $\varepsilon$ for 
different temperatures $T$. The barrier height is 
$E_{\rm B}=1.4$ and the number of energy levels is $M=4$. The 
bath parameters are $\gamma=0.1$ and $\omega_c=10.0$. 
}
\end{figure}

%
Adding a static asymmetry renders one (in our case the left) 
of the two formerly stable potential minima  a metastable minimum. 
The consequences for the spectrum of the {\em bare\/} system are that 
avoided level-crossings occur  
for particular values of the asymmetry,  
see Fig.\  \ref{fig.rateasy1} a.). At such avoided level-crossings 
 tunneling is usually enhanced. This effect is known as resonant 
tunneling. This situation, however, is modified in the presence 
of a moderate to strong damping. 
The case of a strong system-bath coupling is considered in Fig.\  
\ref{fig.rateasy1} b.) where  the  relaxation rate shows peaks 
at particular values of the static bias. Their position strongly 
depends on temperature. At low temperatures 
$T=0.05$ to $T=0.15$  (full lines), we 
observe three relative maxima at $\varepsilon=0$ 
and around $\varepsilon=0.12$ and   $\varepsilon=0.25$. First, we emphasize 
that the quantum relaxation rate {\em initially decreases\/} 
when the bias is increased from zero, i.e., when 
the effective barrier height is {\em decreased\/}. This feature is a typical 
quantum mechanical footprint. In a classical system, the relaxation rate 
 {\em grows\/} when the barrier is {\em lowered\/} \cite{HanggiRMP90}.  
Second, we note that the two peaks at nonzero bias values are shifted 
to smaller values of the asymmetry compared to those 
bias strengths where the avoided level crossings occur, 
 see Fig.\  \ref{fig.rateasy1} a.). This indicates that we are no longer 
 in a weak-coupling regime but encounter already strong incoherent tunneling.  

Increasing the temperature results in a decrease of the amplitude 
of the peak at $\varepsilon=0.25$. This indicates the enhanced  
destruction of the resonant tunneling phenomenon. Moreover, the 
peaks broaden. At an intermediate temperature  $T=0.3$ (dashed line) 
 one characteristic peak occurs at $\varepsilon=0.2$. 
 Its height is smaller than in the 
 low temperature cases which indicates that tunneling is reduced 
 compared to the low temperature case. This is mainly due to the enhanced 
 environmental level broadening. However, we still observe a clear decrease 
 of $\Gamma$ when the bias is increased from zero onwards; therefore we conclude 
 that quantum tunneling still occurs. 
This, however, is no longer observable for the high temperature 
case $T=0.5$ (dotted line). 
A relaxation rate which grows with increasing bias is a signature 
 of classical behavior. 

\begin{figure}[t]
\centerline{
\epsfig{figure=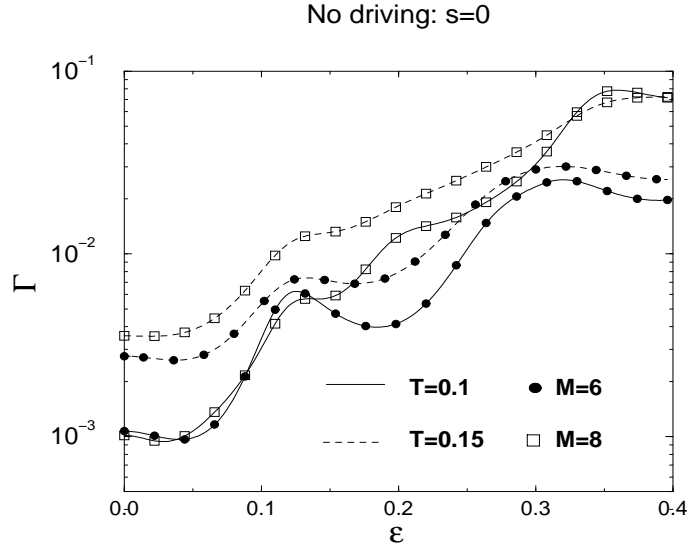,width=90mm,height=75mm,angle=0}}
\caption{ 
\label{fig.rateasy2}
Quantum relaxation rate $\Gamma$ as a function of the static bias 
$\varepsilon$ for four different combinations of the number of levels, i.e., 
$M=6$ ($\bullet$) and $M=8$ ($\Box$),  and different temperatures, i.e., 
 $T=0.1$ (solid line) and $T=0.15$ (dashed line). 
 For the remaining parameters, see Fig.\  \ref{fig.rateasy1}.
}
\end{figure}

The question of the convergence of the rate with an increasing number $M$ 
of energy states is addressed with  Fig.\  \ref{fig.rateasy2}. We show 
four different cases for $M=6$ ($\bullet$) and $M=8$ ($\Box$) and 
$T=0.1$ (full line) and $T=0.15$ (dashed line), respectively. 
The low temperature case $T=0.1$ shows a clear convergence when $M$ is increased 
from $M=6$ to $M=8$ for small asymmetries up to $\varepsilon=0.15$. 
For larger asymmetries, the two results, however, do not agree. This behavior 
can be resolved as follows: For large asymmetries, the left well 
is strongly lifted above the right well. Moreover, the position eigenvalues 
on the left side move towards each other and 
are densely located. This means that the tunneling 
distance of the corresponding transition becomes smaller which in turn reduces 
the effective damping. Then a {\em flow to weak damping\/} occurs and the 
second order gNICA breaks down.  The same explanation holds for the larger  
temperature $T=0.15$ where  the results 
for $M=6$ and $M=8$ show qualitatively a similar behavior up to 
$\varepsilon=0.15$. 

\begin{figure}[t]
\centerline{
\epsfig{figure=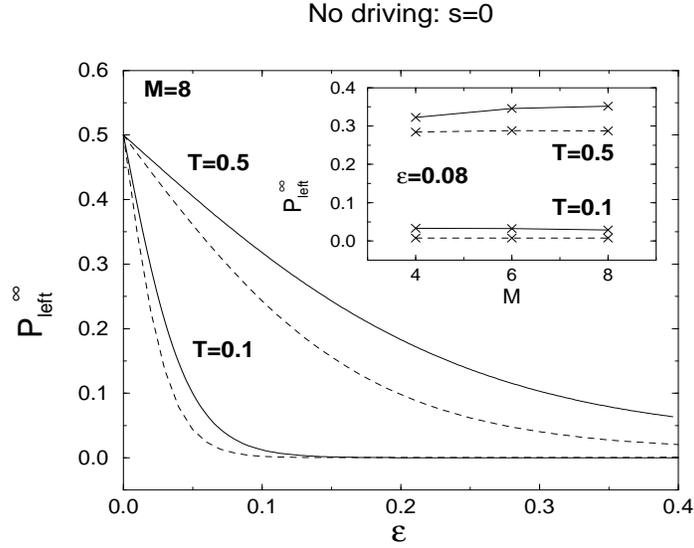,width=90mm,height=75mm,angle=0}
}
\caption{Asymptotic population $P_{\rm left}^{\infty}$ of the left well 
in absence of time-dependent driving 
as a function of the asymmetry $\varepsilon$ for two different 
 temperatures (solid line) for $M=8$. The parameters are: 
 $E_{\rm B} =1.4$, $\gamma=0.1$ and $\omega_c=10.0$. 
 The dashed line marks the results obtained from a Boltzmann equilibrium 
 distribution for the same parameters. Inset: $P_{\rm left}^{\infty}$ 
 vs.\ the number $M$ of energy levels for a fixed asymmetry 
 $\varepsilon=0.08$ (solid line: gNICA in second order; dashed line: 
 with Boltzmann distribution).
\label{fig.pinfasy1}
}
\end{figure}

We investigate in the following the asymptotic population  
$P_{\rm left}^{\infty}$ of the left well determined in Eq.\ (\ref{pleftinf}). 
For the symmetric case, it assumes the value 
 1/2. In presence of a positive asymmetry $\varepsilon > 0$,  
$P_{\rm left}^{\infty}$ is smaller than 1/2 
since the left well is energetically higher. Fig.\ \ref{fig.pinfasy1} shows 
 $P_{\rm left}^{\infty}$ as a function of $\varepsilon$ for two different 
 temperatures (solid line) for $M=8$. The damping constant is chosen to be 
 $\gamma=0.1$. For comparison we additionally 
 show the asymptotic population obtained from a Boltzmann equilibrium 
 distribution for the same parameters (dashed line), 
 i.e., $\bfrho (\infty) = \exp(-{\bf H}_{0} / k_{\rm B} T)$. 
 First, we note that  
 the asymptotic population decreases exponentially with increasing bias. 
 Second, we emphasize that the often made assumption of a 
  Boltzmann equilibrium distribution is only valid for an infinitesimally small 
  coupling of the system to the environment \cite{Kubo85}. 
The depicted results for $M=8$ have already been converged and are not 
distinguishable on our scale from the $M=6$ case. This is indicated in the 
inset of Fig.\  \ref{fig.pinfasy1} for a fixed asymmetry $\varepsilon=0.08$ for 
two different temperatures. Convergence is also found for the entire 
considered parameter range of $\varepsilon$ (not shown).  
The full line again shows the result obtained 
from gNICA in second order while the dashed line marks the results from 
a Boltzmann equilibrium distribution.
%
%
\subsubsection{Dependence on the static bias in presence of  
 external ac-driving}
\label{subsubsec.drivac}
\begin{figure}[t]
\centerline{
\epsfig{figure=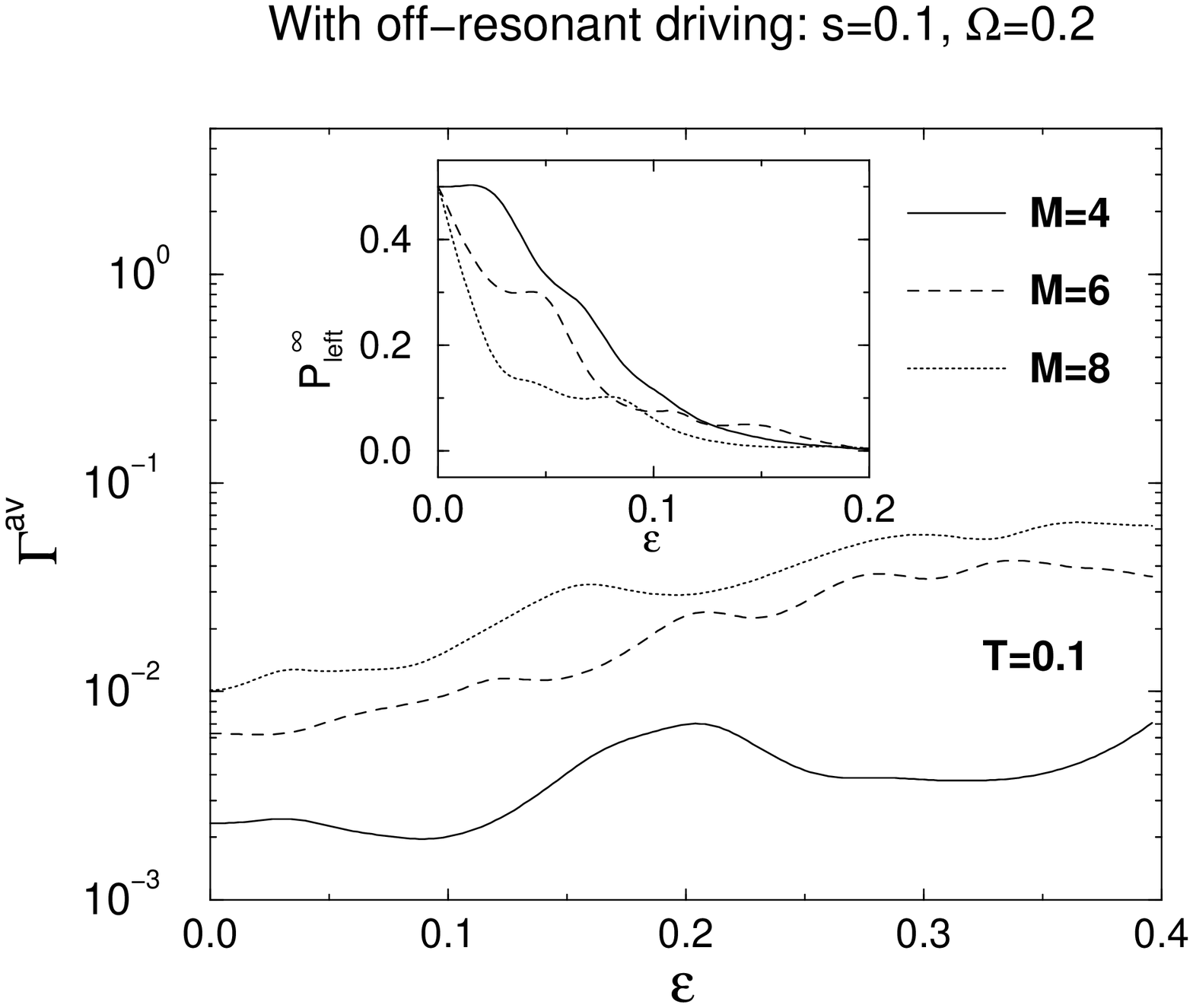,width=90mm,height=75mm,angle=0}
}
\caption{ 
\label{fig.rateasy0}
Averaged quantum relaxation rate $\Gamma^{\rm av}$ 
as a function of the static bias $\varepsilon$ in presence of an 
off-resonant driving with $s=0.1, \Omega=0.2$. Shown 
are the results for three different numbers $M$ of energy levels. The 
parameters are $E_{\rm B}=1.4, T=0.1, \gamma=0.1$ and 
$\omega_c=10.0$. Inset: The corresponding asymptotic populations 
$P_{\rm left}^{\infty}$ of the left well as a function of $\varepsilon$.}
\end{figure}

\begin{figure}[t]
\centerline{
\epsfig{figure=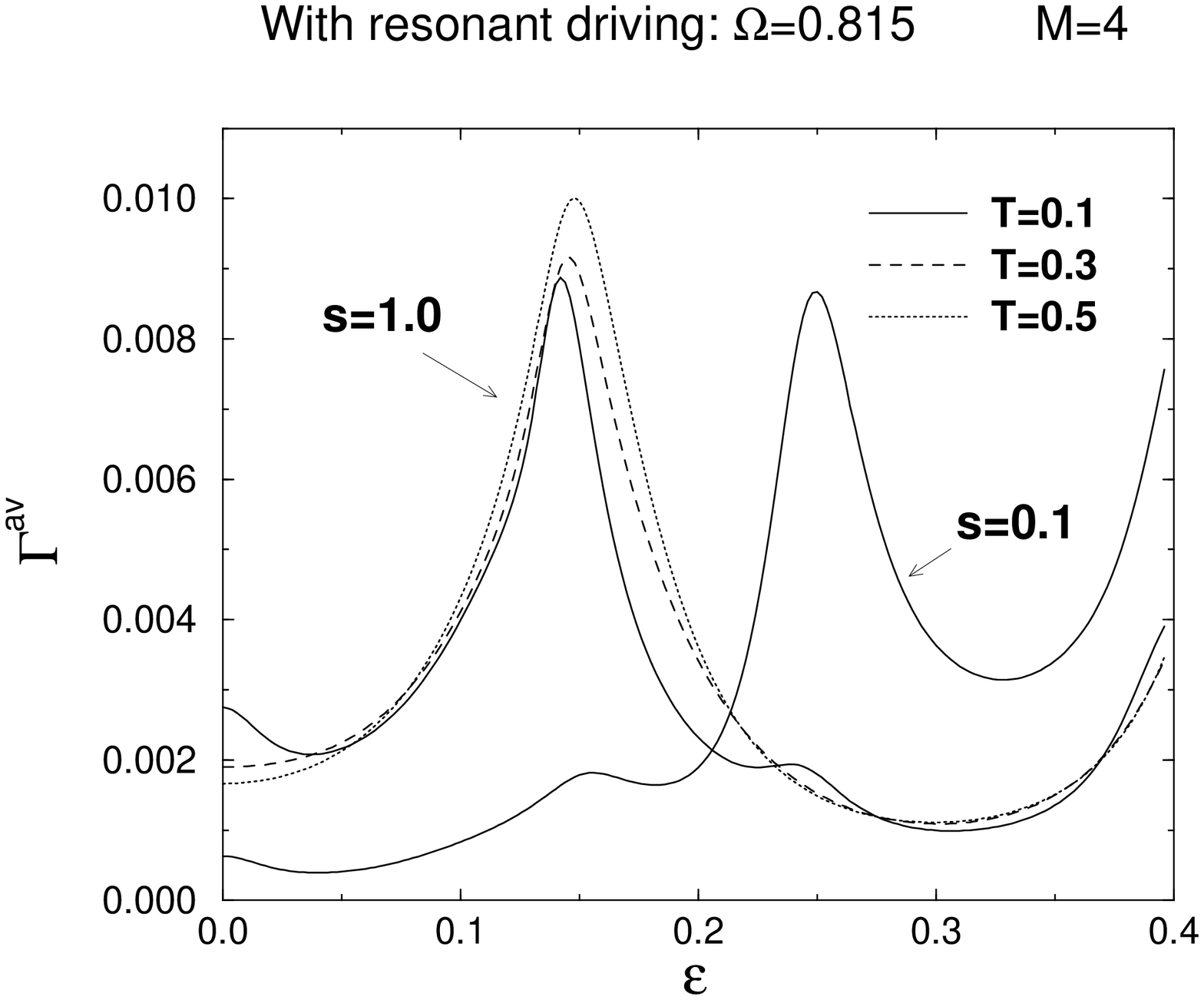,width=90mm,height=75mm,angle=0}
}
\caption{ 
\label{fig.rateasy3}
Averaged quantum relaxation rate $\Gamma^{\rm av}$ 
as a function of the static bias $\varepsilon$ for three 
different temperatures $T=0.1$ (solid line), $T=0.3$ (dashed line) and 
$T=0.5$ (dotted line) for the resonantly 
driven double-doublet system $M=4$. Shown are the cases of 
weak driving $s=0.1$ (for $T=0.1$ only) and 
strong  driving ($s=1.0$). For the remaining 
parameters, cf.\  Fig.\  \ref{fig.rateasy0}. }
\end{figure}

\begin{figure}[t]
\centerline{
\epsfig{figure=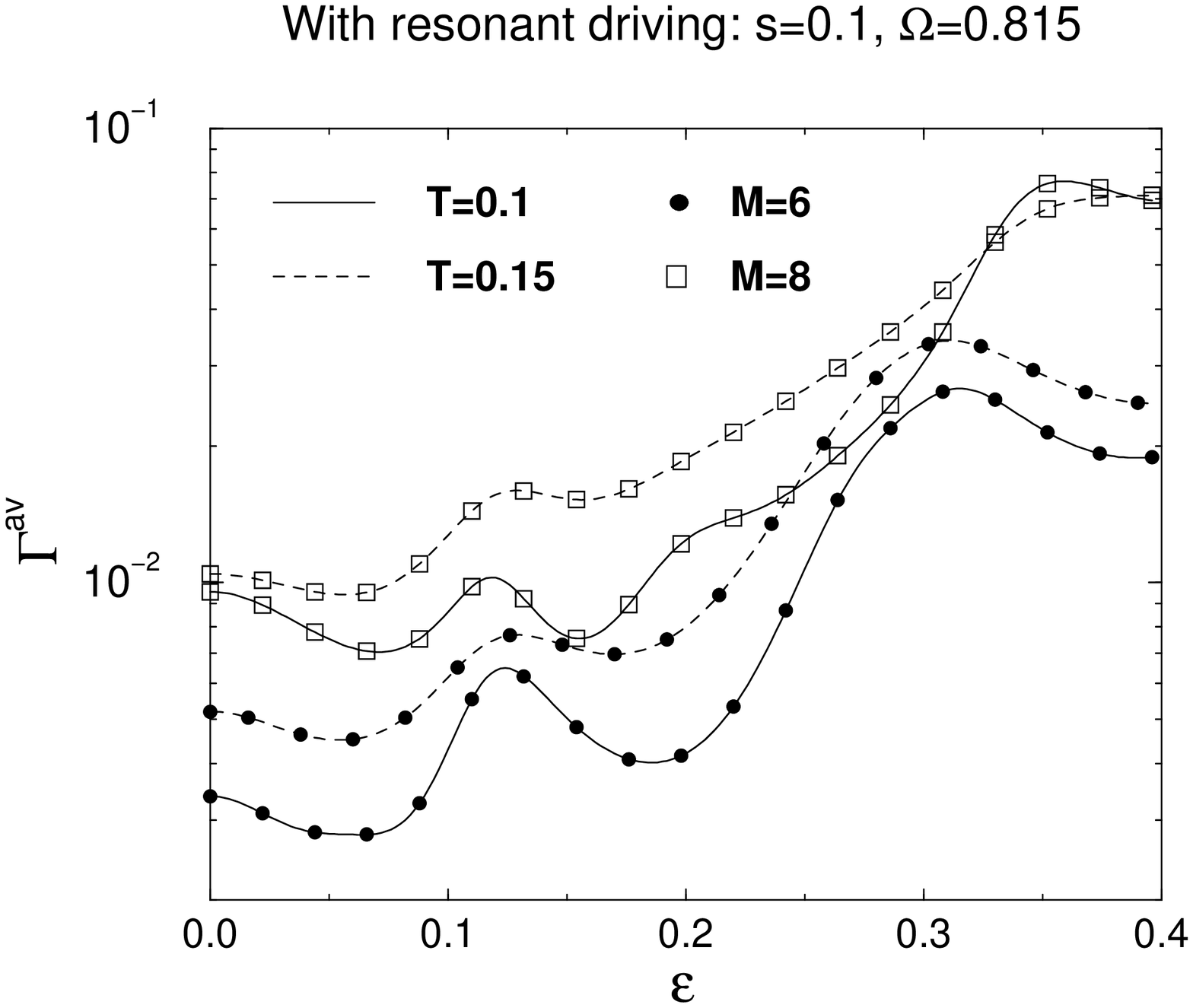,width=90mm,height=75mm,angle=0}
}
\caption{ 
\label{fig.rateasy4}
$\Gamma^{\rm av}$ as a function of the static bias 
$\varepsilon$ for four different combinations of the number of levels, i.e., 
$M=6$ ($\bullet$) and $M=8$ ($\Box$),  and different temperatures, i.e., 
 $T=0.1$ (solid line) and $T=0.15$ (dashed line). 
 For the remaining parameters, see Fig.\  \ref{fig.rateasy0}. 
}
\end{figure}

The influence of a time-dependent periodic driving on the quantum 
relaxation rate is elucidated with 
Figs.\  \ref{fig.rateasy0} - \ref{fig.rateasy4}. 

For the case of off-resonant driving, Fig.\  \ref{fig.rateasy0} exhibits  
a non-monotonic dependence of  
the averaged relaxation rate $\Gamma^{\rm av}$ on the bias. For increasing 
$M$ the results approach each other. However, a complete 
convergence as observed in 
the undriven case (Fig.\  \ref{fig.rateasy2}) is not obtained. 

Tuning the driving frequency $\Omega$ into resonance, the results 
in Fig.\  \ref{fig.rateasy3}  show for a fixed driving strength 
a characteristic peak, being almost independent of the temperature $T$. 
The position of the peak is sensitive to the driving strength. This 
indicates that the population of the upper doublet is mainly the result 
of driving and not due to thermal population. 

Furthermore, we draw the reader's attention to the strongly ($s=1.0$)
driven symmetric case 
$\varepsilon=0$. The low temperature relaxation rate for $T=0.1$ 
is larger than for the two other cases  with higher temperatures. 
This is opposite to the situation {\em without\/} driving, see Fig.\ 
\ref{fig.rateasy1} b.) where $\Gamma$ is {\em smaller\/} for $T=0.1$ compared 
to $T=0.5$. This is a typical footprint of a  quantum effect: The 
resonant driving ($\Omega =0.815$) in the symmetric potential transfers 
population to the upper doublet where tunneling is enhanced because the 
tunneling splitting is large and the temperature is not high enough 
to  destroy coherence completely. 

The problem of convergence of the results with increasing $M$ is 
addressed in Fig.\  \ref{fig.rateasy4} for the resonantly 
driven case. Shown are four  
different combinations for $M=6$ ($\bullet$) and $M=8$ ($\Box$) for two  
temperatures $T=0.1$ (full line) and $T=0.15$ (dashed line). 
We ascertain that no convergence is obtained upon increasing 
$M$. The difference between the results for $M=6$ and $M=8$ for 
this large driving frequency is larger than in the 
case of off-resonant driving, see Fig.\  \ref{fig.rateasy0}. 
Obviously, more than only a few energy eigenstates are necessary 
to describe the resonantly driven double-well potential accurately.  
This is not astonishing, however, because the driving 
frequency is in resonance ($\Omega=0.815$) and many higher energy levels 
are excited. Moreover, we stress that the 
results vary {\em within\/} the same order of magnitude. 
  
\begin{figure}[t]
\centerline{
\epsfig{figure=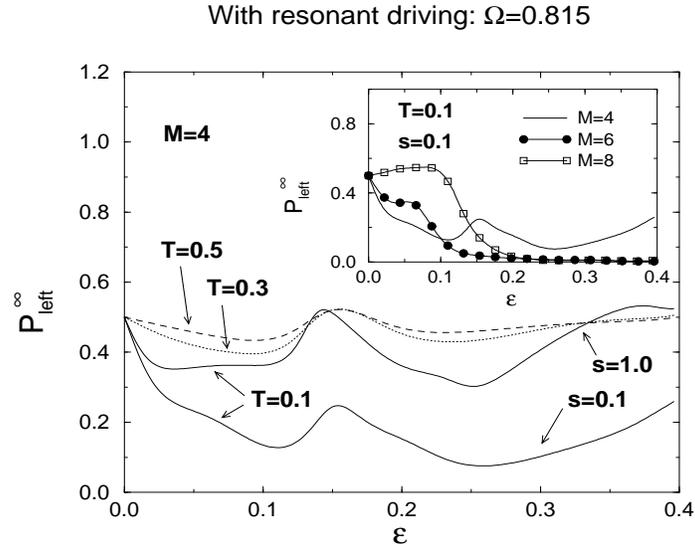,width=90mm,height=75mm,angle=0}
}
\caption{ 
\label{fig.pinfasy2}
Asymptotic population $P_{\rm left}^{\infty}$ of the left well 
as a function of the asymmetry $\varepsilon$ for the  
 temperatures $T=0.1$ (solid line), $T=0.3$ (dotted line) and $T=0.5$ (dashed 
 line) for the double-doublet system $M=4$. 
 Shown are results for strong driving $s=1.0$ and for weak driving 
 $s=0.1$ (for $T=0.1$ only). 
 For the remaining parameters, see 
 Fig.\  \ref{fig.rateasy0}. Inset: $P_{\rm left}^{\infty}$ 
 vs.\ the static bias $\varepsilon$ for a fixed temperature $T=0.1$ 
 for $M=4$ (solid line), $M=6$ ($\Box$) and $M=8$ ($\bullet$). 
}
\end{figure}

The asymptotic population $P_{\rm left}^{\infty}$ of the left well 
is shown in Fig.\  \ref{fig.pinfasy2} as a function of the static bias 
$\varepsilon$ for three different temperatures. The driving frequency is 
in resonance with the interdoublet energy gap. Compared to the 
strict monotonic behavior of the undriven case, see Fig.\  \ref{fig.pinfasy1}, 
the results with strong driving ($s=1.0$) 
show a non-monotonic dependence on the static bias 
with a distinct maximum at $\varepsilon=0.15$  being nearly independent 
of $T$. This maximum has a value of  $P_{\rm left}^{\infty} \approx 0.5$ 
indicating an equal population of both the metastable as well as the stable 
well. The position of this local maximum is invariant under the choice of 
the driving strength, as indicated by the weak driving result for ($s=0.1$). 
We note that a net population inversion can be achieved for the 
parameter combinations $M=4, T=0.1, s=1.0, \varepsilon=0.14$. 
It is interesting to see that 
convergence of $P_{\rm left}^{\infty}$ with increasing $M$ is 
obtained for asymmetries $\varepsilon>0.2$, see inset 
of Fig.\  \ref{fig.pinfasy2}.  For smaller values of the bias the qualitative 
behavior in the case of $M=6$ is similar to the case of $M=8$. 
However, the local maximum around $\varepsilon=0.15$ in the 
double-doublet case $M=4$ vanishes upon increasing $M$. 
%
\subsubsection{Dependence on the driving strength}
\label{subsubsec.drivs}
\begin{figure}[t]
\centerline{
\epsfig{figure=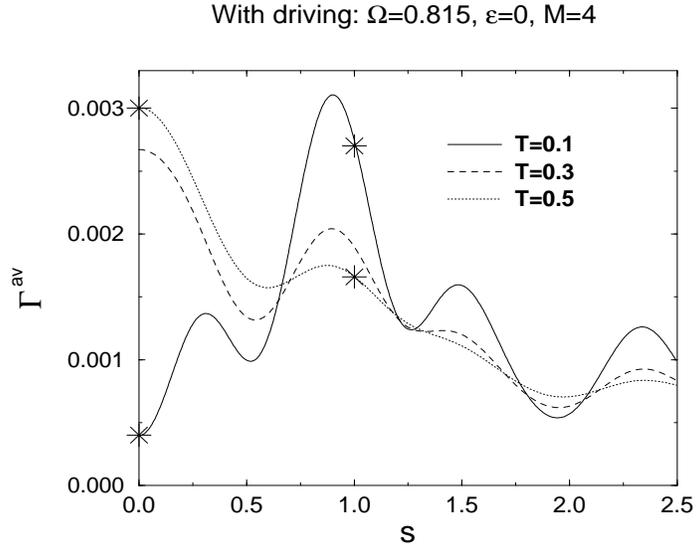,width=90mm,height=75mm,angle=0}
}
\caption{ 
\label{fig.rateampl1}
Averaged quantum relaxation rate $\Gamma^{\rm av}$ 
as a function of the driving amplitude $s$ for three 
different temperatures $T=0.1$ (solid line), $T=0.3$ (dashed line) and 
$T=0.5$ (dotted line) for the driven symmetric ($\varepsilon=0$) 
double-doublet system $M=4$. The static barrier height is 
$E_{\rm B}=1.4$ and the driving frequency is 
$\Omega=\overline{\omega}_0=0.815$.   The 
bath parameters are $\gamma=0.1$ and $\omega_c=10.0$. 
The asterisks $\ast$ mark the findings of an exponential fit to QUAPI results 
(not shown). 
}
\end{figure}

Fig.\ \ref{fig.rateampl1} depicts the averaged rate as a function of  the 
amplitude $s$ of the ac-driving field with resonant driving frequency 
$\Omega=\overline{\omega}_0=0.815$ for the symmetric double-doublet system 
($\varepsilon=0$, $M=4$). Shown are the results for three different temperatures. 
The asterisks $\ast$ mark the results of an exponential fit to QUAPI 
results (not shown) and confirm the validity of our new analytical 
approach.

\begin{figure}[t]
\centerline{
\epsfig{figure=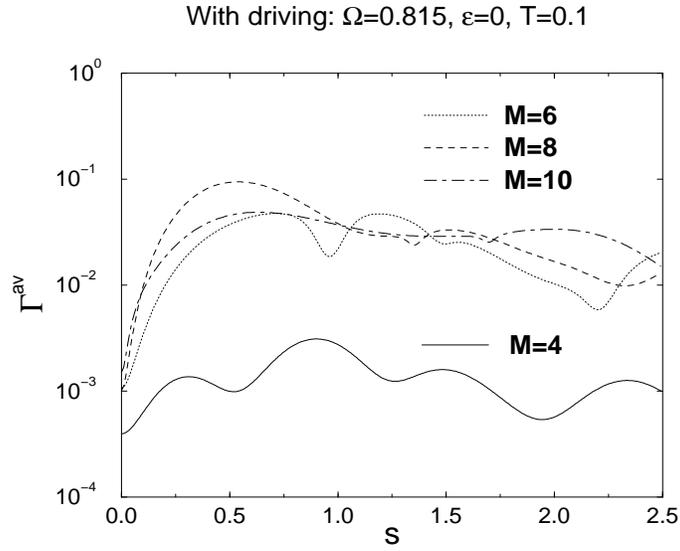,width=88mm,height=75mm,angle=0}
}
\caption{ 
\label{fig.rateampl2}
$\Gamma^{\rm av}$ as a function of the driving amplitude $s$
 for an increasing number of levels. The temperature is fixed to $T=0.1$. 
 For the remaining parameters, see Fig.\  \ref{fig.rateampl1}. 
}
\end{figure}

The averaged rate for the case of a high temperature $T=0.5$ is reduced compared 
to the undriven situation where  $\Gamma^{\rm av}(\equiv\Gamma)$ has a maximum. 
Upon decreasing the temperature, the relative maximum at $s\approx0.9$ grows out to 
a global maximum for $T=0.1$. This resonance is useful for practical applications 
(``Hydrogen subway'', see Section \ref{subsec.exp}) if one desires to 
accelerate the transfer of population from the left to the right well. So not 
only a resonant driving frequency $\Omega=\overline{\omega}_0$ but also a suitably 
chosen driving strength is important to maximize the transfer.  
The behavior of $\Gamma^{\rm av}$ vs.\ the driving amplitude is shown 
in Fig.\  \ref{fig.rateampl2} for an increasing number $M$ of energy states 
(note the logarithmic scale).  
For small driving strengths (up to $s\approx 0.2$), the result
 for $M=10$ does not significantly differ from 
the case for $M=8$, indicating numerical 
convergence. For intermediate to strong driving, 
the differences increase. However, we stress that the results for $M=8$ and $M=10$ 
remain within the same order of magnitude.  
\subsubsection{Dependence on the driving frequency}
\label{subsubsec.drivom}
\begin{figure}[t]
\centerline{
\epsfig{figure=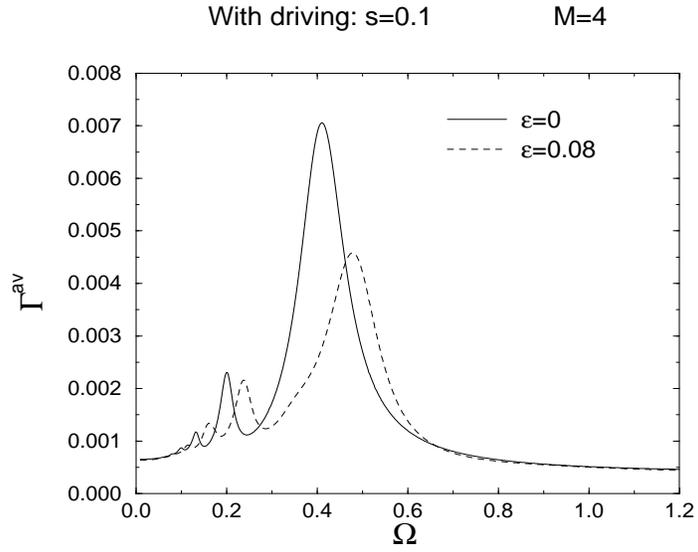,width=90mm,height=75mm,angle=0}
}
\caption{ 
\label{fig.ratefreq1}
Averaged quantum relaxation rate $\Gamma^{\rm av}$ 
as a function of the driving frequency $\Omega$ for the driven 
symmetric  (solid line) and the asymmetric (dashed line) 
  double-doublet system $M=4$.
The static barrier height is 
$E_{\rm B}=1.4$ and the driving strength is 
$s=0.1$.   The 
bath parameters are $T=0.1, \gamma=0.1$ and $\omega_c=10.0$. 
}
\end{figure}

\begin{figure}[t]
\centerline{
\epsfig{figure=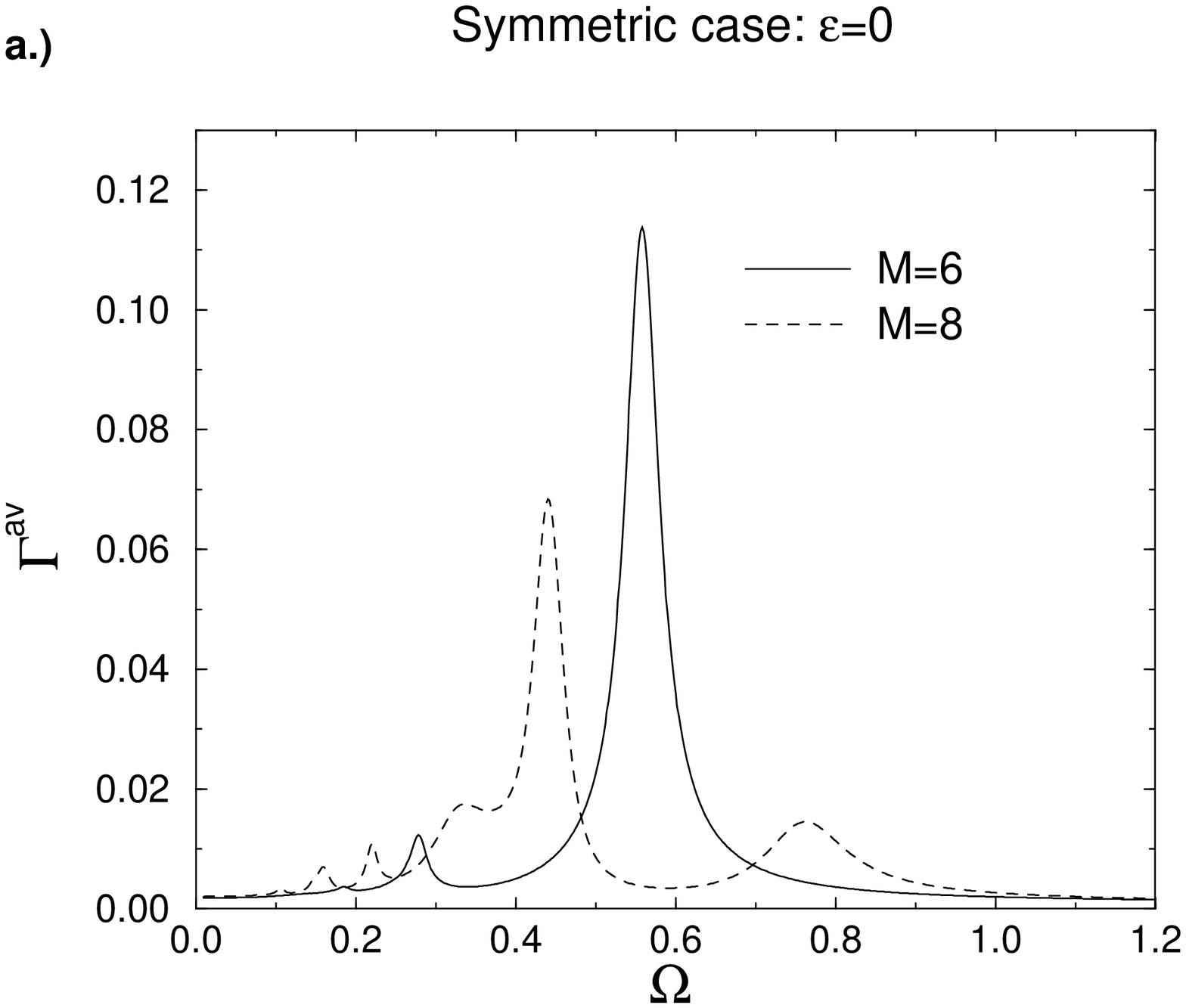,width=90mm,height=75mm,angle=0}
\epsfig{figure=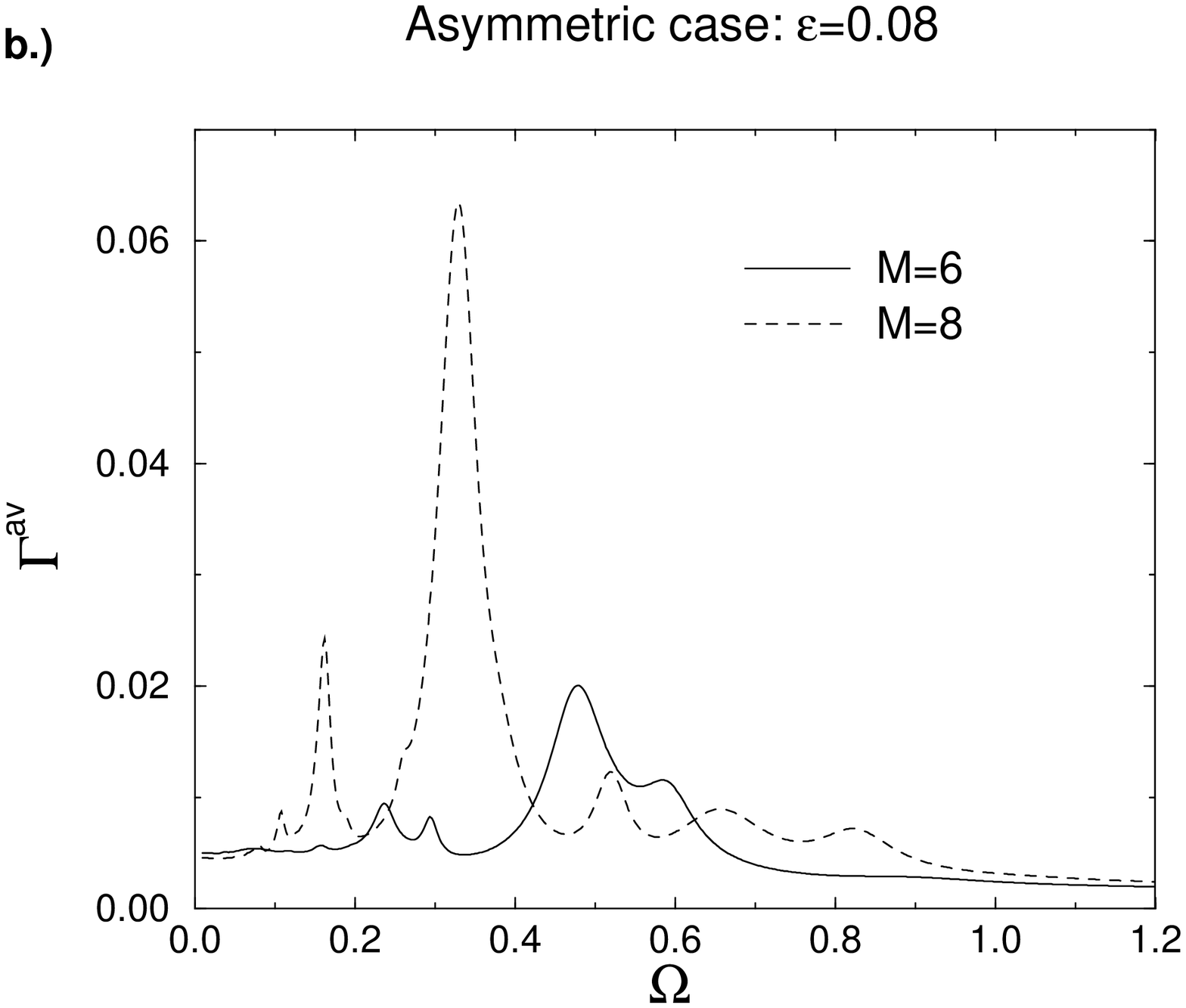,width=90mm,height=75mm,angle=0}
}
\caption{ 
\label{fig.ratefreq2}
$\Gamma^{\rm av}$ as a function of the driving frequency 
$\Omega$ for different numbers $M$ of levels. a.): 
symmetric case $\varepsilon=0$, b.): 
asymmetric case $\varepsilon=0.08$. The remaining parameters are as in 
 Fig.\  
\ref{fig.ratefreq1}.
}
\end{figure}

The dependence of the averaged relaxation rate on the driving frequency is shown in 
Fig.\  \ref{fig.ratefreq1} for the symmetric and the asymmetric double-doublet 
system $M=4$.  The results can be viewed as a scan of the 
spectrum of the driven dissipative double-doublet system. At some 
values of $\Omega$ the transition from the left to the right well is enhanced. 
The intermediate damping constant $\gamma=0.1$ leads to a considerable 
broadening of the energy levels involved in the transitions, as can be 
deduced  
from the rather broad resonance lines. The symmetric (asymmetric) 
case reveals a distinct peak at $\Omega=0.4$ ($\Omega=0.5$)
together with sidebands at the corresponding fractions of $\Omega$. 
The behavior of $\Gamma^{\rm av}$ 
for an increasing number $M$ of energy levels is depicted in 
Fig.\ \ref{fig.ratefreq2} a.) 
for the symmetric case and in   Fig.\ \ref{fig.ratefreq2} b.) for the 
asymmetric case. It can be seen that the additional energy levels yield 
additional resonance lines. However, if one chooses the driving 
frequency sufficiently far from any resonance line, 
convergence can be achieved. 

\begin{figure}[t]
\centerline{
\epsfig{figure=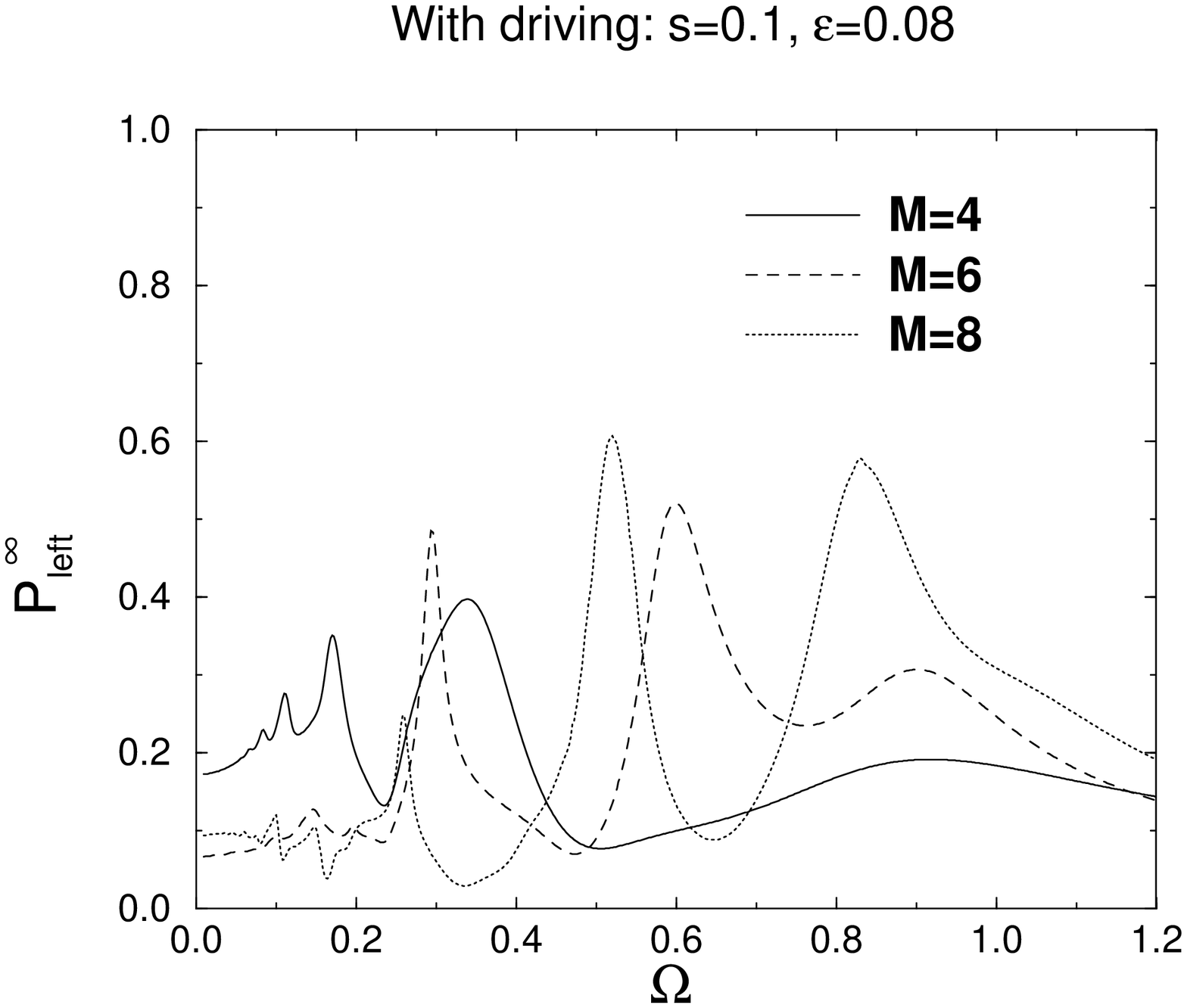,width=90mm,height=75mm,angle=0}
}
\caption{Asymptotic population $P_{\rm left}^{\infty}$ of the left well 
as a function of the driving frequency $\Omega$ for an increasing number $M$ 
of states. The driving amplitude is $s=0.1$ and the 
static bias is $\varepsilon=0.08$. 
The temperature is fixed at $T=0.1$, the damping constant is chosen to be 
 $\gamma=0.1$ and the cut-off is $\omega_c=10.0$. 
\label{fig.pinffreq}
}
\end{figure}

The asymptotic population $P_{\rm left}^{\infty}$ of the left well 
as a function of the driving frequency $\Omega$ is depicted in 
Fig.\ \ref{fig.pinffreq} for an increasing number $M$ of states. 
Clearly, the results do in general not converge with growing $M$. This 
 confirms our result  
that for an accurate description of a strongly driven quantum system 
more than only a few basis states are required.
%
\subsection{Dependence on the bath parameters}
\label{subsec.bath}
%
%
\subsubsection{Influence of temperature}
\label{subsubsec.bathtemp}
\begin{figure}[t]
\centerline{
\epsfig{figure=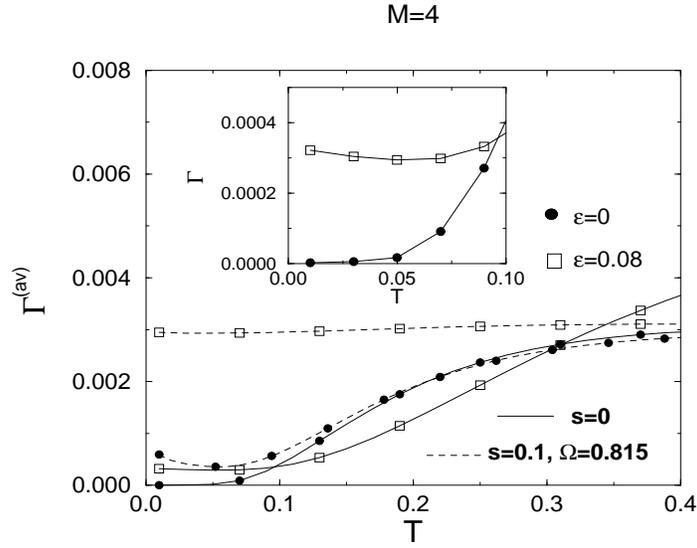,width=90mm,height=75mm,angle=0}
}
\caption{ 
\label{fig.ratetemp1}
(Averaged) quantum relaxation rate $\Gamma^{\rm (av)}$ 
as a function of temperature $T$ for four  
different combinations of the undriven ($s=0$) and the 
resonantly driven ($s=0.1, 
\Omega=0.815$) case without ($\varepsilon=0$) and 
with ($\varepsilon=0.08$) static bias 
for the double-doublet system $M=4$. 
The parameters are  $E_{\rm B}=1.4, \gamma=0.1$ and $\omega_c=10.0$. 
Inset: Enlarged part of the low temperature 
regime for the undriven case $s=0$.  
}
\end{figure}

The dependence of the quantum relaxation rate on temperature is depicted in 
Fig.\  \ref{fig.ratetemp1} for the case of the double-doublet system $M=4$. 
Shown are four cases without ac-driving ($s=0$), with 
resonant ac-driving ($s=0.1, \Omega=0.815$), 
 without bias ($\varepsilon=0$) and with bias ($\varepsilon=0.08$). 
We first concentrate on the undriven case $s=0$ (full line). Interestingly 
enough, we 
find in the low temperature regime (see inset of Fig.\  \ref{fig.ratetemp1}) 
that the rate first {\em decreases\/} when the temperature is {\em increased\/} in 
presence of a static bias. This is a characteristic quantum feature: In 
contrast to a  
classical behavior, the quantum relaxation rate 
first decreases with increasing temperature due to an enhancement of  
decoherence. 
Then, however, the rate starts to increase again as soon as the higher 
doublet becomes thermally 
populated. This typical behavior has also been observed 
experimentally in the context of tunneling of impurities in solids 
\cite{DefectExp,Golding92,Chun93,Cukier95,Noya97,Enss97} 
(see also Section \ref{subsec.exp} and especially Ref.\ \cite{Chun93}). 
For the intermediate temperature regime the comparison between the 
symmetric and 
the asymmetric case reveals another interesting characteristics: One 
could argue 
that the almost linear increase  of the rate with temperature reveals a 
classical Arrhenius behavior. This, however, is not the case. We are still 
in a deep quantum regime, since $\Gamma$ in the asymmetric case is 
{\em smaller\/} than in the symmetric case! This is again a clear sign of quantum 
mechanics since the symmetric potential with $\varepsilon=0$ corresponds to a 
resonant tunneling situation. A finite bias of $\varepsilon=0.08$ implies a 
non-resonant tunneling situation. 
There, the transfer via resonant tunneling is suppressed and the rate becomes 
 smaller. 
In the regime of intermediate to high temperatures, this behavior is   
inverted, i.e., the rate for the asymmetric case with reduced barrier height is 
now larger as compared to the symmetric case. 

In presence of a weak resonant ac-driving $s=0.1, \Omega=0.815$, 
the quantum relaxation rate   
$\Gamma^{\rm av}$  for the symmetric case $\varepsilon=0$  at low temperature 
is larger than the corresponding undriven relaxation rate. 
Coherent excitations to the upper 
doublet where tunneling is enhanced dominate at low temperature. 
 The presence of an additional static bias 
$\varepsilon=0.08$ renders the averaged 
relaxation rate  $\Gamma^{\rm av}$ almost independent of temperature. 

\begin{figure}[t]
\centerline{
\epsfig{figure=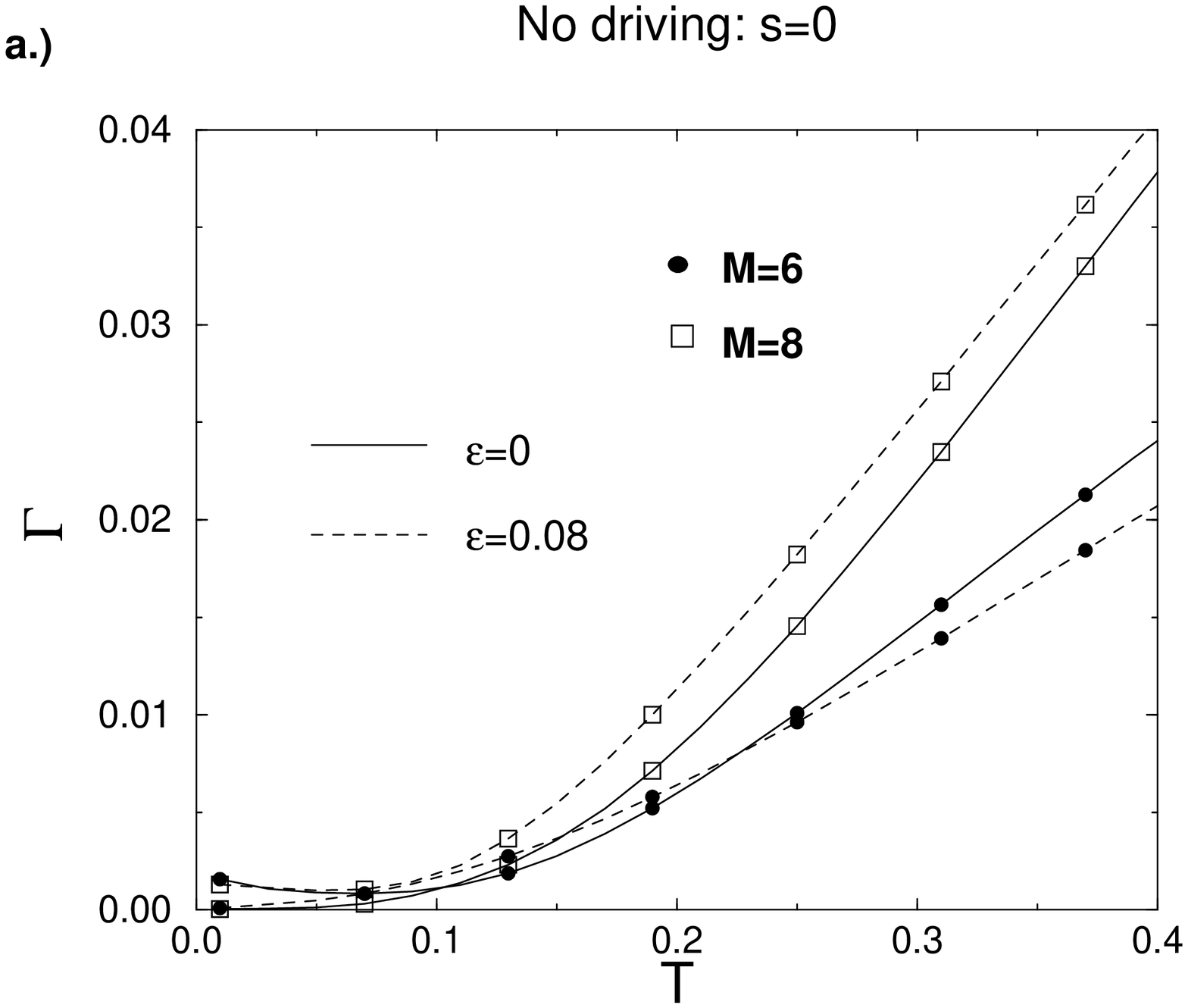,width=90mm,height=75mm,angle=0}
\epsfig{figure=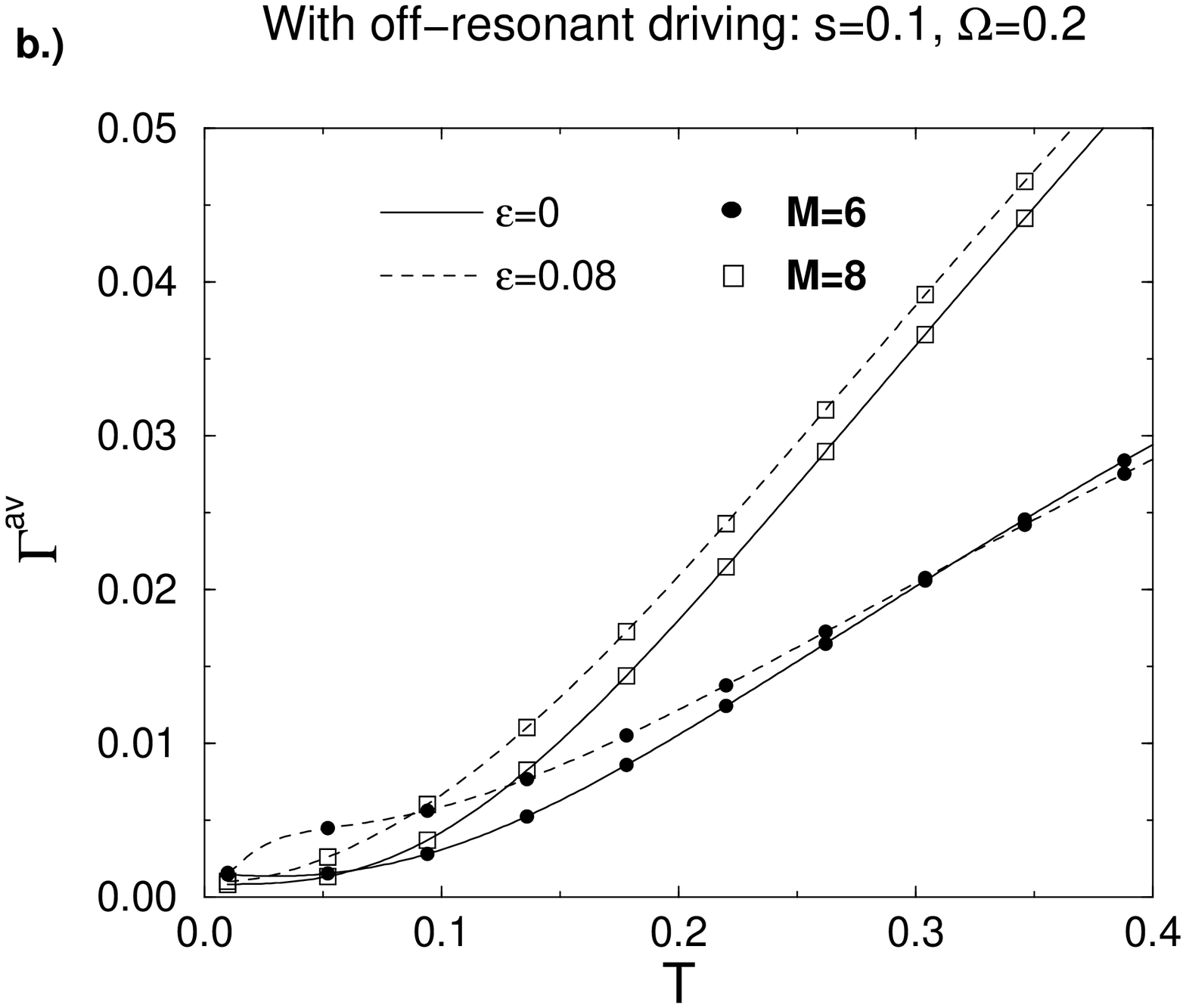,width=90mm,height=75mm,angle=0}
}
\caption{ 
\label{fig.ratetemp2}
$\Gamma^{\rm (av)}$ as a function of temperature  
$T$ for different numbers $M$ of levels. a.): 
no ac-driving $s=0$, b.): 
 with weak off-resonant 
 ac-driving $s=0.1,\Omega=0.2$. 
 The remaining parameters are as in Fig.\  
\ref{fig.ratetemp1}.
}
\end{figure}

The question of convergence of the rate with increasing the number $M$ of 
 energy eigenstates is 
addressed in Fig. \ref{fig.ratetemp2} a.) for the undriven case $s=0$ and in 
Fig. \ref{fig.ratetemp2} b.) for the off-resonantly 
driven case $s=0.1, \Omega=0.2$. 
In absence of ac-driving, a satisfactory convergence between the case 
$M=6$ and $M=8$ 
is achieved in the low temperature regime for both the 
symmetric and the asymmetric 
potential. Clearly for higher temperatures the 
agreement is worse because now the  higher lying energy 
states are not negligible. 
Also in presence of an off-resonant ac-driving, 
 convergence is achieved at low temperatures. 

\begin{figure}[t]
\centerline{
\epsfig{figure=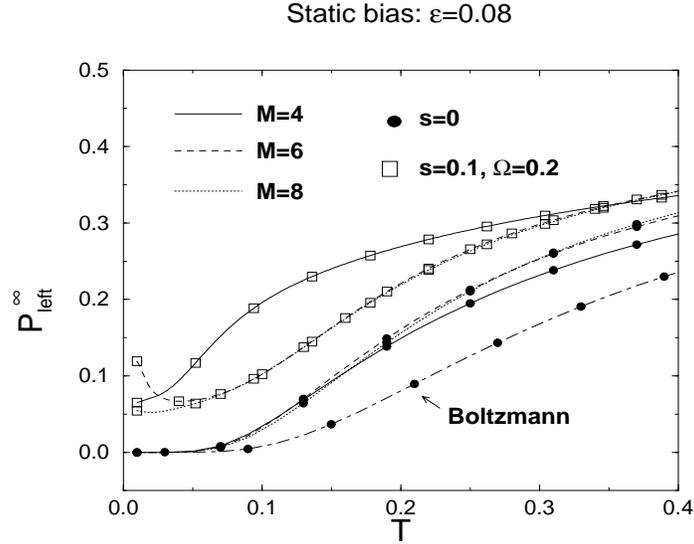,width=90mm,height=75mm,angle=0}
}
\caption{Asymptotic population $P_{\rm left}^{\infty}$ of the left well 
as a function of temperature $T$ for an increasing number $M$ 
of states. The static bias is $\varepsilon=0.08$. 
Shown are the case without ac-driving $s=0$ together with the results 
obtained from a Boltzmann equilibrium distribution (dashed-dotted line) and 
the case with off-resonant ac-driving $s=0.1, \Omega=0.2$.  
The remaining parameters are as in  Fig.\  \ref{fig.ratetemp1}. 
\label{fig.pinftemp}
}
\end{figure}

The asymptotic population $P_{\rm left}^{\infty}$ of the left well 
as a function of the temperature $T$ is shown in 
Fig.\ \ref{fig.pinftemp} for an increasing number $M$ of states. 
The results for  the undriven case $s=0$ reveal a satisfactory convergence 
over the entire temperature regime. For comparison, we also 
depict the asymptotic population stemming from the assumption of a Boltzmann 
equilibrium distribution. A similar argumentation as in Section 
\ref{subsubsec.drivdc} 
(see Fig.\ \ref{fig.pinfasy1}) holds to explain the disagreement.   

\subsubsection{Influence of  damping}
\label{subsubsec.bathgamma}
%
%
%

\begin{figure}[t]
\centerline{
\epsfig{figure=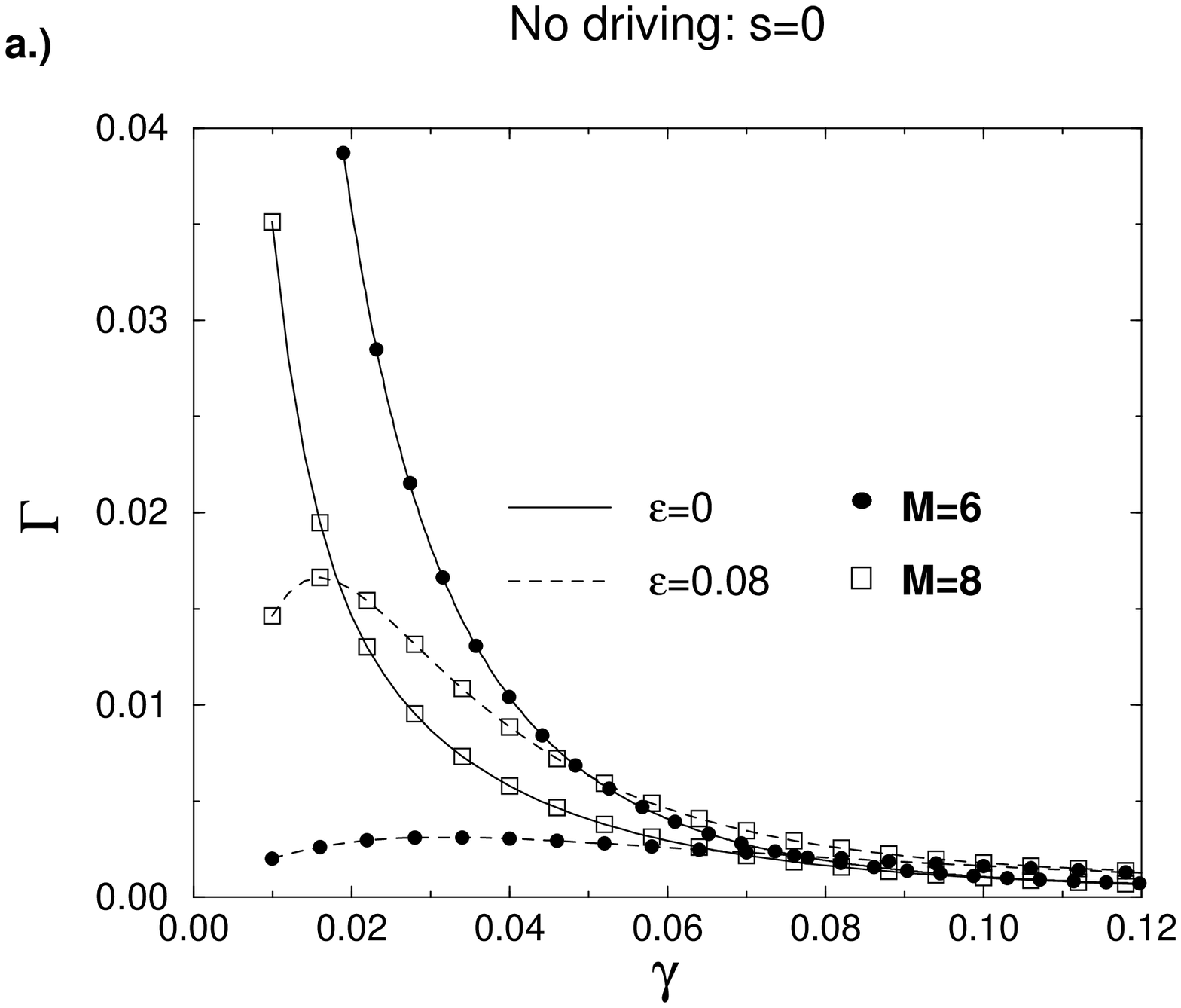,width=90mm,height=75mm,angle=0}
\epsfig{figure=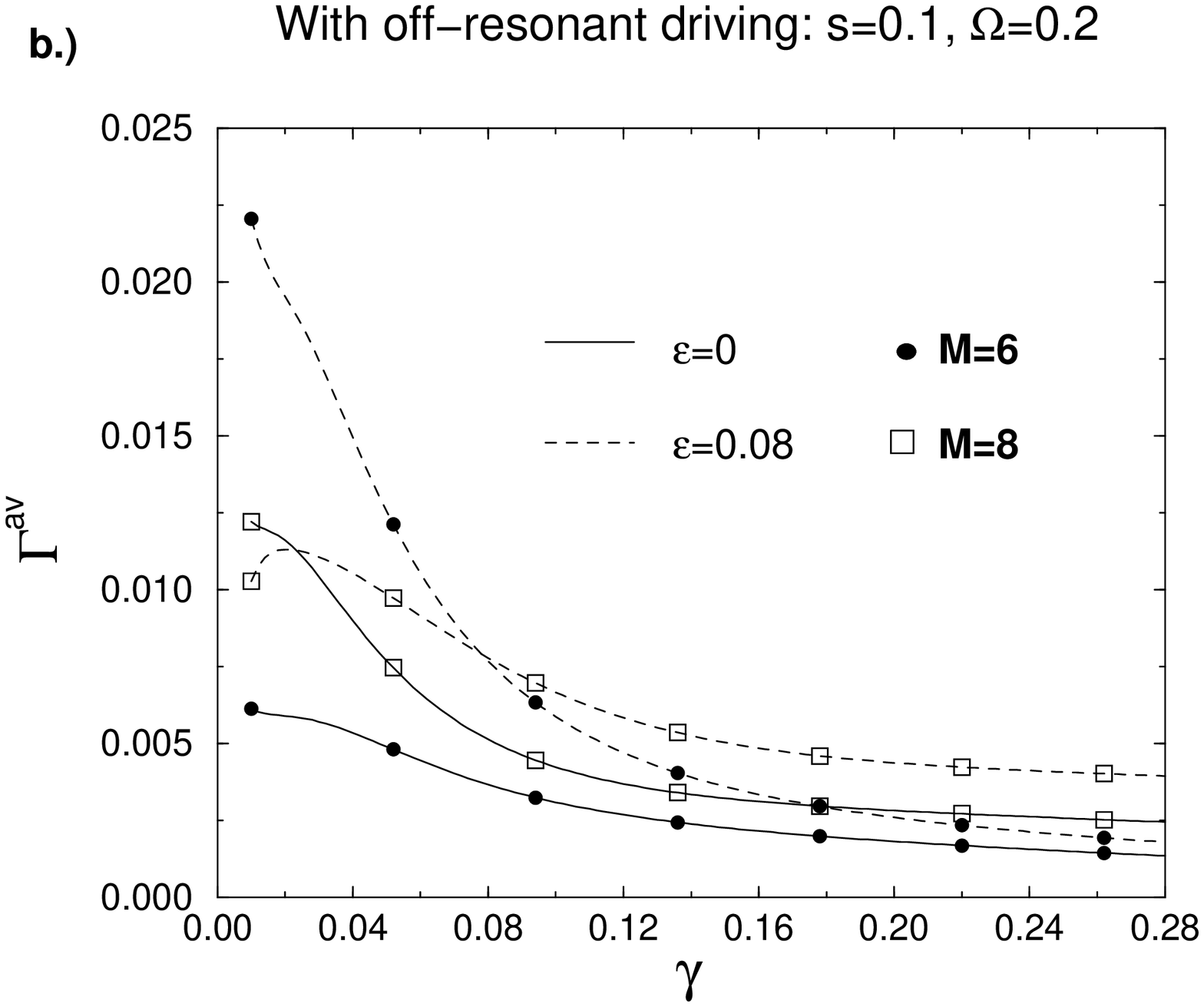,width=90mm,height=75mm,angle=0}
}
\caption{ 
\label{fig.rategamma2}
(Averaged) quantum relaxation rate $\Gamma^{\rm (av)}$, respectively, 
as a function of the damping strength $\gamma$ 
for different numbers $M$ of levels. a.): 
no ac-driving $s=0$, b.): 
 with off-resonant ac-driving $s=0.1,\Omega=0.2$.  Shown are the 
 results for the cases  
without ($\varepsilon=0$) and with ($\varepsilon=0.08$) static bias. 
The parameters are $E_{\rm B}=1.4, 
T=0.1$ and $\omega_c=10.0$. 
}
\end{figure}

%
Fig.\ \ref{fig.rategamma2} depicts the (averaged) rate vs.\ $\gamma$ for 
an increasing number $M$ of states. We find for the undriven case 
$s=0$ in Fig.\  \ref{fig.rategamma2} a.) convergence for $\gamma\ge0.08$ for 
both the symmetric and the asymmetric potential. However, 
for smaller $\gamma$ notable differences occur which indicate that the 
gNICA to second order is not reliable because the effective 
damping becomes too weak  
(effect of flow to weak damping, see discussion in Appendix \ref{app.outlook}). 
The situation is similar in presence of an off-resonant ac-driving $s=0.1, 
\Omega=0.2$, see Fig.\  \ref{fig.rategamma2} b.). The results for the case of a 
resonant driving $\Omega=\overline{\omega}_0$ are qualitatively similar 
(not shown). 

%
\section{Conclusions and outlook} \label{sec.conclusio}

A {\em novel\/} scheme to investigate analytically as well as numerically 
tunneling and vibrational relaxation in a strongly driven bistable potential 
was presented. A necessary first step in our approach is the reduction 
of the system dynamics to the Hilbert space spanned by the $M$ lowest 
energy eigenstates of the static bistable potential. Because the coupling 
to the heat bath is bilinear in the system and bath coordinates, 
the convenient basis to perform calculations consists of the eigenbasis 
of the position operator, i.e., the so-termed discrete variable representation 
(DVR). It is this DVR basis which permitted us to derive a set of 
non-Markovian generalized master equations (GME) for the 
diagonal elements of the reduced density matrix. In the studied regime 
of temperature and damping, the Markovian approximation to the GME 
yields novel analytical results. They agree well, both  with those of the full 
GME and of precise {\em ab-initio} numerical path integral calculations. 
In turn, the {\em quantum relaxation rate\/} could be extracted from 
the Markovian rate matrix. The dependence of the quantum relaxation 
on the five most relevant model parameters, namely bias strength $\varepsilon$, 
driving strength $s$,  driving frequency $\Omega$, temperature $T$ and  
damping $\gamma$, was outlined in detail. 

We have identified several quantum mechanical 
footprints in this strongly damped system. The four most pronounced quantum 
features are: 

(i) In absence of ac-driving we find resonant incoherent tunneling. 
This is demonstrated by striking resonances in the relaxation rate at distinct 
values of the dc-bias. 

(ii) We observe a decrease of the relaxation rate as the effective barrier is 
lowered by a static bias. This finding is due to a reduction of tunneling 
since the energy gaps forming the 
tunneling doublets increase with increasing 
asymmetry. In contrast, the relaxation rate in a classical system always 
increases for a reduced barrier.

(iii) The ac-driven quantum system shows distinct resonances of the 
relaxation rate for particular values of the driving amplitude. 
Especially at low temperature, the relaxation rate is enhanced by driving 
as compared to the undriven case.

(iv) A {\em non-monotonic\/} dependence of the relaxation rate on temperature 
is observed. Increasing the temperature in a classical system always increases 
the rate. However, in a quantum system, a higher temperature induces a 
larger population of the energetically higher lying doublets, where tunneling 
is favored. Increasing temperature further renders the 
quantum system more incoherent and the relaxation via tunneling is again 
hampered. The rate therefore decreases before it grows again due to thermal 
hopping. 

Our analysis furthermore permits to determine 
 the asymptotic population of the left metastable well. 
We have shown explicitly that a Boltzmann equilibrium distribution 
(both in absence and in presence of an ac-field) is {\em not\/} 
attained for the chosen set of parameters. 

The GME in Eq.\  (\ref{thegme}) and its time-inhomogeneous 
Markov approximation in Eq.\  (\ref{themarkovgme}) 
treat the external driving field 
{\em exactly\/}, and are reliable for moderate temperatures and/or  
moderate damping strengths. Indeed, the equations become exact 
for Ohmic dissipation at high temperatures. Thus, our approach 
complements a Redfield-type analysis being appropriate for weak damping. 
In contrast to semiclassical calculations we can consider shallow potential 
barriers substaining only a few doublets below the barrier. Moreover, 
in contrast to semiclassical imaginary-time rate calculations, we are not 
limited by the requirement of thermal equilibrium at adiabatically 
varying external fields. 

A major restriction of the presented method is that the generalized 
non-interacting cluster approximation (gNICA) that we used to obtain 
the GME turns out to be useful {\em in praxi\/} for numerical purposes  only 
 when (i) the number $M$ of levels remains moderately small ($M \le 10$), and 
 when (ii) the truncation of the gNICA kernels in Eq.\  (\ref{kernels}) 
to lowest order in $\Delta_{\mu \nu}^2$ is appropriate,  see 
 Eq.\  (\ref{ddskernel}).  Clearly, 
the number $M$ of levels involved in the dynamics 
increases with increasing temperature and/or for large driving 
strengths $s$, and/or for resonant driving frequencies $\Omega$. 

Taking into account 
a larger number $M$ of basis states implies that the position eigenvalues  
lie more dense in position space (in the limit of infinitely many 
energy eigenstates the distance between neighboring DVR-points is 
infinitesimally small). However, we have observed in the 
preceding Sections  that the square of the distance between two 
DVR-points enters as a prefactor for the twice integrated bath autocorrelation 
function  $S(t) +  i R(t)$  
in the second order kernels in Eq.\  (\ref{ddskernel}) of the generalized 
master equation. This implies that upon increasing $M$ the effective 
damping of each transition is reduced and the multi-level system 
effectively {\em flows to weak damping\/}. For small effective damping 
the noise action does no longer suppress long intervals in the off-diagonal 
states (clusters) and higher than second order transitions start 
to contribute. 

To deal with this effective weak coupling situation (which occurs for large $M$ 
even when the global coupling constant $\gamma=\eta/ {\cal M}$ is not small), 
a procedure similar to the one used by  Zwerger \cite{Zwerger87} and 
Grabert, Ingold and Paul 
\cite{Grabert98} to investigate transport in  Josephson junctions can be used, 
see Appendix \ref{app.outlook}. 

Due to its very general nature, the newly developed analytical technique 
contains a large potential for applications to specific experimental 
situations. Several possible applications for 
experimental systems have been discussed 
in Section \ref{subsec.exp}. 

A prominent question for future work refers to the behavior of the 
crossover to the classical 
regime. Our results should merge into those of the quantum Kramers rate 
\cite{HanggiRMP90} for semiclassical barrier heights, i.e., 
$\Delta U / (\hbar \omega_0) = E_{\rm B} \gg 1$, in the case without 
ac-driving. Once this regime is explored, the method can be  
generalized to time-dependent semiclassical quantum systems. 
However, one has to be aware that for the driven system a chaotic 
dynamics generally occur 
\cite{Kohler00}. We expect that the dynamics in the semiclassical 
regime involves an increasing number $M$ of DVR states. Then the effect of the 
flow to weak damping occurs. The analysis in Appendix \ref{app.outlook} 
represents the starting point for future investigations towards 
this challenge. 

\section*{Acknowledgement}
We thank Prof.\  M.\ Morillo and  
Dr.\ I.\ Goychuk for most helpful discussions. 
This work has been supported by the Deutsche Forschungsgemeinschaft Grant 
No.\ HA 1517/19-1 (P.H., M.T.), in part by the Sonderforschungsbereich 
 486 of the Deutsche Forschungsgemeinschaft (P.H.) and by the 
DFG-Graduiertenkolleg 283.
%
%
%
\begin{appendix}
\section{Scaling to dimensionless quantities}
\label{app.scaling}
For the specific calculations, 
we introduce in this Appendix dimensionless quantities. 
They are obtained by scaling the Hamiltonian in Eqs.\  
(\ref{system}) and (\ref{staticpotential}) and the 
environmental parameters specified in the Eqs.\  (\ref{specdens}) and 
(\ref{ic}). The relations read  
\[
\begin{array}[t]{cclccclcccl}
\tilde{t}&=& \omega_0 t \, , & &
\tilde{q}( \tilde{t} )&=& \sqrt{{\cal M} \omega_0 / \hbar} \, q 
(t=\frac{ \tilde{t}}{\omega_0}) \, , & & 
E_{\rm B} &=&
 \Delta U / \hbar \omega_{0} \, , 
\end{array} 
\]
\[
\begin{array}[t]{cclccclcccl}
\tilde{\varepsilon} & = &\varepsilon 
\sqrt{{\cal M} \omega_0 / \hbar} / \hbar \omega_{0} \, , & & 
\tilde{s} & = &s \sqrt{{\cal M} \omega_0 / \hbar} / \hbar \omega_{0} \, , & & 
\tilde{\Omega} & = & \Omega / \omega_0 \, ,
\end{array} 
\]
\begin{equation} 
\begin{array}[t]{cclccclcccl}
\tilde{\gamma} 
&=& \gamma / \omega_0  \, , & &\tilde{T} &=& \frac{k_B }{\hbar \omega_0} T \, , 
& & \tilde{\omega}_{c} &=&
  \omega_{c} / \omega_0 \, .
\label{dwscale}
\end{array} 
\end{equation}
We omit  all the tildes for the sake of better readability. 
%
\section{The bath correlation function}
\label{app.corr}
In this Appendix we give the explicit expressions for 
the twice integrated bath correlation function $Q(t)=S(t)+iR(t)$ of Eq.\  
(\ref{twintcorr}) with an Ohmic spectral density of 
Eq.\  (\ref{specdens}), see also \cite{Weiss99}.  
%
%
The real part $S(t)$ can be evaluated 
analytically by solving the integral in terms of the $\psi$-function. One  
arrives after some algebra at the exact expression 
\begin{equation} 
S(t) =  \frac{\eta}{\pi} \left\{
-\ln \left| \frac{\Gamma(1+\frac{1}{\hbar \beta \omega_c} + 
\frac{i t}{\hbar \beta})}
{\Gamma(1+\frac{1}{\hbar \beta \omega_c})}\right|^2 + 
\frac{1}{2} \ln (1+\omega_c^2t^2) 
\right\} \label{realcorrexact} 
%
%
%
%
\end{equation} 
%

For finite temperatures $k_{\rm B} T \ll \hbar \omega_c$ 
({\em scaling limit\/}), we obtain with $\Gamma(1+i y) \Gamma(1-i y) 
= \pi y / \sinh(\pi y)$ the expression
\begin{equation} 
S_{{\rm s.l.}}(t) =  \frac{\eta}{\pi} \left\{
-\ln \left[
\frac{\pi t}{\hbar \beta \sinh (\pi t / \hbar \beta)} 
\right]
+ \frac{1}{2} \ln (1+\omega_c^2t^2)
\right\} \, .
%
\label{lowtempcorr}
\end{equation} 
At long times $\omega_c t \rightarrow \infty$ the function 
$S_{{\rm s.l.}}(t)$ behaves like
\begin{eqnarray}
S_{{\rm s.l.}}(t) & \stackrel{\omega_c t 
\rightarrow \infty}{\longrightarrow} & 
\frac{\eta}{\pi} \ln \left( \frac{\hbar \beta \omega_c}{\pi} 
\sinh \frac{\pi t}{\hbar \beta}\right)  
\label{scalinglimits} \\
& & \approx \frac{\eta}{\pi} \left\{
\frac{\pi t}{\hbar\beta} + 
\ln  \left(\frac{\hbar\beta\omega_c}{2\pi} \right)
\right\}
\, , \, \mbox{\hspace{3ex}} k_{\rm B}T \ll \hbar \omega_c \, .
\label{slongtime}
\end{eqnarray}
This illustrates that the 
correlations between the paths  are 
damped out exponentially at long times for a low temperature Ohmic bath. 
%
%
The temperature-independent 
imaginary part $R(t)$ in Eq.\  (\ref{twintcorr}) can be determined 
exactly. We obtain after the integration over $\omega$
\begin{equation}
R(t) = \frac{\eta}{\pi} \arctan (\omega_c t) \, .
\end{equation} 
For long times $\omega_c t \rightarrow \infty$,
the $\arctan$-function approaches the Heaviside function, i.e., 
\begin{equation}
R(t) \stackrel{\omega_c t  
\rightarrow \infty}{\longrightarrow} \frac{\eta}{2} \Theta(t)  \, ,
\label{largeomc}
\end{equation} 
so that the imaginary part $R(t)$ becomes a constant function for all 
times $t$.
\section{Numerical iteration scheme for solving the generalized master equation}
\label{app.numgme}
The generalized master equation (\ref{thegme}) is a {\em set of 
$M$ coupled integro-differential equations with inhomogeneities\/}. For 
the one-dimensional case $M=1$, the 
GME would be of Volterra type. For this case, several standard techniques for 
a numerical treatment are known \cite{Volterra1,NumRec} 
and common numerical libraries \cite{NumRec} supply 
codes for this case. The general $M$-dimensional case is far from being standard 
and we are not aware of available algorithms in the literature. 
The non-locality in time detains us from diagonalizing the kernel rate matrix 
and thereby decoupling the equations. 
We develop in this Appendix a rather simple numerical algorithm 
for solving the general set of $M$ coupled inhomogeneous 
integro-differential equations.

We start by formally integrating the GME (\ref{thegme}) once and choose the 
integration constants such that the initial conditions are fulfilled. 
After interchanging the order of integration, we obtain 
\begin{equation}
\rho_{\mu\mu} (t) = \sum_{\nu=1}^{M} \, \int_{t_0}^t dt' \, 
{\cal K}_{\mu \nu} (t, t') \rho_{\nu \nu} (t') + \int_{t_0}^t dt' \, 
 I_{\mu} (t',t_0) +  \rho_{\mu\mu} (t_0) \, ,
\label{thegmeint} 
\end{equation}
where we have defined the integrated kernels
\begin{equation}
{\cal K}_{\mu \nu} (t, t') \equiv \int_{t'}^t dt'' {\cal H}_{\mu \nu} (t'', t') 
\, .
\label{intkernels}
\end{equation}
In the next step, we iterate the integrated GME from time $t$ to time 
$t+\Delta t$, and split the integrals to obtain  
\begin{eqnarray}
\rho_{\mu\mu} (t+\Delta t) &=& \sum_{\nu=1}^{M} \, \int_{t_0}^{t+\Delta t} 
dt' \, 
{\cal K}_{\mu \nu} (t+\Delta t, t') \, \rho_{\nu \nu} (t') +
 \int_{t_0}^{t+\Delta t} dt' \, 
 I_{\mu} (t',t_0) +  \rho_{\mu\mu} (t_0) \nonumber \\
& = & 
 \sum_{\nu=1}^{M} \, \int_{t_0}^{t} 
dt' \, \int_{t'}^{t} 
dt'' \, 
{\cal H}_{\mu \nu} (t'', t') \,  \rho_{\nu \nu} (t') +
 \int_{t_0}^{t} dt' \, 
 I_{\mu} (t',t_0) +  \rho_{\mu\mu} (t_0) \nonumber \\
 &&\mbox{} + \sum_{\nu=1}^{M} \, \int_{t_0}^{t}  
dt' \, \int_{t}^{t+\Delta t} 
dt'' \, 
{\cal H}_{\mu \nu} (t'', t') \, \rho_{\nu \nu} (t') \nonumber \\
&&\mbox{}+ \int_{t}^{t+\Delta t} dt' \, 
 I_{\mu} (t',t_0) + \sum_{\nu=1}^{M} \, \int_{t}^{t+\Delta t} 
dt' \, 
{\cal K}_{\mu \nu} (t+\Delta t, t') \, \rho_{\nu \nu} (t')  
  \nonumber \\
 &=& \rho_{\mu \mu} (t) + \sum_{\nu=1}^{M} \, \int_{t_0}^{t}  
dt' \, \rho_{\nu \nu} (t') 
\int_{t}^{t+\Delta t} dt'' \, 
{\cal H}_{\mu \nu} (t'', t') \,  \nonumber \\
&&\mbox{}+ \int_{t}^{t+\Delta t} dt' \, 
 I_{\mu} (t',t_0) + \sum_{\nu=1}^{M} \, \int_{t}^{t+\Delta t} 
dt' \, 
{\cal K}_{\mu \nu} (t+\Delta t, t') \,  \rho_{\nu \nu} (t') \, . 
\label{splitintgme} 
\end{eqnarray}
So far, every manipulation was an exact transformation. To proceed, 
we have to invoke an approximation for 
the last term in Eq.\ (\ref{splitintgme}), i.e., that one involving  
${\cal K}_{\mu \nu} (t+\Delta t, t')$. It is this term only which requires 
the knowledge of $\rho_{\nu \nu} (t')$ in the time interval $[t,t+\Delta t]$. 
All the other terms only need  $\rho_{\nu \nu} (t')$ up to time $t$ which 
are known. For that reason, we use in a third step the simplest approximation 
rule for the integral, i.e., the Simpson trapezoid rule, to obtain 
\begin{eqnarray}
\lefteqn{
\sum_{\nu=1}^{M} \, \int_{t}^{t+\Delta t} 
dt' \, 
{\cal K}_{\mu \nu} (t+\Delta t, t') \,  \rho_{\nu \nu} (t') 
\approx} \nonumber \\
 && \sum_{\nu=1}^{M} \frac{\Delta t}{2} \Big\{
{\cal K}_{\mu \nu} (t+\Delta t, t) \,  \rho_{\nu \nu} (t) + 
{\cal K}_{\mu \nu} (t+\Delta t, t+\Delta t) \,  \rho_{\nu \nu} (t+\Delta t)
\Big\} \, .
\label{simpson}
\end{eqnarray}
The corrections are of the order of $\Delta t ^ 3$. 
With Eq.\ (\ref{intkernels}), it follows that 
${\cal K}_{\mu \nu} (t+\Delta t, t+\Delta t) = 0$ and we arrive at the 
final iteration scheme
\begin{eqnarray}
\rho_{\mu\mu} (t+\Delta t) &=& 
 \sum_{\nu=1}^{M} \, \rho_{\nu \nu} (t) \Big\{ 
 \delta_{\mu\nu} + \frac{\Delta t}{2} {\cal K}_{\mu \nu} (t+\Delta t, t)
 \Big\} \nonumber \\
&&\mbox{} + \sum_{\nu=1}^{M} \, \int_{t}^{t+\Delta t}  
dt''  \int_{t_0}^{t} 
dt' \, 
{\cal H}_{\mu \nu} (t'', t') \, \rho_{\nu \nu} (t') 
+ \int_{t}^{t+\Delta t} dt' \, 
 I_{\mu} (t',t_0) \, , \nonumber \\
\label{gmeiteration}
\end{eqnarray}
where the $\delta_{\mu\nu}$ is the Kronecker symbol. We note that this 
iterative procedure requires the knowledge of $\rho_{\nu \nu} (t')$ 
in the time interval $t' \in [t_0,t]$ when propagating from time $t$ 
to time $t+\Delta t$. 
Furthermore, we remark that this 
iterative algorithm is not restricted to the lowest order for the 
kernels ${\cal H}_{\mu \nu} (t, t')$ and the inhomogeneities 
$I_{\mu} (t,t_0)$. Finally, we observe that the integrations from $t$ to 
$t+\Delta t$ for each step can be performed numerically 
 to a very high precision such that the only relevant numerical error 
arises from the splitting in Eq.\ (\ref{simpson}). Practical calculations 
reveal that the time step $\Delta t$ has to  be chosen rather small 
since the problem is similar to a stiff differential equation. In praxi, 
this means a value of the order of $\Delta t = 10^{-2}$ or smaller. 
This rather small value for $\Delta t$ restricts the applicability 
of this very simple and straightforward iteration scheme to 
problems where the decay is not too slow.  More refined 
numerical algorithm are imaginable which could circumvent this 
problem.

Because we are interested in the long-time dynamics, iterations 
up to times $t=5000$ can be necessary. Since the 
kernel matrix elements contain exponentials with asymptotically linearly  
growing exponents (see Eqs.\  (\ref{twintcorr}) and (\ref{slongtime}) in Appendix  
\ref{app.corr}), the memory ranging from time $t$ backwards to time $t_0$ 
can be cut-off after some fixed time span $\tau_r$. Then, the memory is 
only relevant over the time interval $t-\tau_r$ and all exponentially 
small contributions from the time interval $[t_0, t-\tau_r]$ can  
be neglected.
 This accelerates the iteration considerably and avoids 
too large arrays for the storage of the intermediate values  
of $\rho_{\nu \nu} (t')$.  However, lowering the temperature demands 
an increasing memory range $\tau_r$. 

Once the diagonal elements $\rho_{\mu\mu}(t)$ are known, the 
population of the left well $P_{\rm left} (t)$ can be evaluated according to 
Eq.\  (\ref{observable}).  
%
\section{Example: A single path subject to dissipation} 
\label{app.examplepath}
\begin{figure}[t]
\centerline{
\epsfig{figure=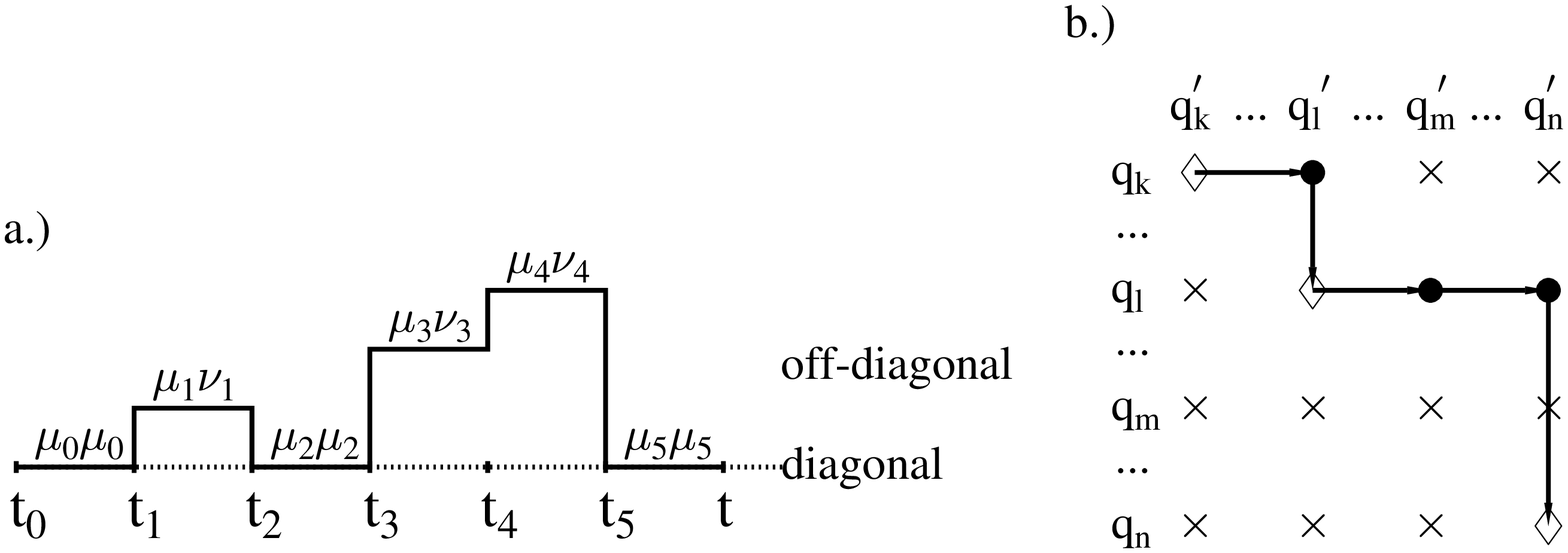,width=150mm,height=60mm,angle=0} 
}
\caption{Example of one path consisting of three sojourns and 
two clusters (see text). a.) Time-resolved representation of the path 
jumping between diagonal (dashed line) and off-diagonal states. 
b.) The same path illustrated in the $(q,q')$-plane of the reduced density 
matrix. The labels $q_{k},q_{k}^{\prime}$ of the reduced density 
matrix are not specified further. Diamonds $\Diamond$ mark the 
visited diagonal states 
and filled circles $\bullet$ mark the visited off-diagonal states. 
 \label{fig.onepathexample}}
\end{figure}

The purpose of this Appendix is to illustrate the general derivation 
of the cluster function, Eq.\  (\ref{fullpsum}), 
within the gNICA scheme introduced 
in Section \ref{sec.gnica}. This approximation scheme is the basis for the 
derivation of the generalized master equation presented in Section 
\ref{sec.gme}. For simplicity, 
we pick one single path subject to dissipation and 
 starting in the diagonal state 
$(\mu_0, \mu_0) = (q_k, q_k^{\prime}=q_k)$ and 
ending after $N=5$ jumps in the diagonal state 
$(\mu_5, \mu_5)=(q_n, q_n^{\prime}=q_n)$. The full 
path is illustrated in Fig.\  \ref{fig.onepathexample} a.) in 
a time-resolved picture and in Fig.\  \ref{fig.onepathexample} b.) as  
a path jumping between the states of the reduced density matrix in 
the $(q,q')$-plane. The path is characterized 
by the sequence of index pairs  
\begin{eqnarray} 
\begin{array}{c*{5}{c@{\: \rightarrow \:}}cl}
   &(\mu_0, \mu_0) & (\mu_1, \nu_1) & (\mu_2, \mu_2)  & 
(\mu_3, \nu_3) & (\mu_4, \nu_4) & (\mu_5, \mu_5) &  \\
 \equiv &(q_k, q_k) & (q_k, q_l^{\prime}) & (q_l, q_l) & 
(q_l, q_m^{\prime}) & (q_l, q_n^{\prime}) & (q_n, q_n) & \!\!.
\end{array}\label{onepathexample}
\end{eqnarray}
It contains three sojourns (from $t_0$ to $t_1$, 
from $t_2$ to $t_3$ and from $t_5$ to $t$), cf.\ 
Fig.\  \ref{fig.onepathexample} a.). Moreover, 
the path 
contains two clusters (from $t_1$ to $t_2$ and from $t_3$ to $t_5$). 
The details of the visited states are illustrated in 
Fig.\  \ref{fig.onepathexample} b.): The diagonal states are marked by 
diamonds $\Diamond$ and the visited off-diagonal states by filled circles 
$\bullet$. For our purpose here, it is only important to distinguish 
between diagonal and off-diagonal states. The specific indices 
of the states are irrelevant.
%

We evaluate now the contribution  ${\cal I}(t)$ 
 of this specific path to the full path sum in Eq.\  (\ref{fullpathint}). 
The product of the factors 
$\Delta_j$ in Eq.\ (\ref{fullpathint}) yields a 
proportionality 
factor which we omit for the moment for simplicity. 
For further convenience, we consider the time-{\em in\/}dependent system, 
implying that the diagonal elements $E_{\mu_j}$, 
see Eq.\  (\ref{diagele}), in the 
Hamiltonian ${\bf H}_{\rm S}^{\rm DVR}$ are constant in time. 
The generalization to time-dependent systems is 
discussed in Section \ref{subsec.generalgme}. 
Therefore, we 
introduce the short-hand notation in Eq.\ (\ref{fullpathint}) 
according to 
\begin{eqnarray}
\Delta E_j &=& E_{\mu_j} - E_{\nu_j}\, , \nonumber \\ 
{\cal F}_{lj}(t_l - t_j) 
&=& \exp \{ \xi_l S(t_l - t_j) \xi_j + i \xi_l R(t_l - t_j) 
\chi_j \} \, . 
\label{shortcutnotation}
\end{eqnarray}
%

The contribution of the specific path to the path sum follows 
from Eq.\ (\ref{fullpathint}) as
\begin{eqnarray}
{\cal I}(t) & = & \int_{t_0}^{t} dt_5 \, \int_{t_0}^{t_5} dt_4 \, \int_{t_0}^{t_4} dt_3 \, 
\int_{t_0}^{t_3} dt_2 \, \int_{t_0}^{t_2} dt_1 \, 
\exp \{ i [\Delta E_1 (t_2-t_1) +\Delta E_3 (t_4-t_3) 
\nonumber \\
&&+\mbox{} \Delta E_4 (t_5-t_4) ]\}  
 {\cal F}_{1,0}(t_1 - t_0) {\cal F}_{2,0}(t_2 - t_0)
 {\cal F}_{2,1}(t_2 - t_1) {\cal F}_{3,0}(t_3 - t_0) \nonumber \\
&&
 \times
 {\cal F}_{3,1}(t_3 - t_1) 
{\cal F}_{3,2}(t_3 - t_2) {\cal F}_{4,0}(t_4 - t_0) 
{\cal F}_{4,1}(t_4 - t_1) {\cal F}_{4,2}(t_4 - t_2) \nonumber \\
&&
 \times
 {\cal F}_{4,3}(t_4 - t_3) {\cal F}_{5,0}(t_5 - t_0) 
{\cal F}_{5,1}(t_5 - t_1) {\cal F}_{5,2}(t_5 - t_2) 
{\cal F}_{5,3}(t_5 - t_3) {\cal F}_{5,4}(t_5 - t_4) \, . \nonumber \\
\label{onepathcontr}
\end{eqnarray}

Eq.\  (\ref{onepathcontr}) is still exact. 
We apply then the generalized non-interacting cluster approximation gNICA 
in different steps. \\ 

\noindent
(i) Neglect of the intercluster correlations. \\
Let us consider the product 
\begin{equation}
{\cal P}_1 \equiv  
{\cal F}_{3,1}(t_3 - t_1) {\cal F}_{3,2}(t_3 - t_2) {\cal F}_{4,1}(t_4 - t_1) 
{\cal F}_{4,2}(t_4 - t_2) {\cal F}_{5,1}(t_5 - t_1) {\cal F}_{5,2}(t_5 - t_2) 
 \, . \label{prod1}
\end{equation}
If we assume a very large cut-off frequency, i.e., $\omega_c \rightarrow 
\infty$, we can approximate the real part $S(t_i-t_j)$ of the 
bath correlation function by its linearized form, Eq.\  (\ref{slongtime}), 
and the imaginary part $R(t_i-t_j)$ by the constant value 
$\eta/2$, see Eq.\ (\ref{largeomc}). Then, ${\cal P}_1$ can easily be brought 
into the form
\begin{eqnarray}
{\cal P}_1 & \approx & \exp 
\left\{
[\xi_3 + \xi_4 + \xi_5] \left[\frac{\eta}{\hbar \beta} (t_3 - t_1) \xi_1 + 
\frac{\eta}{\hbar \beta} (t_3 - t_2) \xi_2 + 
\frac{\eta}{\pi} \ln \left(\frac{\hbar\beta\omega_c}{2 \pi}\right) 
\left(\xi_1+\xi_2\right)\right] \right. \nonumber \\
& & \left. + \left[
\xi_4 \frac{\eta}{\hbar \beta} (t_4 - t_3) 
+ \xi_5 \frac{\eta}{\hbar \beta} (t_5 - t_3) \right]
\left[
\xi_1+\xi_2
\right]
\right\} \nonumber \\
& = & 1 \, .
\end{eqnarray}
In the last step, we have used the neutrality of each cluster, i.e., 
$\xi_1+\xi_2=0$ and $\xi_3 + \xi_4 + \xi_5=0$. 
%
%
%
%
%
%
The overall result is that we can 
set in the influence functional the product  with all the couplings between  
different clusters equal to 1, i.e., we disregard intercluster 
correlations. \\
%

\noindent
(ii) Neglect of cluster-sojourn correlations. \\
Next, we consider the product 
\begin{eqnarray}
\lefteqn{{\cal P}_2 \equiv 
{\cal F}_{3,0}(t_3 - t_0) {\cal F}_{4,0}(t_4 - t_0) 
{\cal F}_{5,0}(t_5 - t_0) = } \nonumber \\ 
& & 
\exp \{ \xi_3 S(t_3 - t_0) \xi_0 + \xi_4 S(t_4 - t_0) \xi_0  + 
\xi_5 S(t_5 - t_0) \xi_0 \nonumber \\ 
& & \mbox + i [ \xi_3 R(t_3 - t_0) \chi_0 + 
\xi_4 R(t_4 - t_0) \chi_0 +\xi_5 R(t_5 - t_0) \chi_0 ] \}
 \,  \label{prodini1}
\end{eqnarray}
describing the interaction of the clusters with the initial state characterized 
by $(\chi_0, \xi_0)$. 
Since we start in a diagonal state, it follows that $\xi_0=0$. Moreover, 
we apply the same argumentation like in (i) for the imaginary part $R(t)$ 
and obtain
%
\begin{equation}
{\cal P}_2 
\approx 
%
\exp \left\{ i \left[ \xi_3 +\xi_4+\xi_5 
\right]
\frac{\eta}{2} \chi_0 \right\}  = 1 
 \, . \label{prodini2}
\end{equation}
The corresponding argumentation holds for the 
third product ${\cal F}_{2,0}(t_2 - t_0) {\cal F}_{1,0}(t_1 - t_0)$. \\

\noindent
Step (i) and (ii) contain all the correlations disregarded within gNICA. 
In contrast, we entirely keep the {\em intra\/}cluster interaction 
${\cal F}_{5,4}(t_5 - t_4)$  as well as the interactions 
of the particular clusters with the directly preceding sojourn, i.e., 
 ${\cal F}_{4,3}(t_4 - t_3)$ and ${\cal F}_{5,3}(t_5 - t_3)$ 
 and ${\cal F}_{2,1}(t_2 - t_1)$. 
After reordering the integrals we obtain
\begin{eqnarray}
{\cal I}^{\rm gNICA}(t) & = & \int_{t_0}^{t} dt_5  \int_{t_0}^{t_5} dt_4  
\int_{t_0}^{t_4} dt_3  
\exp \{ i [\Delta E_3 (t_4-t_3) + \Delta E_4 (t_5-t_4) ]\}  
%
\nonumber \\ 
&& \times 
{\cal F}_{4,3}(t_4 - t_3) 
{\cal F}_{5,3}(t_5 - t_3) {\cal F}_{5,4}(t_5 - t_4)
\nonumber \\
&& \times
\int_{t_0}^{t_3} dt_2  \int_{t_0}^{t_2} dt_1  
\exp \left[ i \Delta E_1 (t_2-t_1) \right]  
%
{\cal F}_{2,1}(t_2 - t_1)
\, . 
\label{onepathcontrnica}
\end{eqnarray}
This expression can be treated more conveniently after a 
Laplace transformation to 
${\cal I}(\lambda) = {\cal L}_t\{{\cal I}(t)\} = \int_{0}^{\infty} 
dt \exp (-\lambda t) 
{\cal I}(t)$. Using the property ${\cal L}_t\{ \int_{t_0}^t dt_5 f(t_5-t_0) \} 
= \frac{1}{\lambda}{\cal L}_{\tilde{t}}\{  f(\tilde{t}) \}$ 
the integration over $t_5$ yields 
the factor $1/\lambda$. The remaining function $f(\tilde{t})$ 
to be Laplace transformed can be recast into the convolutive form 
$f(\tilde{t}) = \int_{t_0}^{\tilde{t}} dt_3 g(\tilde{t} - t_3 ) h(t_3)$ 
with $g(\tilde{t} - t_3) = 
\int_{t_3} ^{\tilde{t}} dt_4 
\exp \{ i [\Delta E_3 (t_4-t_3) + \Delta E_4 (\tilde{t}-t_4) ]\}
{\cal F}_{4,3}(t_4 - t_3) 
{\cal F}_{5,3}(\tilde{t} - t_3) {\cal F}_{5,4}(\tilde{t} - t_4)
\}
$
 and with 
$h(t_3) = 
\int_{t_0}^{t_3} dt_2 
\int_{t_0}^{t_2} dt_1 \, 
\exp [ i \Delta E_1 (t_2-t_1) ]  
{\cal F}_{2,1}(t_2 - t_1) 
$. 
By application of the convolution theorem 
${\cal L}_{\tilde{t}} \{ \int_{t_0}^{\tilde{t}} 
 dt_3 g(\tilde{t}-t_3) h(t_3) \} 
= {\cal L}_{\tilde{t}}\{  g(\tilde{t}) \} \, {\cal L}_{\tilde{t}}\{ 
h(\tilde{t}) \}$ and 
by performing the integration over $t_2$, we obtain the product
\begin{eqnarray}
{\cal I}^{\rm gNICA}(\lambda) &=& {\cal L}_t\{{\cal I}^{\rm gNICA}(t)\} \nonumber \\
&=& \frac{1}{\lambda} {\cal L}_{\tilde{t}} \Bigg\{
\int_{t_0}^{\tilde{t}} dt_4 \, \int_{t_0}^{t_4} dt_3 \, 
\exp \{ i [\Delta E_3 (t_4-t_3) + \Delta E_4 (\tilde{t}-t_4) ]\}
%
\nonumber \\
&& 
\times 
{\cal F}_{4,3}(t_4 - t_3) 
{\cal F}_{5,3}(\tilde{t} - t_3) {\cal F}_{5,4}(\tilde{t} - t_4)
\Bigg\} \nonumber \\
&& \times 
\frac{1}{\lambda} {\cal L}_{\tilde{t}} \left\{
\int_{t_0}^{\tilde{t}} dt_1 \, 
\exp [ i \Delta E_1 (\tilde{t}-t_1) ]  
%
{\cal F}_{2,1}(\tilde{t} - t_1) 
 \right\} \, . 
\label{intgnica2}
\end{eqnarray}
%
The Laplace transforms are the 
 contributions of the full intracluster interactions of the two clusters. 
Performing the Laplace transforms and transforming to the time 
differences $\tau_j = t_{j+1}- t_j$, we finally arrive at the expression
\begin{eqnarray}
{\cal I}^{\rm gNICA}(\lambda) &=&
\frac{1}{\lambda} \int_0^{\infty} d\tau_4 \int_0^{\infty} d\tau_3 
\exp\{ -\lambda (\tau_3+\tau_4)\} \exp\{ i (\Delta E_3 \tau_3+\Delta E_4 \tau_4)  \} 
\nonumber \\
&&\times {\cal F}_{4,3}(\tau_3) \, {\cal F}_{5,3}(\tau_3+\tau_4) \, 
{\cal F}_{5,4}(\tau_4) \nonumber \\
&& \times \frac{1}{\lambda} \int_0^{\infty} d\tau_1 
\exp\{ -\lambda \tau_1 \} \exp\{ i \Delta E_1 \tau_1   \} 
{\cal F}_{2,1}(\tau_1) \, \frac{1}{\lambda} \, ,
\label{onex}
\end{eqnarray}
where the last factor $1/\lambda$ appears after the integration over the 
first sojourn $t_1 - t_0$. 
Eq.\  (\ref{onex}) is an example of a contribution of one specific 
path to the total path sum in Eq.\  (\ref{fullpsum}). 
%
%
%
%
\section{Harmonic well approximation}
 \label{app.harmwell}
 
In this Appendix we give explicit results for
 an approximation for the energy eigenstates in 
the wells of the double-well potential 
(\ref{staticpotential}) without external forces, i.e., $\varepsilon=0$ and 
$s=0$. The scaling of this Hamiltonian is performed according to the 
standard procedure described in the previous Appendix 
\ref{app.scaling}, see eq.\  
(\ref{dwscale}). In the literature \cite{DekkerGroup}, 
often the assumption is made that the 
energy eigenvalues and the localized states in the two wells 
can be approximated by those of a harmonic oscillator potential whose 
minimum coincides with the single well minimum. The localized states of the 
double-well potential are linear combinations of the symmetric and the 
antisymmetric energy eigenstate corresponding to one doublet. 
They have been introduced 
in Section \ref{subsec.dds}, Eq.\  (\ref{locbasis}) for the case of the 
double-doublet system and a generalization to more than two doublets is 
straightforward. 

The eigenenergies and the energy eigenstates of a spatially shifted 
harmonic potential are given in dimensionless units by
\begin{equation}
{\cal E}_n = n+1/2, \mbox{\hspace{3ex}} n=0, 1, \dots,
\end{equation}
and
\begin{equation}
\psi_n(q) = \langle n|q\rangle = (2^n\, n! \sqrt{\pi})^{-1/2} 
\exp \left\{ -\frac{1}{2} (q-q_0)^2 \right\} 
H_n(q-q_0), \mbox{\hspace{3ex}} n=0, 1, \dots,
\end{equation}
where $q_0=\pm \sqrt{8 E_{\rm B}}$ is the position of the minima of the 
double-well potential with barrier height $E_{\rm B}$ and $H_n(q)$ 
are the Hermite polynomials. 

\begin{figure}[t]
\centerline{
\epsfig{figure=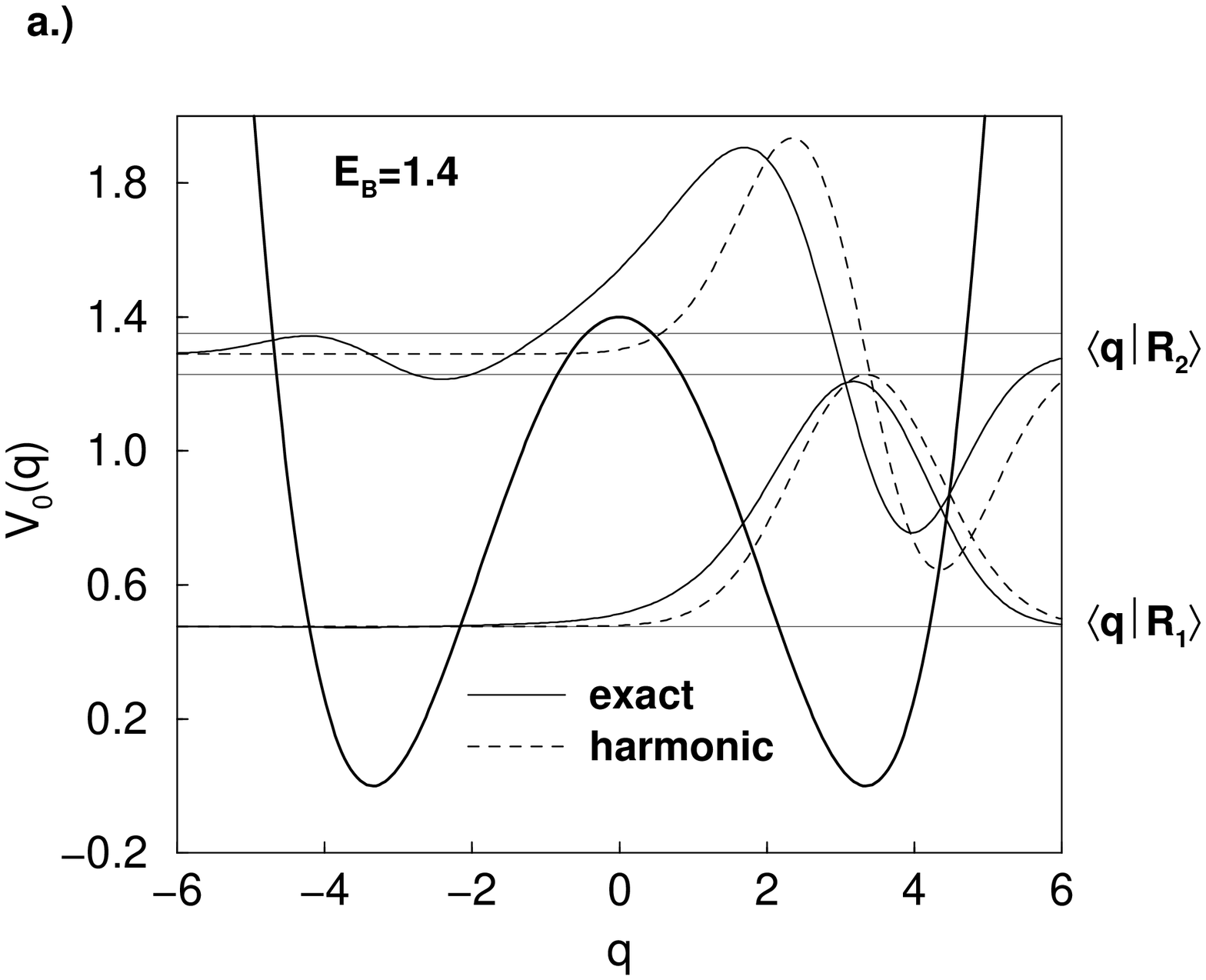,width=90mm,height=75mm,angle=0}
\epsfig{figure=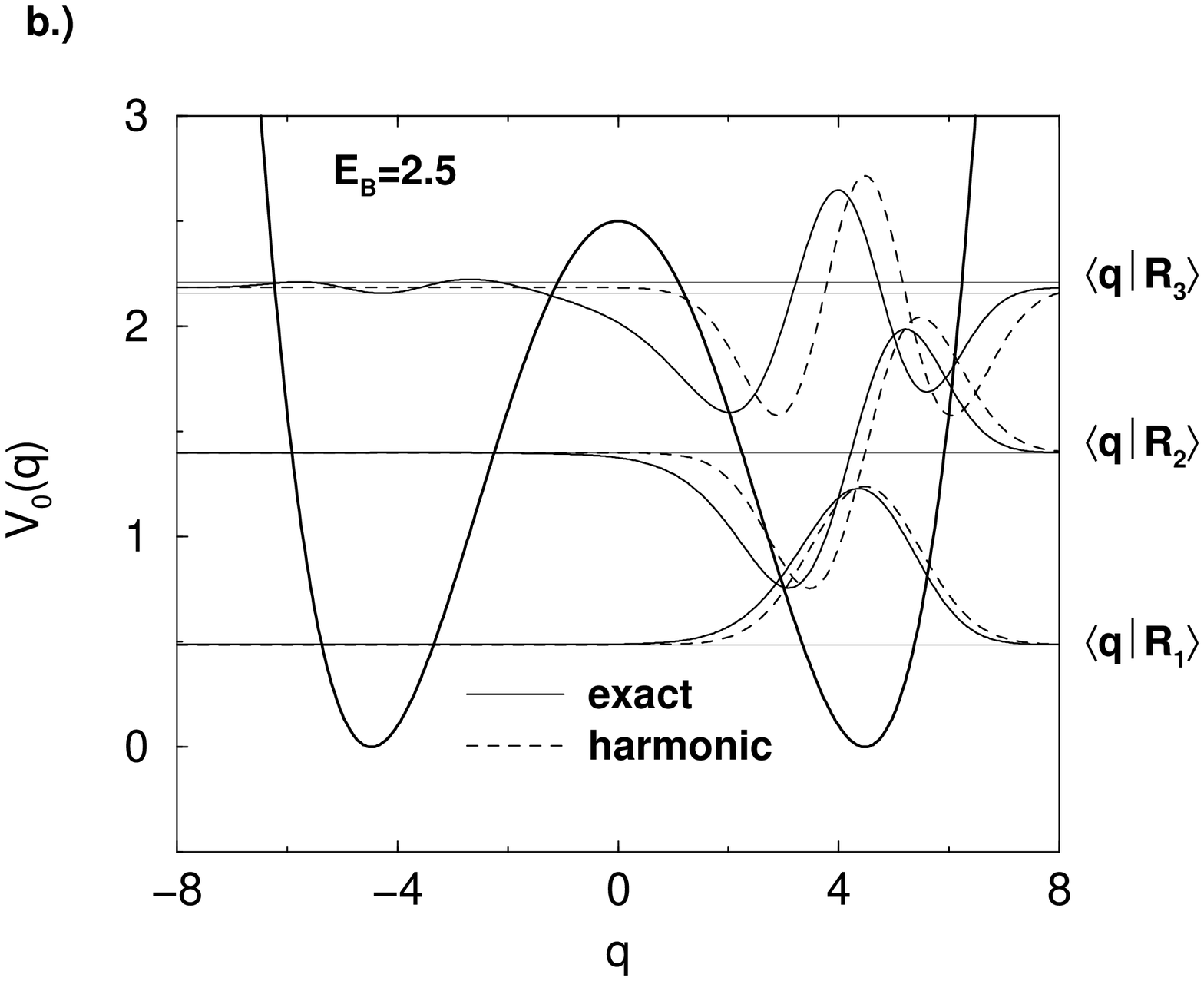,width=90mm,height=75mm,angle=0}
}
\caption{ 
\label{fig.harmwell}
a.) Exact localized states $\langle q| R_1\rangle$ and 
$\langle q| R_2\rangle$ (full lines) of the unperturbed system Hamiltonian 
(\ref{staticpotential}) in position representation 
for a barrier height of $E_{\rm B}=1.4$.  The dashed lines depict the 
results of the harmonic well approximation. The horizontal lines mark the 
exact eigenenergies of the double-well potential occurring always in pairs 
(doublets, the intradoublet spacing between the 
lower lying states is not visible on this scale). 
Note that the harmonic states are energetically shifted to the position 
of the  exact localized states for graphical reasons. 
b.) Same for a barrier height $E_{\rm B}=2.5$.
}
\end{figure}

Fig.\ \ref{fig.harmwell} shows the results of this approximation for two
cases of a barrier height $E_{\rm B}$ in the deep quantum regime, i.e., 
$E_{\rm B}=1.4$ (Fig.\ \ref{fig.harmwell} a.) with two doublets below 
the barrier and $E_{\rm B}=2.5$ (Fig.\ \ref{fig.harmwell} b.) with 
three doublets below the barrier. For comparison the exact 
(numerically obtained) wave functions are depicted. 
In both cases, the agreement is not convincing, as expected. 
Increasing the barrier 
height improves the agreement for the lower lying states. The states lying 
closed to the barrier top show a noticeable disagreement. 
Especially the part of the wave function which penetrates the barrier  and 
which is in turn associated with tunneling is underestimated. 

We note that for graphical reasons, the harmonic 
states are positioned at the {\em exact\/} eigenenergies of the double-well 
potential. However, in the approximation the harmonic eigenenergies 
are also shifted compared to the exact one (see scaling on the ordinate). 
To be specific, the interdoublet energy gap in the case of $E_{\rm B}=1.4$ 
is $\overline{\omega}_0=0.815$ in contrast to 
$\overline{\omega}_0=\omega_0=1$ for the corresponding harmonic potential. 
For the case of $E_{\rm B}=2.5$ the lower interdoublet splitting is 
$\overline{\omega}_{0,1}=0.892$ and the upper interdoublet splitting is 
$\overline{\omega}_{0,2}=0.805$. These values have also to be compared with 
$\overline{\omega}_0=\omega_0=1$ for the corresponding harmonic potential. 

The deviations up to $20 \%$ are certainly not astonishing since the 
condition for the applicability of the harmonic well approximation 
is a rather high barrier, i.e., $E_{\rm B} \gg 1$. 
This is also the relevant condition where semiclassical methods are 
applicable to calculate quantum relaxation rates \cite{HanggiRMP90}. 
They are rather simple compared to existing approaches 
\cite{DekkerGroup}.  
For barrier heights of the order of the interdoublet spacing, this 
approximative treatment of the eigenenergies and the wave functions 
is not applicable. 

\section{Flow to weak damping}
\label{app.outlook}
In this Appendix we describe the main steps of a proposal of how to deal with 
the effect of the flow to weak damping. A detailed investigation 
is still in progress. This part is to be viewed as 
a starting point for future work (see Section \ref{sec.conclusio}) 
and should only point out that this 
behavior can also be treated {\em within\/} the formalism of real-time 
path integrals. 

%
To deal with this effective weak-coupling situation, 
the twice integrated 
bath autocorrelation function $S(t) +  i R(t)$ in the asymptotic 
limit of the scaling limit, i.e., Eqs. (\ref{slongtime}) and (\ref{largeomc}) 
of Appendix \ref{app.corr}, may be used. The deviations from the exact form 
of the autocorrelation function are small for high temperatures and/or 
weak damping. We use these  
approximative expressions in the kernels in Eq.\ (\ref{kernels}) for the 
generalized master equation and obtain for the 
discrete influence phase (see also Eq.\  (\ref{discrfeynman})) 
\begin{eqnarray}
\phi_{\rm FV}[\chi,\xi] & = & 
- \sum_{l=1}^{N} \sum_{j=0}^{l-1} \xi_{l} 
S(t_l-t_j) \xi_{j}  
- i \sum_{l=1}^{N} \sum_{j=0}^{l-1} \xi_{l} 
R(t_l-t_j) \chi_{j} \nonumber \\
& = & -\frac{\eta}{\hbar \beta} \sum_{l=1}^{N} \sum_{j=0}^{l-1} \xi_{l} (t_l-t_j) \xi_{j} 
- \frac{\eta}{\pi} \ln \left(\frac{\hbar\beta\omega_c}{2\pi}\right) 
\sum_{l=1}^{N} \sum_{j=0}^{l-1} \xi_{l} \xi_{j} \nonumber \\
& & \mbox{} - i \frac{\eta}{2}  \sum_{l=1}^{N} \sum_{j=0}^{l-1} \xi_{l} \chi_{j} \, .
\end{eqnarray}
We  define 
\begin{equation}
p_j \equiv \sum_{l=1}^j \xi_l \label{defpj}, 
\end{equation}
and note that each cluster has a cumulative path weight of zero, 
see Eq.\  (\ref{clusterneutrality}), 
i.e., $p_N=0$. This allows for an elementary 
rearrangement of the double sums in order to obtain with $\tau_j = t_j - t _{j-1}$ 
the expressions 
\begin{eqnarray}
\sum_{l=1}^{N} \sum_{j=0}^{l-1} \xi_{l} (t_l-t_j) \xi_{j} & = & 
\sum_{j=1}^{N} \tau_j p_j^2 \, , 
\mbox{\hspace{5ex}} \sum_{l=1}^{N} \sum_{j=0}^{l-1} \xi_{l} \xi_{j}  = 
-\frac{1}{2}\sum_{j=1}^N\xi_j^2 \, , \nonumber \\
\sum_{l=1}^{N} \sum_{j=0}^{l-1} \xi_{l} \chi_{j} & = & 
- \sum_{j=1}^{N-1} \chi_j p_j \, . \label{rearrangedsums}
\end{eqnarray}
Inserting these equations into the kernels, Eq.\  (\ref{kernels}), 
it follows that 
\begin{eqnarray}
{\cal H}_{\mu \nu} (t,t') & = & 
\sum_{N=2}^{\infty} 
\sum_{\stackrel{ \{ \mu_j \nu_j \} }{\mu_j \ne \nu_j}} 
\prod_{j=1}^{N} (-1)^{\delta_{j}} \left(\frac{i}{2}\right)^N  
{\Delta}_{j-1} 
\left(\frac{2\pi}{\hbar\beta\omega_c}\right)^{\xi_j^2 \eta/ 2\pi \hbar} 
\exp \left(-i \frac{\eta}{2 \hbar} (-1)^{\delta_j} \xi_j p_j\right) \nonumber \\
& & \times \int_{t'}^{t} dt_{N} \int_{t'}^{t_{N}} dt_{N-1} \dots 
\int_{t'}^{t_{2}} dt_{1} \nonumber \\
& & \times 
\exp \left\{i \, \sum_{j=0}^{N-1} \left\{
\int_{t_{j}}^{t_{j+1}} dt''  [ E_{\mu_{j}}(t'') - E_{\nu_{j}}(t'')] 
-\frac{\eta}{\hbar\beta} p_{j+1}^2 \tau_{j+1} 
\right\} \right\} \, , 
\label{weakkernels}
\end{eqnarray}
where $\delta_j=0$ $(1)$ for a vertical (horizontal) jump. 

In order to evaluate the series of integrals in Eq.\  (\ref{weakkernels}), 
we use the 
following technique: The upper limit $t$ of the first integral 
is replaced by $\infty$ and  for compensation the step 
function $\Theta(t-t_N)$ is added to the integrand. Then, the order of 
integration is interchanged and the integrals are 
transformed to difference coordinates $\tau_j=t_j -t _{j-1}$. Like in the 
previous Sections, it is assumed that the driving frequency $\Omega$ is large and 
averaging over the driving period is appropriate. In a final step, 
the $\Theta$-function in the integrand is represented as a complex 
integral according to 
\begin{equation}
\Theta(t-t_N)=\frac{1}{2\pi i} \int_{-i \infty-\varepsilon}^{+i \infty-\varepsilon} 
\, d\lambda \; \frac{1}{\lambda} \exp \left[ \lambda 
\left(t-\sum_{j=0}^{N-1} \tau_{j+1}-t'\right) \right]
\, . \label{Thetaex}
\end{equation}
The complex integration over $\lambda$ can afterwards be carried out with the 
help of the calculus of residues for the residue at $\lambda=0$.  
After this tedious 
but straightforward procedure, one arrives at the final result for the 
averaged Markovian approximated rate matrix elements
\begin{eqnarray}
\Gamma_{\mu \nu}^{\rm av} & = & 
\sum_{N=2}^{\infty} 
\sum_{\stackrel{ \{ \mu_j \nu_j \} }{\mu_j \ne \nu_j}} 
\prod_{j=1}^{N} (-1)^{\delta_{j}} \left(\frac{i}{2}\right)^N  
{\Delta}_{j-1} 
\left(\frac{2\pi}{\hbar\beta\omega_c}\right)^{\xi_j^2 \eta/ 2\pi \hbar} 
\exp \left(-i \frac{\eta}{2 \hbar} (-1)^{\delta_j} \xi_j p_j\right) \nonumber \\
& & \times \int_{0}^{\infty} d\tau \, 
J_0\left( p_j \frac{2s}{\Omega} \sin\left(\frac{\Omega\tau}{2}\right)  \right) 
\exp \left\{-\left(i \,  [ F_{\mu_{j}}- F_{\nu_{j}}]  
-\frac{\eta}{\hbar\beta} p_{j}^2\right) \tau \right\}  \, , 
\label{weakrate}
\end{eqnarray}
where $J_0(x)$ is the zeroth Bessel function. 

It is not possible to treat this complicated expression analytically. If one 
has to determine explicit results, the help of the computer is needed to 
evaluate the  sum over all configurations $\{ \mu_j \nu_j \}$. Then, in 
a first  step, all paths belonging to a fixed order $N$ are created numerically 
by means of recursive programming. In the next step the sum over all the 
occurring paths has to be evaluated before one has to go to the next 
order by increasing 
$N$. Finally, convergence with respect to $N$ has to be obtained. 

We summarize this Appendix by concluding that also the effect of the flow 
to weak damping can be treated by real-time path integrals although the 
expressions become much more involved. Eq.\  (\ref{weakrate}) constitutes 
the starting point for the study of the cross-over to the classical regime. 
One has to be aware, however, that the driven problem is far from being 
trivial since even a chaotic dynamics may occur.
\end{appendix}
%

\newpage



%

\newpage

\newpage

%


\newpage



%

%
%

%
%

%
%
%
\newpage


\begin{thebibliography}{10}

\bibitem{Leggett8187}A.\ O.\ Caldeira and  A.\ J.\ Leggett,
 Phys.\ Rev.\ Lett.\  {\bf 46}, 211 (1981); 
 A.\ J.\ Leggett, S.\ Chakravarty, A.\ T.\ Dorsey, M.\ Fisher, 
A.\ Garg, and W.\ Zwerger, Rev.\ Mod.\ Phys.\  {\bf 59}, 1 (1987); 
{\bf 67}, 725 (E) (1995).

\bibitem{HanggiRMP90} P.\ H\"anggi, P.\ Talkner, and M.\ Borkovec, 
 Rev.\ Mod.\ Phys.\ {\bf 62}, 251 (1990).

\bibitem{Weiss99}  U.\ Weiss, {\em Quantum Dissipative Systems} 
 (World Scientific, Singapore, 1993; 2nd edition 1999).

\bibitem{VCHBuch} Th.\ Dittrich, P.\  H\"anggi, G.-L.\ Ingold, B.\ Kramer, 
G.\ Sch\"on, and W.\ Zwerger, {\em Quantum Transport and Dissipation}
 (Wiley-VCH, Weinheim, 1998).

\bibitem{Grifoni98}M.\ Grifoni and P.\ H\"anggi,  
Phys.\ Rep.\  {\bf 304}, 230 (1998).

\bibitem{Grossmann91}F.\ Grossmann, T.\ Dittrich, P.\ Jung, and P.\ H\"anggi, 
Phys.\ Rev.\ Lett.\  {\bf 67}, 516 (1991).

\bibitem{Oelschlaegel93}Th.\ Dittrich, B.\ Oelschl\"agel, and P.\ H\"anggi, 
 Europhys.\ Lett.\  {\bf 22}, 5 (1993).

\bibitem{Loefstedt94}R.\ L\"ofstedt and S.\ N.\ Coppersmith,  
 Phys.\ Rev.\ Lett.\  {\bf 72}, 1947 (1994); Phys.\ Rev.\ E {\bf 49}, 
 4821 (1994). 
 
\bibitem{Makarov95}D.\ E.\ Makarov and N.\ Makri, Phys.\ Rev.\ B {\bf 52}, 
R2257 (1995); Phys.\ Rev.\ E {\bf 52}, 5863 (1995). 

\bibitem{Grifoni96a}M.\ Grifoni and P.\ H\"anggi,  Phys.\ Rev.\ Lett.\  
{\bf  76}, 1611 (1996); M.\ Grifoni, L.\  Hartmann, S.\ Berchtold, 
and P.\ H\"anggi,  Phys.\ Rev.\ E {\bf  53}, 5890 (1996);  
M.\ Grifoni and P.\ H\"anggi,  Phys.\ Rev.\ E   
{\bf  54}, 1390 (1996).

\bibitem{Thorwart97a}M.\ Thorwart and P.\ Jung, 
Phys.\ Rev.\ Lett.\ {\bf 78}, 2503 (1997).

\bibitem{Pareek97}T.\ P.\ Pareek, M.\ C.\ Mahato, and A.\ M.\ Jayannavar, 
Phys.\ Rev.\ B {\bf 55}, 9318 (1997). 

\bibitem{Thorwart98a} M.\ Thorwart, P.\ Reimann, P.\ Jung, 
and R.\ F.\ Fox, Chem.\ Phys.\ {\bf 235}, 61 (1998). 

\bibitem{HanggiSRRMP}L.\ Gammaitoni, P.\  H\"anggi, P.\  Jung, and F.\  Marchesoni, 
 Rev.\ Mod.\  Phys.\ {\bf 70}, 223 (1998).

\bibitem{Goychuk99}I.\ Goychuk and P.\ H\"anggi,  Phys.\ Rev.\ E 
{\bf  59}, 5137 (1999); I.\ Goychuk and P.\ H\"anggi,  New J.\ Phys.\  
{\bf  1}, 14.1-14.14 (1999). 

\bibitem{Wellens00}Th.\ Wellens and A.\ Buchleitner, 
Phys.\ Rev.\ Lett.\ {\bf 84}, 5118 (2000). 

\bibitem{Thorwart98b}M.\ Thorwart, P.\ Reimann, P.\ Jung, and R.F. Fox, 
Phys.\ Lett.\ A {\bf 239}, 233 (1998). 

\bibitem{ChemPhysDriv97} {\em Dynamics of Driven Quantum Systems}, 
Special issue of Chemical Physics, 
edited by W.\ Domcke, P.\ H\"anggi and D.\ Tannor, Chem.\ Phys.\ {\bf 217} (2,3), 
117-416 (1997).  

\bibitem{Viola99}L.\ Viola, E.\ Knill,  and S.\ Lloyd,
Phys.\ Rev.\ Lett.\  {\bf 82}, 2417 (1999).

\bibitem{Thorwart00}M.\ Thorwart, L.\ Hartmann, I.\ Goychuk, and P.\ H\"anggi,  
 J.\ Mod.\ Opt.\ {\bf 47}, 2905 (2000).


\bibitem{QUAPI}N.\ Makri, {\em Real time path integrals with quasi-adiabatic 
propagators: Quantum Dynamics of a system coupled to a 
harmonic bath}, in {\em Time-Dependent Quantum Molecular Dynamics},
Proceedings of a NATO Advanced Research Workshop 1992 in Snowbird,
 Utah, NATO ASI Series B: Physics, Vol.\ 299, edited by J.\ Broeckhove and L.\ 
 Lathouwers (Plenum Press, New York, 1992); 
 D.\ E.\ Makarov and N.\ Makri, Chem.\ Phys.\ Lett.\  {\bf 221}, 482 (1994); 
 N.\ Makri, J.\ Math.\ Phys.\  {\bf 36} 2430 (1995); 
 N.\ Makri and D.\ E.\ Makarov, J.\ Chem.\ Phys.\  {\bf 102}, 4600 (1995); 
 {\bf 102}, 4611 (1995). 

\bibitem{Thorwart00a} M.\ Thorwart, M.\ Grifoni, and P.\ H\"anggi, 
 Phys.\ Rev.\ Lett. {\bf 85}, 860 (2000).



\bibitem{MacroTunnel95} 
{\em Quantum Tunneling of Magnetization}, Proceedings of the NATO Advanced 
Research Workshop 1994 in Grenoble, France, NATO ASI Series E: Applied Sciences,
 Vol.\  301, edited  by L.\ Gunther and B.\ Barbara 
 (Kluwer, Dordrecht, 1995).

\bibitem{vanHemmen86}J.\ L.\ van Hemmen and A.\ S\"ut\H{o}, 
Europhys.\ Lett.\  {\bf 1}, 481 (1986); Physica B  {\bf 141}, 37 (1986).

\bibitem{Mn12} J.\ R.\ Friedman, M.\ P.\ Sarachik, J.\ Tejada, and R.\  
Ziolo, Phys.\ Rev.\ Lett.\ {\bf 76}, 3830 (1996); 
%
J.\ M.\ Hern{\'a}ndez, X.\ X.\ Zhang, F.\ Luis, 
J.\ Bartolom{\'e}, J.\ Tejada, and R.\ Ziolo, Europhys.\ Lett.\ {\bf 35}, 
301 (1996); 
%
L.\ Thomas, F.\ Lionti, R.\ Ballou, D.\ Gatteschi, R.\ Sessoli, and B.\ Barbara, 
Nature {\bf 383}, 145 (1996); 
%
E.\ M.\  Chudnovsky, Science {\bf 274}, 938 (1996); 
%
J.\ M.\ Hern{\'a}ndez, X.\ X.\ Zhang, F.\ Luis, J.\ 
Tejada, J.\ R.\ Friedman, M.\ P.\ Sarachik, and R.\ Ziolo, 
Phys.\ Rev.\ B {\bf 55}, 5858 (1997); 
%
J.\ A.\ A.\ J.\ Perenboom, J.\ S.\ Brooks, S.\ Hill, 
T.\ Hathaway, and N.\ S.\ Dalal, Phys.\ Rev.\ Lett.\ {\bf 58}, 330 (1998);
%
G.\ Bellessa, N.\ Vernier, B.\ Barbara, and D.\ Gatteschi, 
Phys.\ Rev.\ Lett.\ {\bf 83}, 416 (1999);
%
W.\ Wernsdorfer, R.\ Sessoli, and D.\ Gatteschi, 
 Europhys.\ Lett.\ {\bf 42}, 254 (1999).

\bibitem{Fe8}C.\  Sangregorio, T.\ Ohm, C.\ Paulsen, R.\ Sessoli, 
 and D.\ Gatteschi,  Phys.\ Rev.\ Lett.\ {\bf 78}, 4645 (1997); 
%
J.\ Tejada, X.\ X.\ Zhang, E.\ del Barco, 
J.\ M.\ Hern{\'a}ndez, and 
E.\ M.\ Chudnovsky, Phys.\ Rev.\ Lett.\ {\bf 79}, 1754 (1997); 
%
W.\ Wernsdorfer and R.\ Sessoli,  
Science {\bf 284}, 133 (1999);
%
W.\ Wernsdorfer, T.\ Ohm, C.\ Sangregorio, R.\ Sessoli, 
D.\ Mailly, and C.\ Paulsen,  
Phys.\ Rev.\ Lett.\ {\bf 82}, 3903 (1999);
%
W.\ Wernsdorfer, I.\ Chiorescu, R.\ Sessoli, D.\ Gatteschi, and 
D.\ Mailly, Physica B: Cond.\ Mat.\ {\bf 284} --  {\bf 288}, 1231 (2000).

\bibitem{Wernsdorfer00}W.\ Wernsdorfer, R.\ Sessoli, A.\ Caneschi, 
D.\ Gatteschi, and A.\ Cornia, Europhys.\ Lett.\ {\bf 50} (4), 552 (2000).

%
\bibitem{Devoret88}J.\ Clarke, A.\ N.\ Cleland, M.\ H.\ Devoret, D.\ Est\`{e}ve, and 
J.\ M.\ Martinis, Science {\bf 239}, 992 (1988).

\bibitem{Han91}S.\ Han, J.\ Lapointe, and J.\ E.\ Lukens, 
Phys.\ Rev.\ Lett.\ {\bf 66}, 810 (1991).

\bibitem{Rouse95}R.\ Rouse, S.\ Han, and J.\ E.\ Lukens, 
 Phys.\ Rev.\ Lett.\ {\bf 75}, 1614 (1995).

\bibitem{Han96}S.\ Han, R.\ Rouse, and J.\ E.\ Lukens, 
 Phys.\ Rev.\ Lett.\ {\bf 76}, 3404 (1996).

\bibitem{Silvestrini97}P.\ Silvestrini, V.\ G.\ Palmieri, B.\ Ruggiero, and 
M.\ Russo, 
 Phys.\ Rev.\ Lett.\ {\bf 79}, 3046 (1997).

\bibitem{Han00}S.\ Han, R.\ Rouse, and J.\ E.\ Lukens,  
Phys.\ Rev.\ Lett.\  {\bf 84}, 1300 (2000).  

\bibitem{Friedmann00}J.\ R.\ Friedmann, V.\ Patel, W.\ Chen, S.\ K.\ Tolpygo, and 
 J.\ E.\ Lukens,  
Nature  {\bf 406}, 43 (2000); G.\ Blatter; {\em ibid.\/} {\bf 406}, 25 (2000).   

\bibitem{Mooij99}J.\ E.\ Mooij,  T.\ P.\  Orlando, L.\  Levitov, L.\ Tian,  
C.\ H.\ van der Waal and S.\ Lloyd, Science  {\bf 285}, 1036 (1999).  

\bibitem{vanderWal00}C.\ H.\ van der Wal, A.\ C.\ ter Haar, 
F.\ K.\ Wilhelm, R.\ N.\ Schouten, 
C.\ J.\ P.\ M.\ Harmans, T.\ P.\ Orlando, S.\ Lloyd and J.\ E.\ 
Mooij, Science  {\bf 290}, 773 (2000). 




\bibitem{DefectExp}  A.\ W\"urger, {\em From Coherent Tunneling to Relaxation:  
Dissipative Quantum Dynamics of Interacting Defects}
 (Springer, Berlin, 1997); 
%
 P.\ Esquinazi (ed.), {\em  
Tunneling Systems in Amorphous and Crystalline Solids} 
(Springer, Berlin, 1998). 

\bibitem{Golding92}B.\  Golding, N.\ M.\ Zimmerman, and S.\ N.\ Coppersmith,  
 Phys.\ Rev.\ Lett.\ {\bf 68}, 998 (1992).

\bibitem{Chun93}K.\ Chun and N.\ O.\ Birge, 
 Phys.\ Rev.\ B {\bf 48}, R11500 (1993).

\bibitem{Cukier95}R.\ I.\ Cukier, M.\ Morillo, K.\ Chun, 
and N.\ O.\ Birge, 
Phys.\ Rev.\ B {\bf 51}, 13767 (1995).

\bibitem{Noya97}J.\ C.\ Noya, C.\ P.\ Herrero, and R.\ Ram{\'{\i}}rez, 
Phys.\ Rev.\ Lett.\  {\bf 79}, 111 (1997).

\bibitem{Enss97}C.\ Enss and S.\ Hunklinger, 
Phys.\ Rev.\ Lett.\  {\bf 79}, 2831 (1997).


\bibitem{ChemExp}B.\ H.\ Meier, F.\ Graf, and R.\ R.\ Ernst,  
J.\ Chem.\ Phys.\  {\bf 76}, 767 (1982); 
%
S.\ Nagaoka, T.\ Terao, F.\ Imashiro, A.\ Saika, and 
N.\ Hirota, J.\ Chem.\ Phys.\  {\bf 79}, 4694 (1983);
%
A.\ Stoeckli, A.\ Furrer, Ch.\ Schoenenberger, 
B.\ H.\ Meier, R.\ R.\ Ernst, and I.\ Anderson,  
Physica B  {\bf 136}, 161 (1986); 
%
A.\ St\"ockli, B.\ H.\ Meier, R.\ Kreis, R.\ Meyer, and 
R.\ R.\ Ernst, J.\ Chem.\ Phys.\   {\bf 93}, 1502 (1990); 
%
A.\ J.\ Horsewill, P.\ J.\ McDonald, and D.\ Vijayaraghavan,   
J.\ Chem.\ Phys.\   {\bf 100}, 1889 (1994);
%
A.\ J.\ Horsewill and A.\ Ikram,    
Physica B {\bf 226}, 202 (1996).

\bibitem{Doslic98}N.\ Do\v{s}li\'{c}, O.\ K\"uhn, J.\ Manz, and 
 K.\ Sundermann, J.\ Phys.\ Chem.\  A {\bf 102}, 9645 (1998).

\bibitem{Naundorf99}H.\ Naundorf, K.\ Sundermann, and O.\ K\"uhn, 
Chem.\ Phys.\  {\bf 240}, 163 (1999).


\bibitem{LouisellHaake73}W.\ H.\ Louisell, {\em Quantum Statistical Properties 
of Radiation}  (Wiley, New York, 1973); F.\ Haake, {\em Statistical treatment of open systems 
by generalized master equations}, in {\em Quantum Statistics in Optics and 
Solid-State Physics}, Vol.\ 66 of {\em Springer Tracts in Modern Physics}, 
ed.\ by G.\ H\"ohler (Springer, Berlin, 1973).

\bibitem{Kohler97} S.\ Kohler, Th.\ Dittrich, and P.\ H\"anggi, 
Phys.\ Rev.\ E {\bf 55}, 300 (1997).

\bibitem{SilbeyGroup}P.\ E.\ Parris and R.\ Silbey, 
 J.\ Chem.\ Phys.\  {\bf 83}, 5619 (1985);
%
D.\ R.\ Reichman and R.\ J.\ Silbey, J.\ Phys.\ Chem.\ {\bf 99}, 2777 (1995).

\bibitem{MorilloGroup}M.\ Morillo and R.\ I.\ Cukier, 
Phys.\ Rev.\ B {\bf 54}, 13962 (1996); 
%
M.\ Morillo, C.\ Denk, and R.\ I.\ Cukier, 
Chem.\ Phys.\  {\bf 212}, 157 (1996); 
%
R.\ I.\ Cukier, C.\ Denk, and M.\ Morillo, 
Chem.\ Phys.\  {\bf 217}, 179 (1997).

\bibitem{DekkerGroup}H.\ Dekker, 
 Phys.\ Rev.\ A {\bf 44}, 2314 (1991); 
%
Physica A {\bf 175}, 485 (1991); {\bf 176}, 220 (1991); {\bf 178}, 289 (1991); 
{\bf 179}, 81 (1991); {\bf 210}, 507 (E) (1994).


\bibitem{Larkin86}A.\ I.\ Larkin and Yu.\ N.\ Ovchinnikov, 
Zh.\ Eksp.\ Teor.\ Fiz.\ {\bf 91}, 318 (1986) [Sov.\ Phys.\ JETP {\bf 64}, 
185 (1986)].

\bibitem{Larkin88}A.\ I.\ Larkin, Yu.\ N.\ Ovchinnikov, and A.\ Schmid, 
Physica B {\bf 152}, 266 (1988).

\bibitem{Ovchi94}Yu.\ N.\ Ovchinnikov and A.\ Schmid, 
Phys.\ Rev.\ B {\bf 50}, 6332 (1994).

\bibitem{Silvestrini90}P.\ Silvestrini, Yu.\ N.\ Ovchinnikov, 
 and R.\ Cristiano, 
 Phys.\ Rev.\ B {\bf 41}, 7341 (1990).

\bibitem{Silvestrini96a}P.\ Silvestrini, B.\ Ruggiero, and Yu.\ N.\ 
 Ovchinnikov, 
Phys.\ Rev.\ B {\bf 53}, 67 (1996).

\bibitem{Silvestrini96b}P.\ Silvestrini, B.\ Ruggiero, and Yu.\ N.\ 
 Ovchinnikov, Phys.\ Rev.\ B {\bf 54}, 1246 (1996).

\bibitem{Vernon63}R.\ P.\ Feynman and F.\ L.\ Vernon, Jr.,
 Ann.\ Phys.\ (N.Y.) {\bf 24}, 118 (1963).

\bibitem{Harris65}D.\ O.\ Harris, G.\ G.\ Engerholm, and W.\ D.\ Gwinn, 
 J.\ Chem.\ Phys.\ {\bf 43}, 1515 (1965).

\bibitem{ThorwartPRE00} M.\ Thorwart, P.\ Reimann, and P.\ H\"anggi, 
 Phys.\ Rev.\ E {\bf 62}, 5808 (2000).

\bibitem{Winterstetter97}M.\ Winterstetter and U.\ Weiss, 
Chem.\ Phys.\  {\bf 217}, 155 (1997).

\bibitem{Egger94}R.\ Egger, C.\ H.\ Mak, and U.\ Weiss, 
 Phys.\ Rev.\ E  {\bf 50}, R655 (1994).
 
\bibitem{NumRec}The NAG Fortran Library,
 Mark: 18A, Implementation IBM RISC System/6000 AIX, Precision: Double 
 (NAG, Oxford, 1999); 
 W.\ H.\ Press, S.\ A.\ 
Teukolsky, W.\ T.\  Vetterling, and 
B.\ P.\ Flannery, {\em Numerical Recipes in FORTRAN}, 2nd ed. (Cambridge UP, 
Cambridge, 1992). 

\bibitem{Gradshteyn}I.\ S.\ Gradshteyn and I.\ M.\ Ryzhik, 
{\em Tables of Integrals, Series and Products} (Academic Press, London, 
1965).

\bibitem{Kubo85}R.\ Kubo, M.\ Toda, N.\ Hashitsume, {\em Statistical 
Physics II}, Vol.\ 31 of Springer Series in Solid-State Sciences 
(Springer, Berlin, 1985).

\bibitem{Zwerger87}W.\ Zwerger, Phys.\ Rev.\ B {\bf 35}, 4737 (1987).

\bibitem{Grabert98}H.\ Grabert, G.-L.\ Ingold, and B.\ Paul, 
Europhys.\ Lett.\ {\bf 44}, 360 (1998).


\bibitem{Volterra1}H.\ Brunner, P.\ J.\ van der Houwen, {\em The 
numerical solution of Volterra Equations}, CWI Monograph 3 (North Holland, 
Amsterdam, 1985); M.\ A.\ Goldberg (ed.), {\em Solution Methods for 
Integral Equations} (Plenum Press, New York, 1979).

\bibitem{Kohler00}S.\ Kohler, R.\ Utermann, P.\ H\"anggi, and Th.\ Dittrich, 
Phys.\ Rev.\ E {\bf 58}, 7219 (1998); 
P.\ H\"anggi, S.\ Kohler, and Th.\ Dittrich, in 
{\em Statistical and Dynamical Aspects of Mesoscopic Systems\/}, 
Lecture Notes in Physics, Vol.\ 547, 
 D.\ Reguera, G.\ Platero, L.\ L.\ Bonilla, J.\ M.\ Rub\'{\i} (eds.) 
 (Springer, Berlin, 2000).

\end{thebibliography}
\end{document}